\documentclass[aps,prb]{revtex4}
\usepackage{amsmath,amssymb}
\usepackage{graphicx}
\usepackage{dcolumn}
\usepackage{bm}
\usepackage{color}
\usepackage{relsize}
\newcommand{\mbb}{\mathbb}
\newcommand{\mc}{\mathcal}

\newcommand{\tet}{\texttt}
\newcommand{\pr}{\partial}

\begin{document}
\title{Exploring non-trivial band structure, Berry curvature, spin polarizations and spin currents in $d$-wave altermagnets tailored by anisotropic optical fields}
\author{
Andrii Iurov$^{1}$\footnote{E-mail contact: aiurov@mec.cuny.edu, theorist.physics@gmail.com},
Liubov Zhemchuzhna$^{1,2,3}$\footnote{E-mail contact: lzhemchuzhna@mec.cuny.edu, lzhemchuzhna@fordham.edu},
Tiyhearah Danner-Jackson$^{1}$
}

\affiliation{
$^{1}$Department of Physics and Computer Science, Medgar Evers College of City University of New York, Brooklyn, NY 11225, USA\\ 
$^{2}$Department of Physics \& Engineering Physics, Fordham University, Bronx, NY 10458, USA\\
$^{3}$Department of Physics and Astronomy, Hunter College of the City University of New York, 695 Park Avenue, New York, New York 10065, USA\\ 
}
\date{\today}

\begin{abstract}
The subject of the present paper is a detailed theoretical investigation of the energy spectrum and bandgaps, as well as topological and collective properties and linear response, in $d$-wave altermagnets in the presence of an off-resonance optical dressing field. We consider the altermagnets with both $d_{x^2-y^2}$ and $d_{xy}$ pairing symmetries and focus on anisotropic dressing fields applied to an anisotropic and non-linear electron Hamiltonian. We have uncovered several crucial properties of the resulting electron-dressed states; specifically, we found that a finite bandgap is opened by linearly polarized irradiation, a phenomenon not observed in Dirac materials. Some of the crucial properties of the electron dressed states in the presence of the linearly polarized light can be uncovered only in the second-order perturbation expansion, which is often omitted. We found that introducing an anisotropic driving field leads to several subtle yet important changes in the Edelstein susceptibilities of altermagents, enabling the fine-tuning of their spin polarizations. We calculate the Berry curvature for various types of altermagnets and obtain closed-form analytical expressions for circularly polarized irradiation. We demonstrate that the optical driving field can generate finite Berry curvature in the absence of altermagnetic order. All these results are expected to become a crucial contribution to the rapidly developing fields of spintronics and device physics. 
\end{abstract}
\maketitle

\section{Introduction} 
\label{sec1}

Altermagnetism is a recently discovered magnetic phase (a distinct state of matter) that is neither ferromagnetic nor conventional antiferromagnetic, yet combines the most crucial features of both.\,\cite{vsmejkal2020crystal, negi2025mnte, vsmejkal2022beyond, tamang2025altermagnetism,  liu2026symmetry,fender2025altermagnetism, herasymchuk2025electric} Specifically, its net magnetic moment is zero, yet the electronic band structure shows spin splitting -- a property which is usually observed in ferromagnets. \,\cite{song2025altermagnets, tamang2025altermagnetism} Thus, altermagnetism introduces a revolutionary paradigm in magnetic materials by breaking the traditional divisions and merging characteristics traditionally attributed to all the known types of magnetic ordering. The magnetic moments of altermagnets on different sublattices are antiparallel, which leads to the cancellation of the net magnetization. Still, the crystal symmetry is such that the electronic states on the two sublattices are not equivalent, producing momentum-dependent spin polarization\,\cite{gomonay2024structure} (see also Fig. 1 ($a$)). This leads to large spin splitting of the electron energy subbands without net magnetization and to efficient spin-resolved transport, which could result in potential applications in low-power spintronics, spin control, and high-temperature devices. The spin splitting of the band structure of the altermagnets is symmetry-protected, which demonstrates a direct relation to the topological phases of the electronic structure. The spin-split band structure without spin-orbit coupling on the basis of microscopic multipole moments outlined the preliminary versions of the altermagnetism model. \,\cite{hayami2019momentum,hayami2020bottom}

\par
The exist a noticeable list of experimentally confirmed altermagnets,\cite{he2025evidence, tschirner2023saturation, zhou2025manipulation, choi2026exploring} specifically a layered intercalated transition-metal dichalcogenide, CoNb$_4S$e$_8$, KV$_2$Se$_2$O, RbV$_2$Te$_2$O \,\cite{fedchenko2024observation,lee2024broken,reimers2024direct} and thin films of manganese telluride, MnTe, \,\cite{amin2024nanoscale,negi2025mnte,gonzalez2023spontaneous} which have clearly exhibited altermagnetic band-structure properties. Researchers have also employed first-principles calculations, such as density functional theory, to propose a wide variety of materials that, in principle, should exhibit alternamagnetism, including FeS-based semiconductors\,\cite{tamang2025altermagnetism}. Importantly, several ordinary materials (oxides, pnictides, etc.) actually demonstrate altermagnetic symmetry, but we only recently realized it. This opens a large library of materials for new technological applications, and gaining new knowledge about their electronic and magnetic structure is crucial. Also, CrSb has been predicted and increasingly studied as an altermagnetic semiconductor.\,\cite{chen2025exposing, yu2025neel}
\par 

Spin–orbit coupling (SOC) \,\cite{galitski2013spin,usman2026resonant,pandita2026magnetotransport,Mojarro2025MajoranaKagome} plays an important role in altermagnets since the spin splitting arises from crystal and magnetic symmetries. Often induced by external electrostatic voltage gate,  it could modify and  hybridize spin texture, as well as enable anomalous Hall responses, spin–orbital textures and  related phenomena.\,\cite{naka2025altermagnetic,cheong2025altermagnetism,wang2025spin,roig2025quasisymmetry} Also, there are topological spin persistance currents,\,\cite{islam2023properties,sarkar2025spin,kapri2025spin} orbital magnetization \,\cite{tamang2023orbital} related to spin-orbit coupling.

\par
The unusual types of spin polarizations discussed above have significant implications for superconductivity.\,\cite{mazin2023altermagnetism, sato2017topological, mazin2023altermagnetism, leraand2025phonon, rasmussen2025inherent} Altermagnetic order may enable directional singlet pairing, generating spin mixtures, which allow topological superconducting phases due to the coupling between symmetry-protected spin structures. In altermagnetic systems, superconductivity coexists with other types of pairing states. This  superconductivity could be very stable since altermagnets do not exhibit a net macroscopic magnetization.\,\cite{fukaya2025superconducting, monkman2026persistent, hadjipaschalis2025majoranas}  
		
\medskip 
Floquet engineering is a dynamical, tunable, and controllable modification of the key electronic properties -- mostly band structure and the corresponding band gaps -- in various two-dimensional materials and electronic surface states.\,\cite{Oka2009PhotovoltaicHall, oka2019floquet, topp2019topological,Mojarro2020FloquetKekule}. It is achieved by applying a high-frequency, periodic optical driving field.\,\cite{goldman2014periodically} In contrast to strong, ionizing irradiation, we consider high-frequency dressing with energies (frequencies) well above any characteristic energy of the system, such as the Fermi energy. Theoretically, these new bound electron-photon states, or quasiparticles, are considered a very new kind of field-matter structures and are modeled by time-periodic perturbations of the coherent dressing. 

\par
Once the dressing field is applied, the band structure of the considered materials is modified in a non-trivial way.\,\cite{mojarro2021optical} In principle,  the results of applying the driving field substantially depend on its polarization. The most well-known result is opening a sizable bandgap by circularly polarized light \,\cite{iurov2020klein, kibis2010metal, mojarro2020dynamical, ibarra2019dynamical, iurov2024floquet,Tamang2021Floquet}
A linearly polarized light leads to the creation or modification of existing anisotropy \,\cite{iurov2017exploring, zhou2023floquet} and asymmetric Klein tunneling \,\cite{roslyak2010unimpeded, iurov2013photon}  in various materials with the Dirac cone. All essential transport properties, such as inverse realization time and components of the Fermi velocities of a material, are changed \,\cite{iurov2020quantum,islam2018driven,iurov2022optically,kristinsson2016control} Importantly, plasmons and other collective effects \,\cite{iurov2017controlling, horing2016low, ross2025dynamical} could be also substantially modified by the optical driving field\,\cite{constant2016all} Applying external dressing filled with different polarizations could also substantially modify the RKKY  interaction \,\cite{oriekhov2020rkky} in altermagnets.\,\cite{yarmohammadi2025anisotropic}

\par 
Most of the crucial aspects of Floquet theory and Floquet engineering for graphene and other Dirac materials have been successfully experimentally realized, including both the induction of non-equilibrium band structures and the emergence of topological properties. The most crucial and widely used methods for the experimental investigation of Floquet states are pump–probe and time- and angle-resolved photoemission spectroscopy (tr-ARPES). These experiments demonstrated most crucial results in the theory of electron dressed states, including light-induced band renormalization and the opening of dynamical gaps at the Dirac points\,\cite{wang2026observation, merboldt2025observation}, Floquet band crossings, accompanied by coherent Floquet sidebands\,\cite{Chen2026few,  wang2013observation} A detailed comparison between the theoretical and experimental results are provided in Ref.~[\onlinecite{sentef2015theory}]. Recently, ultrafast optical driving was applied to engineer transient Berry curvature and anomalous Hall responses.\,\cite{hubener2017creating,  beaulieu2024berry}

\par 
Apart from changes in the energy spectrum and band gaps, optical driving fields have become powerful and widely used tools for inducing and controlling topological phases in graphene and related materials. Opening a gap results in breaking the time-reversal symmetry and creating a Haldane-like Chern insulator with light-tunable band topology \cite{lindner2011floquet,oka2009floquet}, higher-order topology, and dynamical band inversion in transition metal dichalcogenides \,\cite{zeng2015optical, choe2016understanding}. Periodic driving provides a versatile nonequilibrium tool for designing topological band structures \cite{cayssol2013floquet} and many-body phase transitions in low-dimensional materials under such a dressing field.\,\cite{iurov2019peculiar,dey2018photoinduced, tamang2023probing, paul2026emergent,iurov2022floquet}. It also modifies the topological properties of its Fermi surface, known as the Lifshitz phase transition.\,\cite{iorsh2024floquet, mohan2018interplay}

\par 
Finally, optical dressing field could be applied to generate desirable structures of spin polarization, spin texture, topology, and spin-orbit coupling \,\cite{ghorashi2025dynamical, fu2026floquet, liu2026light, zhu2025floquet}, RKKY interaction \cite{yarmohammadi2025anisotropic,li2026rkky, ke2024floquet} in various low-dimensional lattices, including altermangets.  

\par 
\medskip 
The remaining part of the present paper is organized as follows: we begin with presenting a complete formalism for the low-energy Hamiltonian, electronic states, and the corresponding energy spectrum of altermagnets with $d_{x^2-y^2}$ and $d_{xy}$ pairing symmetries in Section \ref{sec2}. We also discuss the proposed experimental setup for the two-dimensional altermagnetic material with gate-voltage-induced spin-orbit coupling and an external off-resonance high-frequency dressing field. Section \ref{sec3} is concerned with presenting our results for the electron dressed states for altermagnet exposed to various types of an isotropic dressing field with $\beta \neq 1$. We discuss in detail all the novel properties in contrast to the previously considered circularly polarized light. The detailed derivations of all these electronic states, energy spectrum, and band gaps are presented in Appendix \ref{apb}. Next, we present detailed numerical results for different types of energy dispersions, along with their constant-energy cuts (showing the angular dependence of the obtained spectrum), in Section \ref{sec4}. Finally, in section \ref{sec5}, we provide the final concluding remarks, emphasizing the significance of our findings. The research outlook and potential future directions are discussed. A comprehensive, unified expression for calculating energy dispersions for all types of spin-1/2 Hamiltonians, which is also applicable to qubit states together with the derivation of the Berry curvature, is presented and explained in Appendix \ref{apa}.

\section{General formalism for altermagnetic electronic states: Hamiltonian, dispersions and spin polarization}
\label{sec2}

\begin{figure} 
\centering
\includegraphics[width=0.49\textwidth]{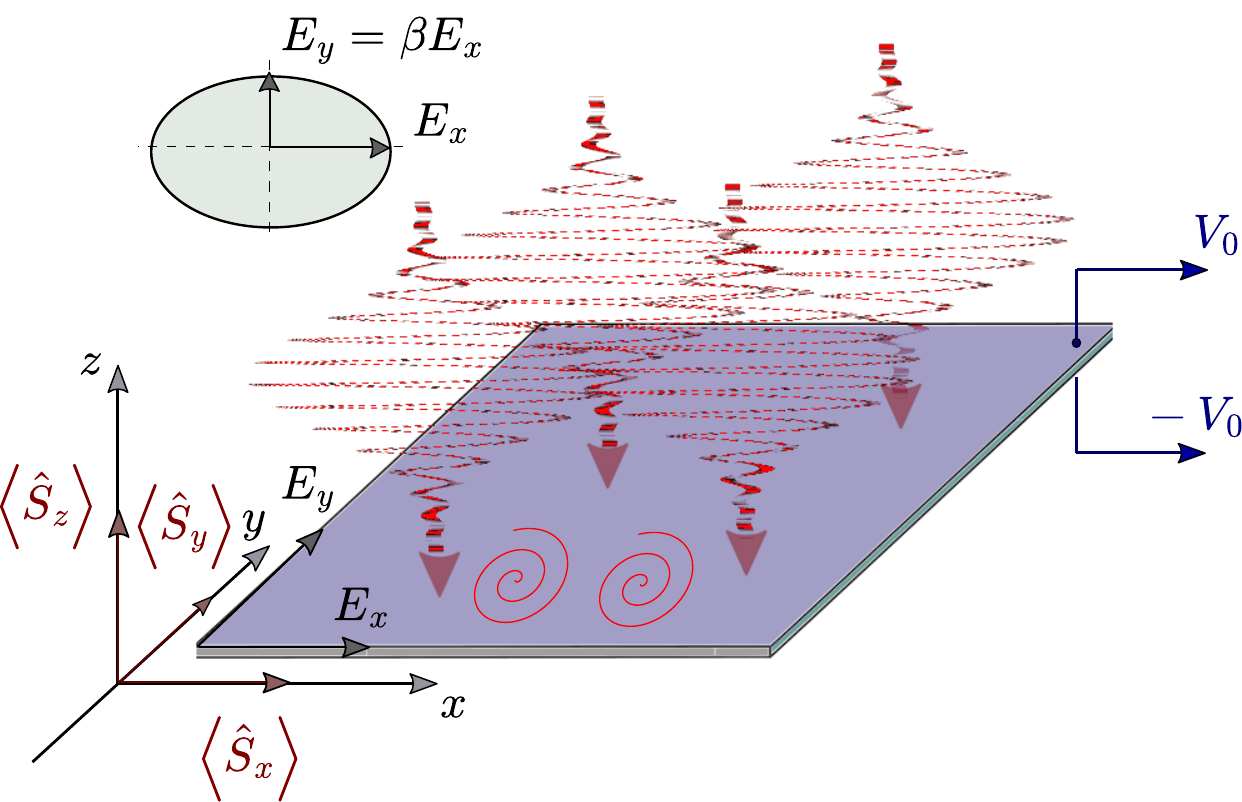}
\caption{(Color online) 
The schematics for a setup of a two-dimensional $d$-wave altermagnetic material in the presence of an off-resonance dressing field with different polarizations (elliptical; circular, and linear, as limiting cases). The electrostatic gating provided by the two electrodes enables Rashba spin-orbit coupling.}
\label{FIG:1}
\end{figure}

We begin by considering a $d$-wave altermagnet with spin-orbit coupling, which is also exposed to an off-resonance optical driving field. The schematics of a possible experimental setup for the considered system are presented in Fig. \ref{FIG:1}. Here, the spin-orbit interaction and the inversion symmetry breaking appear as a result of the applied gate voltage. 

\medskip 
The considered electronic and spin structure is effectively described by a two-band model with the following Hamiltonian 
\begin{equation}
\label{mainHam}
\hat{\mc{H}}_{1,2}^{(0)}(\vec{\bf k}) = \frac{(\hbar k)^2}{2 m} \, \hat{\Sigma}_0^{(2)} + \mc{H}_{1,2}^{(A)}(\vec{\bf k}) \,  \hat{\Sigma}_z^{(2)} + r_{SO} \,
\left( k_x \,  \Sigma_y^{(2)} - k_y \,  \Sigma_x^{(2)} \right) \, , 
\end{equation}
Here, are the $\hat{\Sigma}_{0,x,y,z}^{(2)}$ are $2 \times 2$ Pauli matrices (specifically, $\hat{\Sigma}_0^{(2)}$ is the $2 \times 2$ unit matrix which corresponds to an energy shift), $\vec{\bf k} = (k_x,k_y)$ is the  two-dimensional electron wave vector and $\Theta_{\vec{\bf k}} =\arctan (k_y/k_x)$ is the angle associated with the wave vector; $m$ is the electron mass. The strength of the gate-induced Rashba spin-orbit coupling is given by $r_{SO}$. 

\par 
Following Refs.~[\onlinecite{Hayami2023TRSB}] and [\onlinecite{yarmohammadi2026spin}], we distinguish two types $\mc{H}_{1,2}$ of altermagnets corresponding to $d_{x^2 - y^2}$ and $d_{xy}$ pairing symmetry (or wave symmetry) according to their $\hat{\Sigma}_z^{(2)}$-term:

\begin{equation}
\label{altt1}
\mc{H}_{1}^{(A)}(\vec{\bf k}) = \frac{\hbar^2 \mathfrak{A}_1}{2 m}  \, \left( k_x^2 - k_y^2 \right) 
\end{equation}
and 

\begin{equation}
\label{altt2}
\mc{H}_{2}^{(A)}(\vec{\bf k}) = \frac{\hbar^2 \mathfrak{A}_2}{2 m}  \, k_x k_y \, . 
\end{equation}

The $d_{x^2-y^2}$-type altermagnets are viewed as showing more significant altermagnetic order, but we will consider both of them. Specifically, we want to know how the energy dispersions of the two types of altermagnets are modified in the presence of an optical driving field.  

\par 

We could in principle obtain one type of altermagnets from the others by performing a combined rotation in both momentum and spin space. The two parameters $\mathfrak{A}_1$ and $\mathfrak{A}_2$ represent the strengths of altermagnetic order for both types of pairing symmetries. Obviously, altermagnetism would not exist if $\mathfrak{A}_1 \longrightarrow 0$  and $\mathfrak{A}_2 \longrightarrow 0$. Normally, we assume $0 < \mathfrak{A}_1 \leq 1$ and $\mathfrak{A}_1 < 1$. Here, we will consider both symmetry types and investigate their optically induced energy spectra.  

\medskip

\begin{figure} 
\centering
\includegraphics[width=0.49\textwidth]{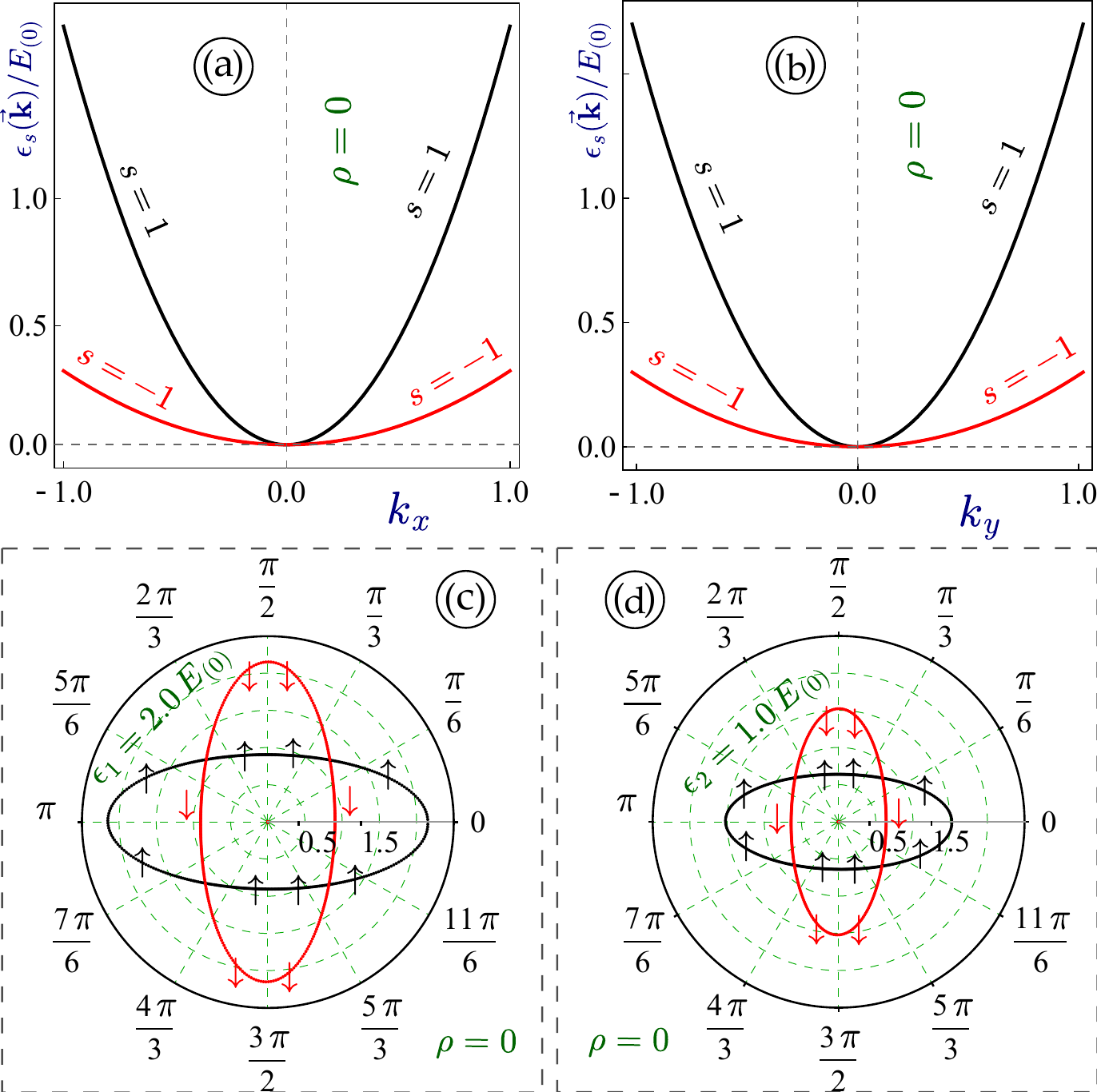}
\caption{(Color online). Energy spectrum of a $d$-wave altermagnet with $d_{x^2-y^2}$ pairing symmetry in the absence of the spin-orbit coupling ($\rho=0$). Panels $(a)$ and $(b)$ demonstrate the spin-resolved ($s=\pm1$) anisotropic energy dispersions $\epsilon_s(\vec{\bf k})$ obtained from Eq.~\eqref{genen1} as the functions of the $k_x$- and $k_y$-  components of the wave vector $\vec{\bf k}$. The black and red lines correspond to the values of spin index $s=\pm 1$. Plots $(c)$ and $(d)$  represent the constant-energy cuts corresponding to $\epsilon_1 = 2.0\,E_{(0)}$ and $\epsilon_1 = 1.0\, E_{(0)}$. Here, the red and black curves are related to the positive and negative out-of-plane spin polarizations $\text{sign}\Big[ \Big\langle \hat{S}_z \Big\rangle \Big] = \text{sign}\Big[ \Big\langle \Psi_s(\vec{\bf k}) \Big\vert \hat{\Sigma}^{(2)}_z \Big\vert \Psi_s(\vec{\bf k}) \Big\rangle \,\Big] = \pm 1$,  which are in principle not equivalent to the values of the spin index $s=\pm 1$.}
\label{FIG:2}
\end{figure}

\begin{figure} 
\centering
\includegraphics[width=0.49\textwidth]{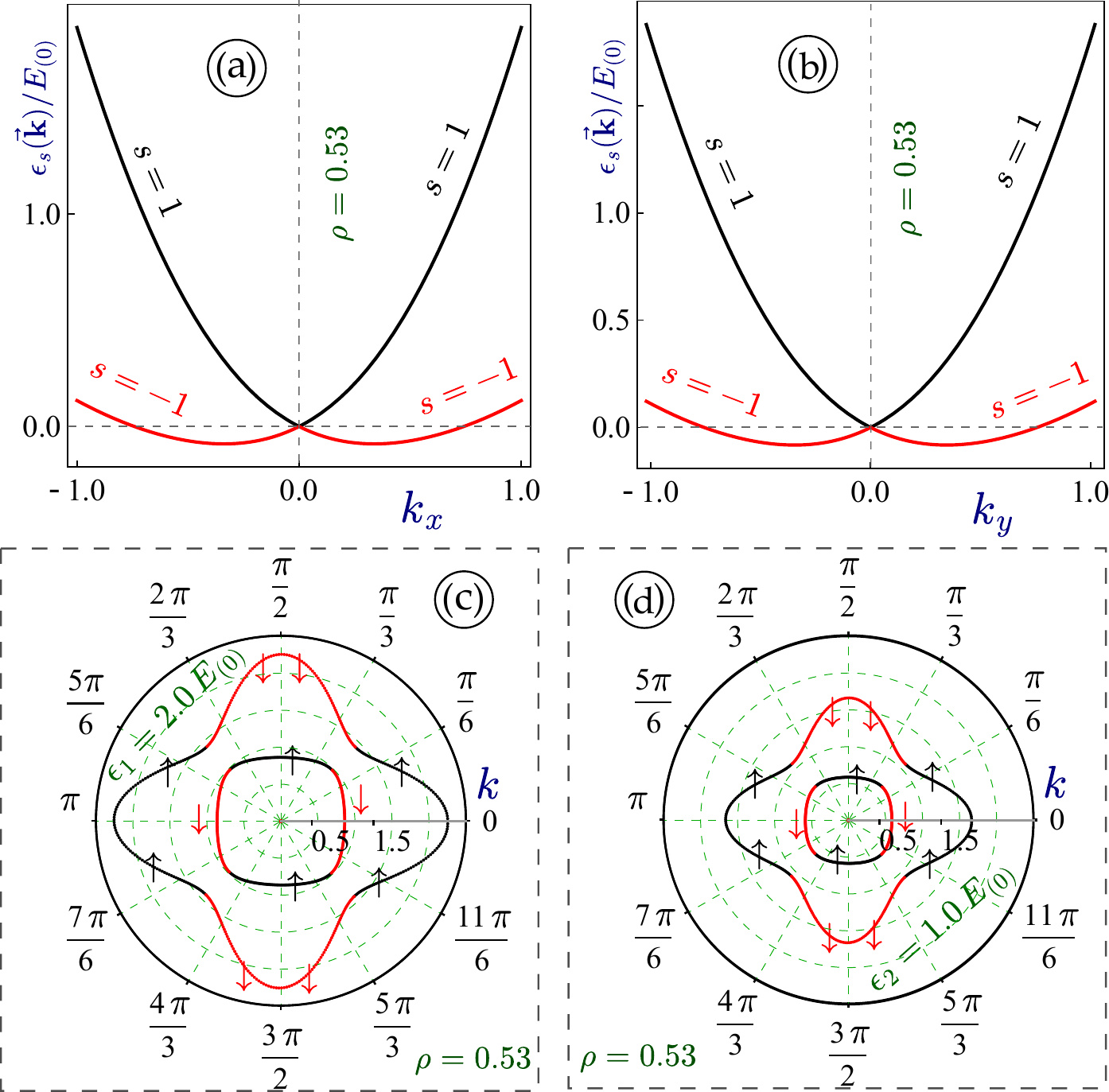}
\caption{(Color online) Energy spectrum of a $d$-wave altermagnet with $d_{x^2-y^2}$ pairing symmetry in the presence of the spin-orbit coupling ($\rho=0.53$). Panels $(a)$ and $(b)$ demonstrate the spin-resolved ($s=\pm1$) anisotropic energy dispersions $\epsilon_s(\vec{\bf k})$ obtained from Eq.~\eqref{genen1} as the functions of the $k_x$- and $k_y$-  components of the wave vector $\vec{\bf k}$. The black and red lines correspond to the values of spin index $s=\pm 1$. Plots $(c)$ and $(d)$  represent the constant-energy cuts corresponding to $\epsilon_1 = 2.0\,E_{(0)}$ and $\epsilon_1 = 1.0\, E_{(0)}$. Here, the red and black curves are related to the positive and negative out-of-plane spin polarizations $\text{sign}\Big[ \Big\langle \hat{S}_z \Big\rangle \Big] = \text{sign}\Big[ \Big\langle \Psi_s(\vec{\bf k}) \Big\vert \hat{\Sigma}^{(2)}_z \Big\vert \Psi_s(\vec{\bf k}) \Big\rangle \,\Big] = \pm 1$,  which are in principle not equivalent to the values of the spin index $s=\pm 1$.}
\label{FIG:3}
\end{figure}

\begin{figure} 
\centering
\includegraphics[width=0.49\textwidth]{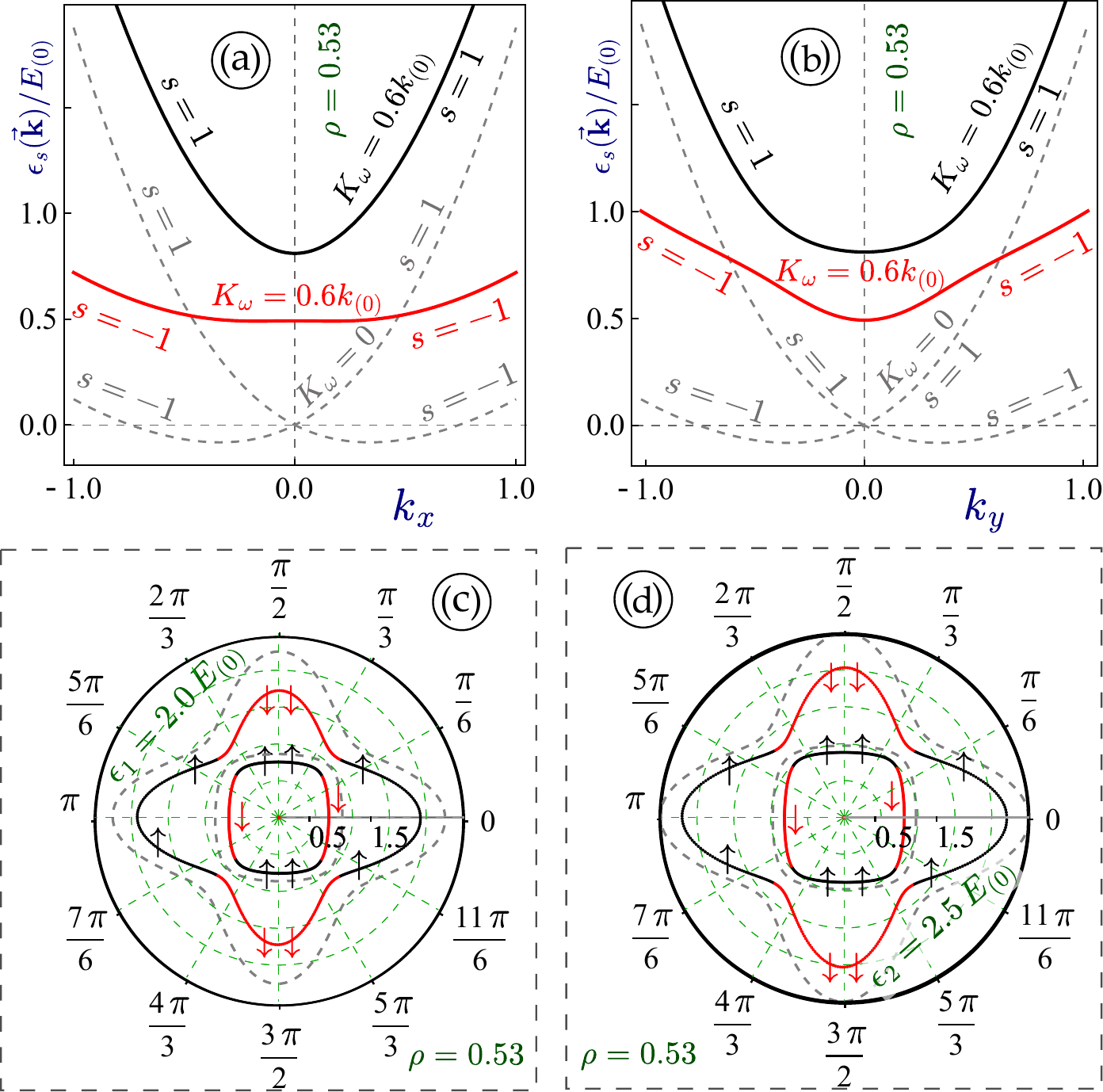}
\caption{(Color online) Energy spectrum of a $d$-wave altermagnet with $d_{x^2-y^2}$ pairing symmetry in the presence of the spin-orbit coupling ($\rho=0.53$) and an off-resonance dressing field with effective coupling parameter $\mc{K}_\omega = 0.6\,k_{(0)}$ and $\beta  = 0.9$ (nearly circular polarization). Panels $(a)$ and $(b)$ demonstrate the spin-resolved ($s=\pm1$) anisotropic energy dispersions $\epsilon_s(\vec{\bf k})$ obtained from Eq.~\eqref{genen1} as the functions of the $k_x$- and $k_y$-  components of the wave vector $\vec{\bf k}$. The black and red lines correspond to the values of spin index $s=\pm 1$. Plots $(c)$ and $(d)$  represent the constant-energy cuts corresponding to $\epsilon_1 = 2.0\,E_{(0)}$ and $\epsilon_1 = 1.0\, E_{(0)}$. Here, the red and black curves are related to the positive and negative out-of-plane spin polarizations $\text{sign}\Big[ \Big\langle \hat{S}_z \Big\rangle \Big] = \text{sign}\Big[ \Big\langle \Psi_s(\vec{\bf k}) \Big\vert \hat{\Sigma}^{(2)}_z \Big\vert \Psi_s(\vec{\bf k}) \Big\rangle \,\Big] = \pm 1$,  which are in principle not equivalent to the values of the spin index $s=\pm 1$.}
\label{FIG:4}
\end{figure}

\begin{figure} 
\centering
\includegraphics[width=0.49\textwidth]{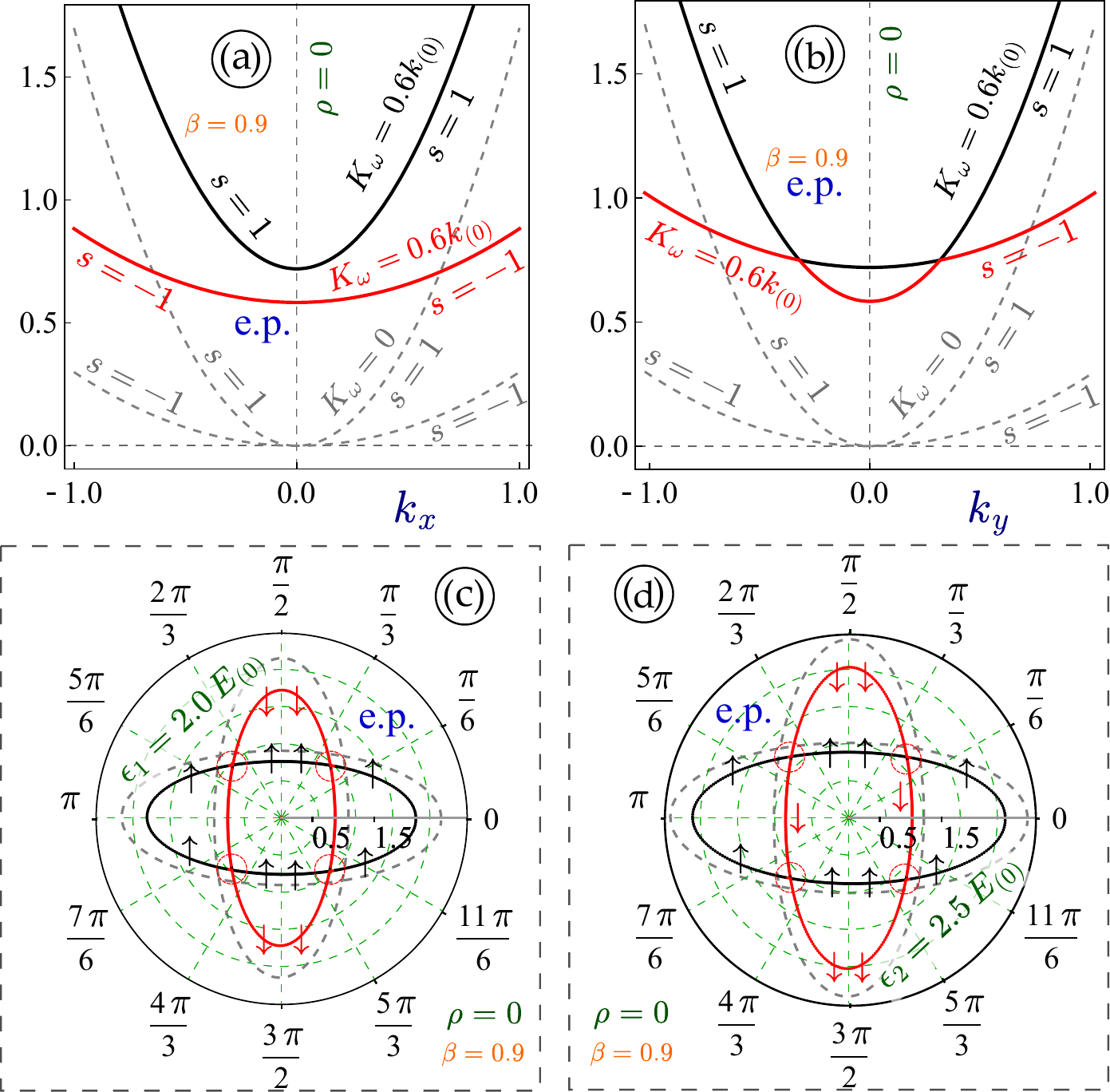}
\caption{(Color online) Energy spectrum of a $d$-wave altermagnet with $d_{x^2-y^2}$ pairing symmetry in the absence of the spin-orbit coupling ($\rho=0$) but exposed to an off-resonance dressing field with effective coupling parameter $\mc{K}_\omega = 0.6\,k_{(0)}$ and $\beta  = 0.9$ (nearly circular polarization). Panels $(a)$ and $(b)$ demonstrate the spin-resolved ($s=\pm1$) anisotropic energy dispersions $\epsilon_s(\vec{\bf k})$ obtained from Eq.~\eqref{genen1} as the functions of the $k_x$- and $k_y$-  components of the wave vector $\vec{\bf k}$. The black and red lines correspond to the values of spin index $s=\pm 1$. Plots $(c)$ and $(d)$  represent the constant-energy cuts corresponding to $\epsilon_1 = 2.0\,E_{(0)}$ and $\epsilon_1 = 1.0\, E_{(0)}$. Here, the red and black curves are related to the positive and negative out-of-plane spin polarizations $\text{sign}\Big[ \Big\langle \hat{S}_z \Big\rangle \Big] = \text{sign}\Big[ \Big\langle \Psi_s(\vec{\bf k}) \Big\vert \hat{\Sigma}^{(2)}_z \Big\vert \Psi_s(\vec{\bf k}) \Big\rangle \,\Big] = \pm 1$,  which are in principle not equivalent to the values of the spin index $s=\pm 1$.}
\label{FIG:5}
\end{figure}

\begin{figure} 
\centering
\includegraphics[width=0.49\textwidth]{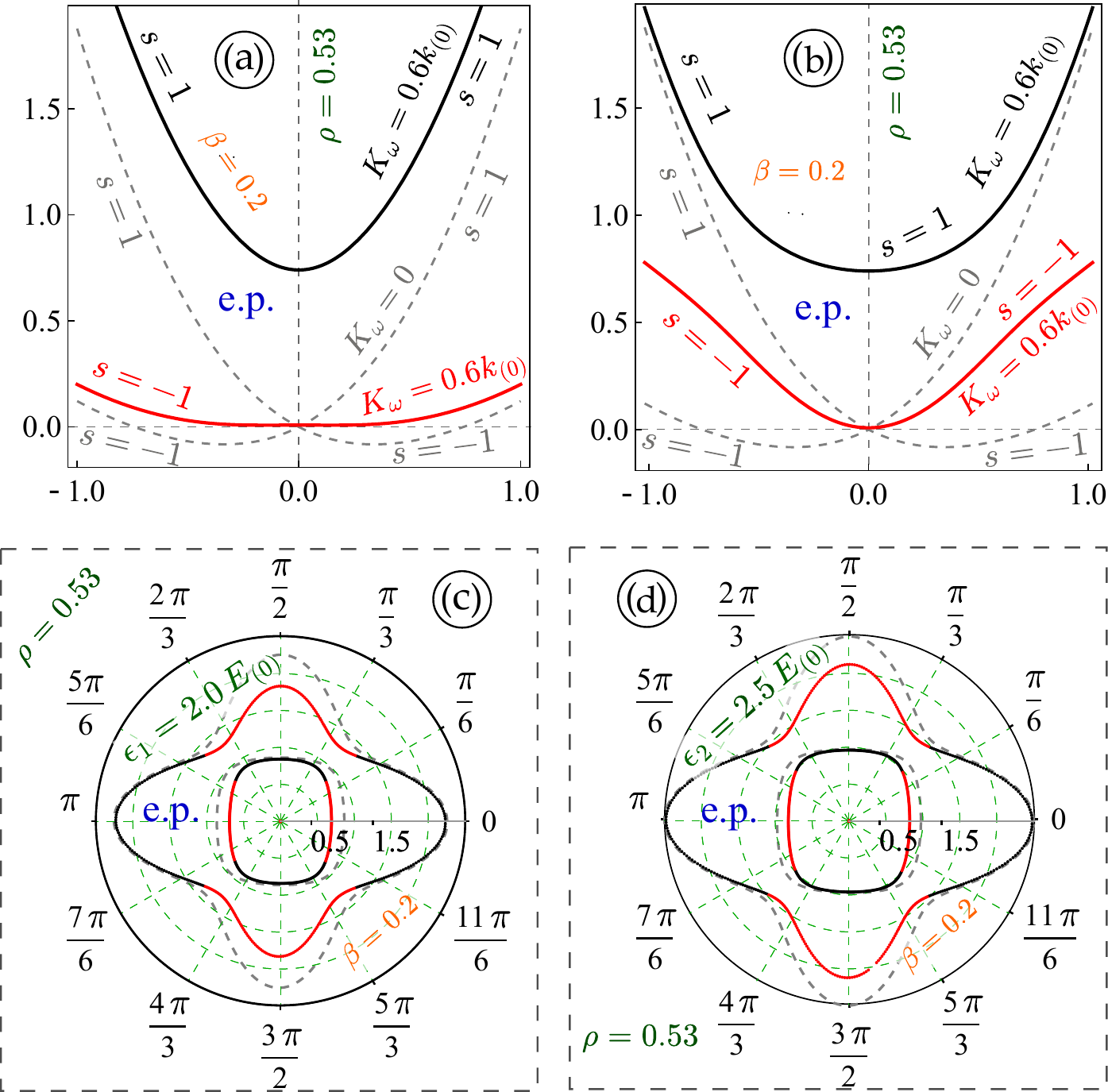}
\caption{(Color online) Energy spectrum of a $d$-wave altermagnet with $d_{x^2-y^2}$ pairing symmetry in the presence of the spin-orbit coupling ($\rho=0.53$) and an off-resonance dressing field with effective coupling parameter $\mc{K}_\omega = 0.6\,k_{(0)}$ and $\beta  = 0.2$ (close to the linear polarization). Panels $(a)$ and $(b)$ demonstrate the spin-resolved ($s=\pm1$) anisotropic energy dispersions $\epsilon_s(\vec{\bf k})$ obtained from Eq.~\eqref{genen1} as the functions of the $k_x$- and $k_y$-components of the wave vector $\vec{\bf k}$. The black and red lines correspond to the values of spin index $s=\pm 1$. Plots $(c)$ and $(d)$  represent the constant-energy cuts corresponding to $\epsilon_1 = 2.0\,E_{(0)}$ and $\epsilon_1 = 1.0\, E_{(0)}$. Here, the red and black curves are related to the positive and negative out-of-plane spin polarizations $\text{sign}\Big[ \Big\langle \hat{S}_z \Big\rangle \Big] = \text{sign}\Big[ \Big\langle \Psi_s(\vec{\bf k}) \Big\vert \hat{\Sigma}^{(2)}_z \Big\vert \Psi_s(\vec{\bf k}) \Big\rangle \,\Big] = \pm 1$,  which are in principle not equivalent to the values of the spin index $s=\pm 1$.}
\label{FIG:6}
\end{figure}

\begin{figure} 
\centering
\includegraphics[width=0.49\textwidth]{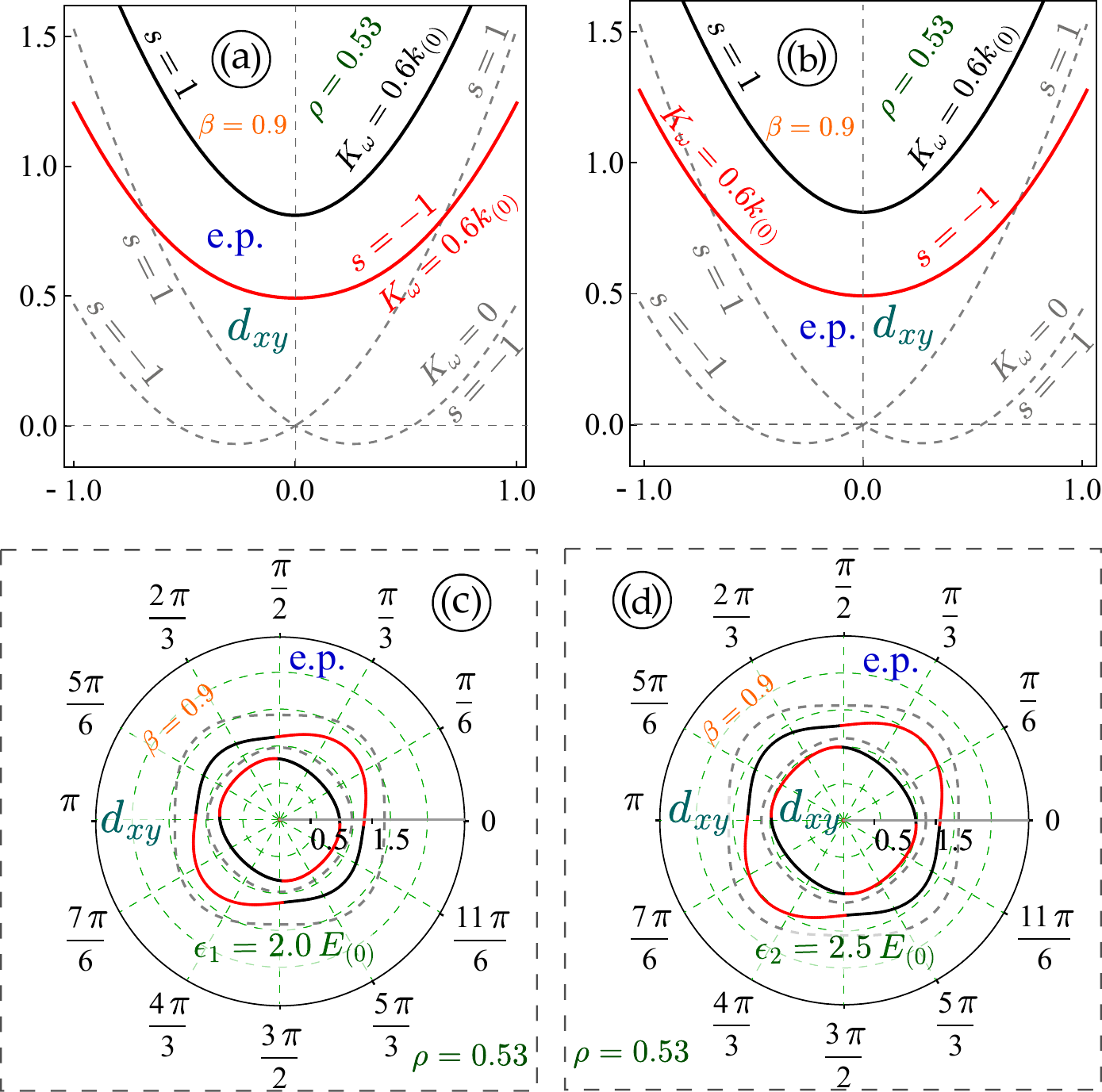}
\caption{(Color online)  Energy spectrum of a $d$-wave altermagnet with $d_{xy}$ pairing symmetry in the presence of the spin-orbit coupling ($\rho=0.53$) and an off-resonance dressing field with effective coupling parameter $\mc{K}_\omega = 0.6\,k_{(0)}$ and $\beta  = 0.9$ (nearly circular polarization). Panels $(a)$ and $(b)$ demonstrate the spin-resolved ($s=\pm1$) anisotropic energy dispersions $\epsilon_s(\vec{\bf k})$ obtained from Eq.~\eqref{genen1} as the functions of the $k_x$- and $k_y$-  components of the wave vector $\vec{\bf k}$. The black and red lines correspond to the values of spin index $s=\pm 1$. Plots $(c)$ and $(d)$  represent the constant-energy cuts corresponding to $\epsilon_1 = 2.0\,E_{(0)}$ and $\epsilon_1 = 1.0\, E_{(0)}$. Here, the red and black curves are related to the positive and negative out-of-plane spin polarizations $\text{sign}\Big[ \Big\langle \hat{S}_z \Big\rangle \Big] = \text{sign}\Big[ \Big\langle \Psi_s(\vec{\bf k}) \Big\vert \hat{\Sigma}^{(2)}_z \Big\vert \Psi_s(\vec{\bf k}) \Big\rangle \,\Big] = \pm 1$,  which are in principle not equivalent to the values of the spin index $s=\pm 1$.}
\label{FIG:7}
\end{figure}

\begin{figure} 
\centering
\includegraphics[width=0.49\textwidth]{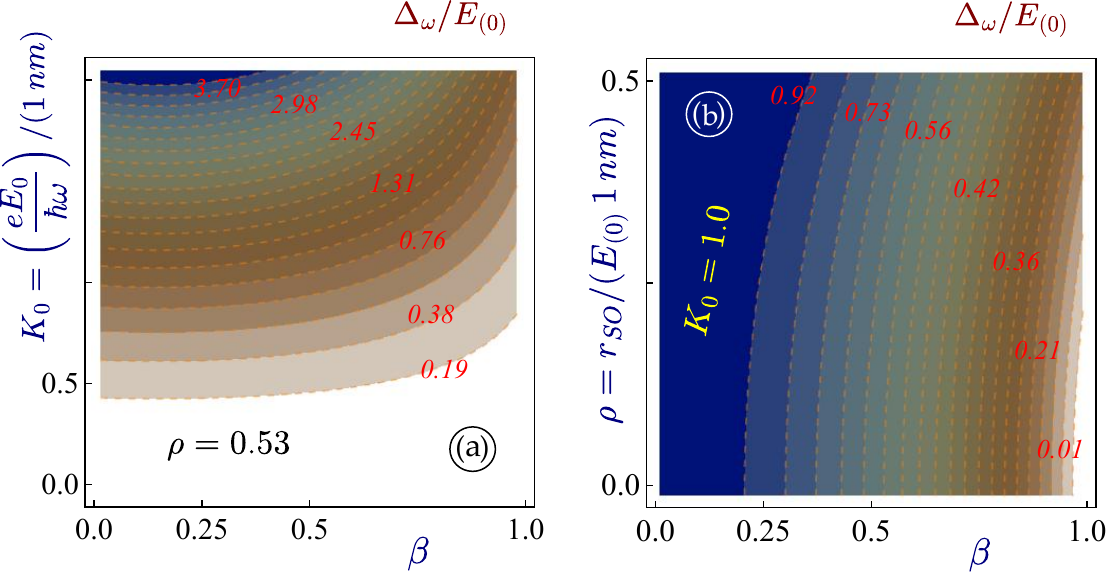}
\caption{(Color online) The energy bandgap $\Delta_0(\rho, \beta \, \vert \, \mc{K}_\omega, \omega)$ between the two spin-resolved bands ($s=\pm1$) in a $d$-wave altermagnet with $d_{x^2-y^2}$ pairing symmetry in the presence of the spin-orbit coupling and an off-resonance dressing field of different polarizations. Panel $(a)$ demonstrates a contour plot of the bandgap $\Delta_0$ as a function of  ratio of field strengths along the two axes of the polarization ellipse $beta$ and effective coupling parameter $\mc{K}_\omega$. Plot $(b)$ represents the bandgap $\Delta_0$ as a function of parameter $\beta$ (the ratio of field strengths along the two axes of the polarization ellipse) and the strength of the spin orbit coupling $\rho = r_{SO}/k_{(0)}$, as labeled.}
\label{FIG:8}
\end{figure}

\begin{figure} 
\centering
\includegraphics[width=0.49\textwidth]{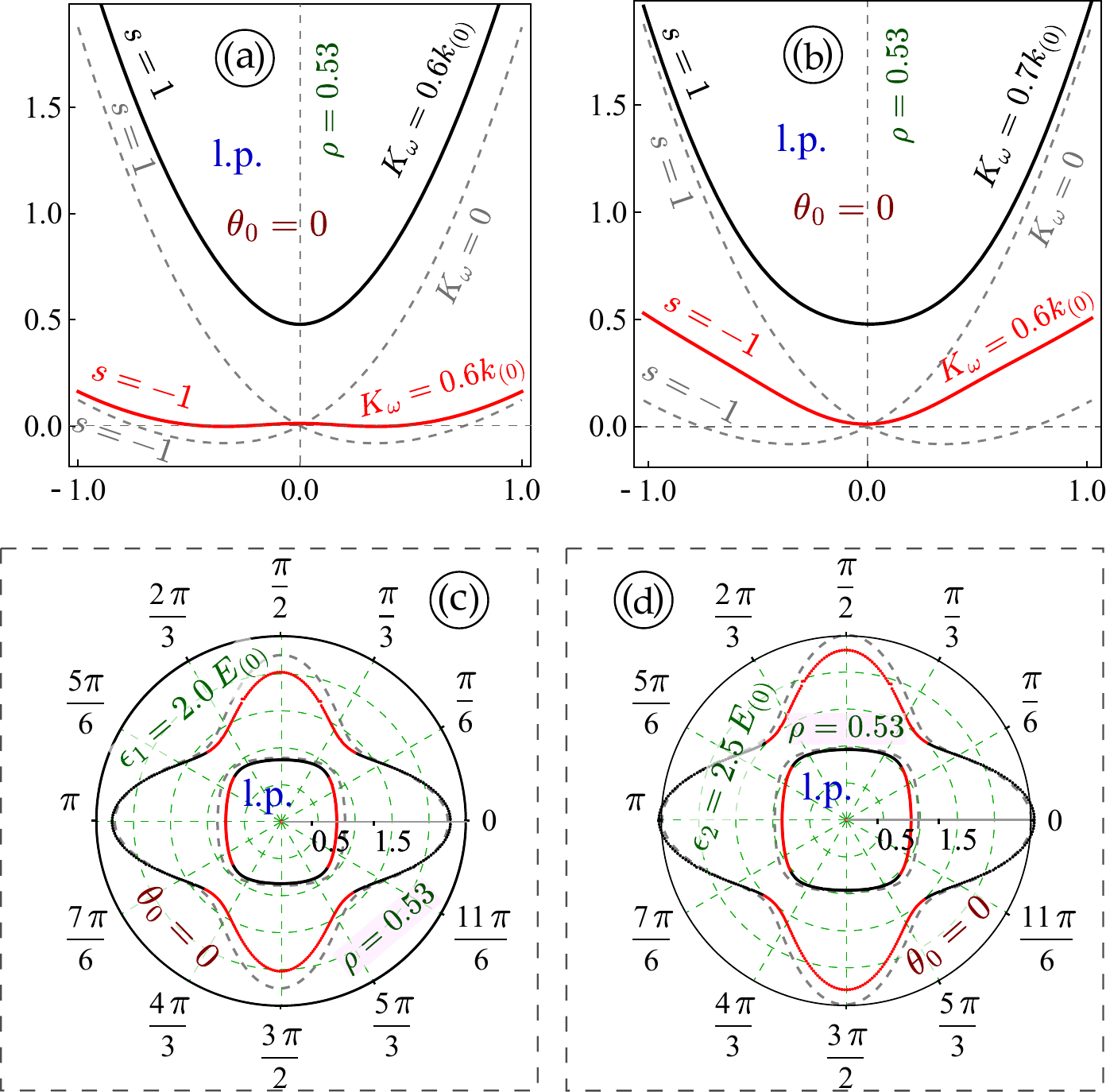}
\caption{(Color online)  Energy spectrum of a $d$-wave altermagnet with $d_{x^2-y^2}$ pairing symmetry in the presence of the spin-orbit coupling ($\rho=0.53$) and an off-resonance {\it linearly polarized} ($\beta  = 0$) dressing field with effective coupling parameter $\mc{K}_\omega = 0.6\,k_{(0)}$. The direction of the linear polarization of the imposed radiation is along the $x-$axis ($\theta_0 =0$). Panels $(a)$ and $(b)$ demonstrate the spin-resolved ($s=\pm1$) anisotropic energy dispersions $\epsilon_s(\vec{\bf k})$ obtained from Eq.~\eqref{genen1} as the functions of the $k_x$- and $k_y$-components of the wave vector $\vec{\bf k}$. The black and red lines correspond to the values of spin index $s=\pm 1$. Plots $(c)$ and $(d)$ represent the constant-energy cuts corresponding to $\epsilon_1 = 2.0\,E_{(0)}$ and $\epsilon_1 = 1.0\, E_{(0)}$. Here, the red and black curves are related to the positive and negative out-of-plane spin polarizations $\text{sign}\Big[ \Big\langle \hat{S}_z \Big\rangle \Big] = \text{sign}\Big[ \Big\langle \Psi_s(\vec{\bf k}) \Big\vert \hat{\Sigma}^{(2)}_z \Big\vert \Psi_s(\vec{\bf k}) \Big\rangle \,\Big] = \pm 1$,  which are in principle not equivalent to the values of the spin index $s=\pm 1$.}
\label{FIG:9}
\end{figure}

\begin{figure} 
\centering
\includegraphics[width=0.49\textwidth]{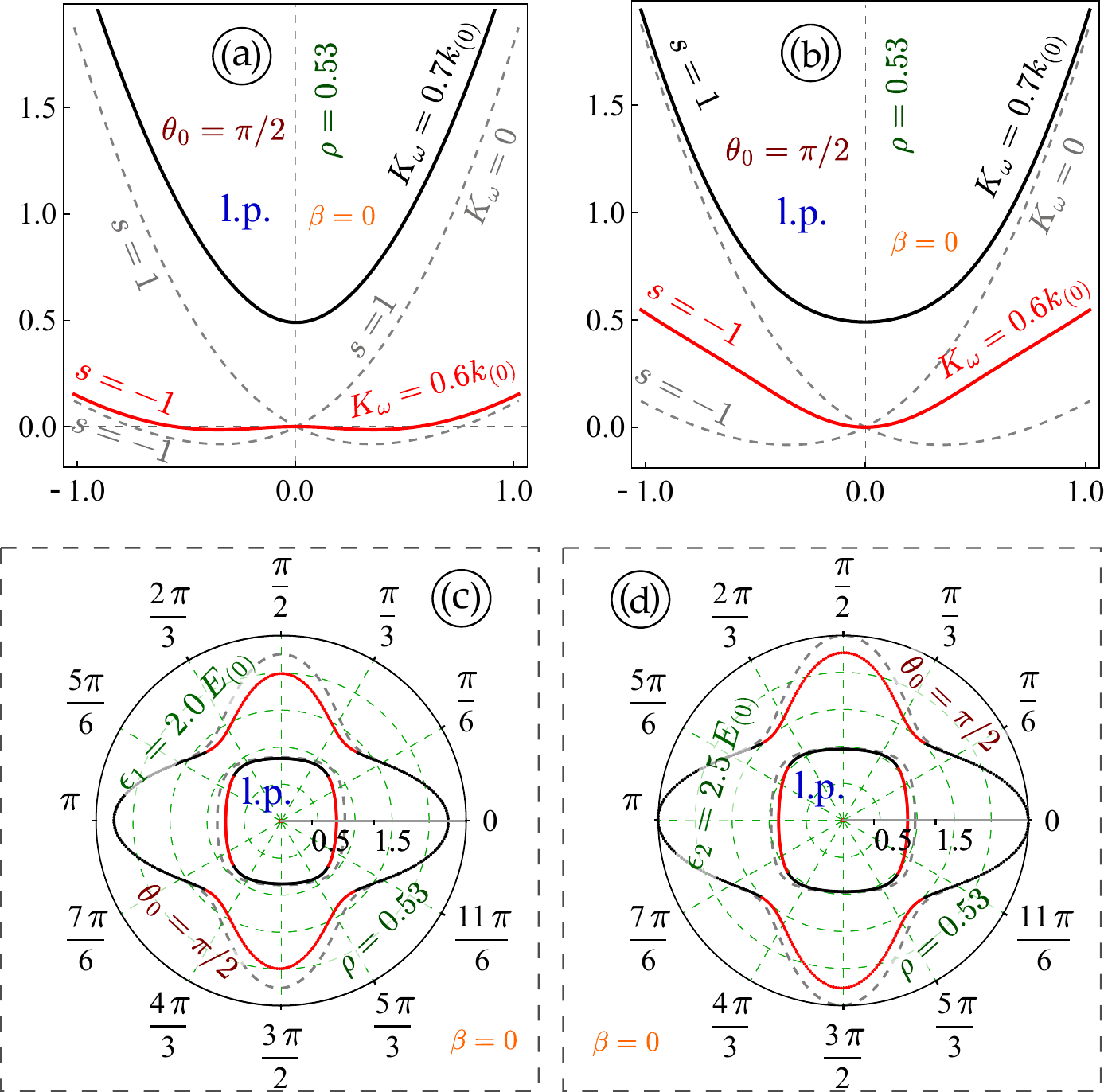}
\caption{(Color online)  Energy spectrum of a $d$-wave altermagnet with $d_{x^2-y^2}$ pairing symmetry in the presence of the spin-orbit coupling ($\rho=0.53$) and an off-resonance {\it linearly polarized} ($\beta  = 0$) dressing field with effective coupling parameter $\mc{K}_\omega = 0.6\,k_{(0)}$. The direction of the linear polarization of the imposed radiation is along the $y-$axis ($\theta_0 =\pi/2$). Panels $(a)$ and $(b)$ demonstrate the spin-resolved ($s=\pm1$) anisotropic energy dispersions $\epsilon_s(\vec{\bf k})$ obtainedfrom Eq.~\eqref{genen1} as the functions of the $k_x$- and $k_y$-components of the wave vector $\vec{\bf k}$. The black and red lines correspond to the values of spin index $s=\pm 1$. Plots $(c)$ and $(d)$  represent the constant-energy cuts corresponding to $\epsilon_1 = 2.0\,E_{(0)}$ and $\epsilon_1 = 1.0\, E_{(0)}$. Here, the red and black curves are related to the positive and negative out-of-plane spin polarizations $\text{sign}\Big[ \Big\langle \hat{S}_z \Big\rangle \Big] = \text{sign}\Big[ \Big\langle \Psi_s(\vec{\bf k}) \Big\vert \hat{\Sigma}^{(2)}_z \Big\vert \Psi_s(\vec{\bf k}) \Big\rangle \,\Big] = \pm 1$,  which are in principle not equivalent to the values of the spin index $s=\pm 1$.}
\label{FIG:10}
\end{figure}

The Rashba spin-orbit coupling term, no matter how small, leads to noticeable changes in the spin polarizations and spin structures of the reconsidered material. In the absence of the term,  we observe complete inversion symmetry $k_x \leftrightarrow -k_x$ and $k_y \leftrightarrow -k_y$. Since only $k_x^2$ and $k_y^2$ terms are present in the Hamiltonian \eqref{mainHam}, each component of the spin,  specifically, $z$, is concerned for each of their spin-resolved $s-$ dependent subbands, meaning that each of these sub bands exactly corresponds to either a positive or a negative direction of the out-of-plane spin component. The energy subbands could be clearly classified by this spin direction, which results in the fact that each of the subbands is shown as completely spin-down (red curves) or completely spin-up (black curves) in Fig.~\ref{FIG:2}. Apparently, we don't observe such a clear separation in Fig.~\ref{FIG:3} when spin-orbit coupling is present. Now the spin is coupled to the electron momentum (known as spin-momentum locking), and we observe the different directions of the out-of-plane spin component (black and red pieces) on a single energy band. 

\medskip 
Importantly, the energy dispersions corresponding to all types of spin-1/2 Hamiltonians, which definitely include our Eq.~\eqref{mainHam}, could be presented in a unified, general, and a very straightforward form. As demonstrated in Appendix \ref{apa}, one can easily compose the expression for the energy dispersions for both irradiated and non-irradiated materials without doing any additional calculations.  

\par
 We'll start with the general form of spin-1/2 Hamiltonian. Namely, any $2 \times 2$ matrix could be presented as a linear combination of all the Pauli matrices $\vec{\Sigma}^{(2)}_i$, $i=0,x,y,z$, including a unit Matrix $\hat{\Sigma}^{(2)}_0$

\begin{equation}
\label{PHam1}
\mc{H}_1^{(0)}(\vec{\bf k}) = \sum\limits_{i=1}^{3} \vec{\bf {V}}_i (\vec{\bf k}) \, \vec{\Sigma}^{(2)}_i = 
 V_0(\vec{\bf k})  \Sigma^{(2)}_0 + \sum\limits_{i=1}^{3} \vec{\bf {V}_i} (\vec{\bf k})  \,  \vec{\Sigma}^{(2)}_i \,  
\end{equation}

We introduce a vector $\vec{\bf {V}_i} (\vec{\bf k}) = \Big[ V_1(\vec{\bf k}), V_2(\vec{\bf k}), V_3(\vec{\bf k}) \Big]$ as 

\begin{equation}
\vec{\bf {V}}_i (\vec{\bf k})  = \frac{\hat{{\bf n}}(\vec{\bf k})}{\vert {\bf V}_i (\vec{\bf k}) \vert} 
 = \frac{\hat{{\bf n}}(\vec{\bf k})}{\sqrt{\sum\limits_{i=1}^{3}  {V}_i^2 (\vec{\bf k})}} \, . 
\end{equation} 

It could be easily verified that the corresponding energy dispersions are immediately presented as 

\begin{equation}
\label{genen1}
\varepsilon_{s=\pm 1}(\vec{\bf k}) = V_0(\vec{\bf k}) + s\,  \vert \vec{\bf V} (\vec{\bf k}) \vert = V_0(\vec{\bf k}) + 
s\, \sqrt{\sum\limits_{i=1}^{3}  {V}_i^2 (\vec{\bf k})} \, . 
\end{equation}

The corresponding wave functions of Hamiltonian \eqref{mainHam}, which are mathematically equivalent to the qubit states in quantum computing and quantum information processing, are also obtained in our Appendix \ref{apa}.

\section{Electron-photon dressed states and modified energy spectrum}
\label{sec3}

In this section, we aim to derive closed-form analytical expressions for the energy dispersions of the quasiparticles represented by the electron-photon dress-states in a two-dimensional $d$-wave altermagnet. Such hybrid states are obtained if a two-dimensional material is irradiated with an off-resonance dressing field with the frequency $\hbar \omega$ much exceeding any characteristic energy of the considered electronic system, such as the Fermi energy (chemical potential): $\hbar \omega \backsim 1\,eV \gg E_F$ .  

\medskip 
The effect of the dressing field is given by a canonical substitute of the electron momentum $k_{x,y} \longrightarrow k_{x,y} - e/\hbar A_{x,y}$. We consider the most general case of an elliptically polarized irradiation with the vector potential given by the equation

\begin{equation}
\mc{A}^{(E)}(t) = \frac{E_0}{\omega} \,  \left(
\begin{array}{c}
\cos (\omega t) \\
\beta \sin (\omega t)
\end{array}
\right) \, .
\end{equation}
We see that the two components of the vector potential \eqref{Ael01} are separated by the phase difference of $\pi/2$. Here, $\beta$ is the ratio of field strengths along the two axes of the polarization ellipse. Vector potential \eqref{Ael01} becomes equivalent to the that of a circularly polarized light in the limiting case of $\beta \longrightarrow 1$, and the linearly polarized light is obtained as $\beta \longrightarrow 0$:

\begin{equation}
\label{Ael01}
\mc{A}^{(L)}(t) = \frac{E_0}{\omega} \,  \left(
\begin{array}{c}
1\\
0
\end{array}
\right) \, \cos (\omega t)  \, .  
\end{equation}
Vector potential \eqref{Ael01}, as well as the resulting electronic states, definitely depend on the direction of the light polarization (unless $\beta = 1$, in which case we're dealing with circularly polarized light), which is assumed to be along the $x-$axis. 

\par 

Importantly, we can select an arbitrary direction of the linear polarization by applying the following transformation

\begin{equation}
\mc{A}^{(L)}(t) = \frac{E_0}{\omega} \, \left[ \hat{\mbb{R}}(\theta_0) \, \left(
\begin{array}{c}
1\\
0
\end{array}
\right) \, \right] \, \cos (\omega t) =  \frac{E_0}{\omega} \, \left(
\begin{array}{c}
\cos \theta_0 \\
\sin \theta_0
\end{array}
\right) \, \cos (\omega t) \, ,  
\end{equation}
where 

\begin{equation}
 \hat{\mbb{R}}(\theta_0)  = \left[
\begin{array}{cc}
\cos \theta_0 & - \sin \theta_0 \\
\sin \theta_0 &  \cos \theta_0 
\end{array}
\right]
\end{equation}
is the rotation matrix for angle $\theta_0$.

\medskip 
By making use of the canonical substitution for the electron momentum $k_{x,y} \longrightarrow k_{x,y} - e/\hbar A_{x,y}$, we obtain the time-dependent Hamiltonian in the presence of the driving field. To solve the current eigenvalue problem, we have to rely on perturbation theory. Importantly, there are several approaches to calculating the dressed states. For the terms which depend on the components of the wave vector $\vec{\bf k}$ linearly, the obtained Hamiltonian consists of the initial terms and additional time-dependent (but not $\vec{\bf k}$-dependent) interaction terms. 

\par 
Our Hamiltonian \eqref{mainHam} contains multiple terms with non-linear $\vec{\bf k}$-dependence, and one possible approach to obtaining an effective time-independent (time-averaged) Floquet Hamiltonian could be a van Vleck frequency expansion, which leads to the following expression

\begin{equation}
\label{VVexp01}
\mc{H}^{(D)}(\vec{\bf k}) = \mc{H}_0^{(F)}(\vec{\bf k}) + (\hbar \omega)^{-1} \, \left[\mc{H}_{-1}^{(F)}(\vec{\bf k}), \mc{H}_{+1}^{(F)}(\vec{\bf k})  \right] + 
1/2 \, (\hbar \omega)^{-2} \, \left[\mc{H}_{-1}^{(F)}(\vec{\bf k}), \,\left[ \mc{H}_{0}^{(F)}(\vec{\bf k}),  \mc{H}_{+1}^{(F)}(\vec{\bf k}) \right] \, \right] ... \, .
\end{equation}
Thus, van Vleck expansion results in a time-independent Hamiltonian and the energy spectrum with modified band gaps and components of the electron group velocities. In most cases, taking only the linear expansion term $ (\hbar \omega)^{-1} \, \left[\mc{H}_{-1}^{(F)}(\vec{\bf k}), \mc{H}_{+1}^{(F)}(\vec{\bf k})  \right] $  into account would be sufficient. However, as we will see soon, in a crucial case of linearly polarized radiation, the first-order expansion term vanishes, and we need to consider the next term.

\medskip
For the considered problem, it seems natural to choose $L_{(0)} = 1$nm as the unit of length, meaning that all our wave vectors will be measured in terms of $k_{(0)}  = 1/L_{(0)} = 1$ nm$^{-1}$ = $10^9$m$^ {-1}$. Next, we choose our unit of energy as $E_{(0)} =  \hbar^2/(2 m_e) \, k_{(0)}^2 = 6.058 \cdot 10^{-21} \, J = 37.86$meV,  which results in the most straightforward form of the $\hat{\Sigma}_0^{(2)}$-term, now given only as $k^2$. Parameter $\mathfrak{A}$, which quantifies the strength of an altermagnetic order, is dimensionless and is normally assumed to be $\mathfrak{A} \leq 1$. A relatively small spin-orbit coupling induced by the extremal voltage gate is typically given by $r_{SO} \leq 0.2$eV$\AA  = 3.2 \cdot  10^{-30}$J$\cdot$m so that we can introduce a dimensionless spin-orbit coupling coefficient as $r \leq 0.525 \,k_{(0)}/E_{(0)}$.   

\par 
Next, we introduce the unit for the vector potential of the imposed irradiation. As expected, $A_{x,y} \backsim  E_e/\omega$, where the off-resonance dressing field is normally applied in the terahertz regime, which results in $\omega \backsim 1.5 \cdot 10^{15}$Hz so that $\hbar \omega \backsim 1$eV$\gg E_{(0)}$. The typical amplitude of the electric field is $E_e \leq 5.0$V/nm so that the change of the electron momentum by the canonical substitution is given by  $e A/(\hbar) = E/\omega  \leq 5.0\,k_{(0)}$. Therefore, we are going to introduce and further employ a dimensionless electron-light coupling parameter $K_\omega = e E_0/(\hbar \omega)$ as the measure of the strength of the electron-light interaction.

\medskip
Now we are in position to present and discuss our results for the energy spectrum of the dressed states in an altermagnet in the presence of an off-resonance dressing optical field which we have derived in our Appendix \ref{apb}. 

\par  
First, we consider a $d_{x^2-y^2}$ altermagnet and the most general elliptically polarized irradiation. In this case, we obtain

\begin{eqnarray}
\label{Ell11}
\hat{\mc{H}}_{1}^{(E)}(\vec{\bf k}) &  = & \mc{H}_{[n=0]}^{(F)}(\vec{\bf k})  + \beta \, \big[ \rho \, \mathcal{K}_\omega \big]^2  \Sigma_z^{(2)} -
\beta \, \mathfrak{A}_1 \, \rho \, \mathcal{K}_\omega^2 \, \Big[ k_y \hat{\Sigma}_x^{(2)}  +  k_x \hat{\Sigma}_y^{(2)} \Big] \, , 
\end{eqnarray}
where the zero-order Floquet Hamiltonian $\mc{H}_{[n=0]}^{(F)}(\vec{\bf k})$

\begin{eqnarray}
\nonumber 
\mc{H}_{[n=0]}^{(F)}(\vec{\bf k}) &=& \mc{H}_{1} (\vec{\bf k}) + \frac{1+\beta^2}{2} \, \mathcal{K}_\omega^2 \, \Sigma_0^{(2)} +
\frac{1-\beta^2}{2} \, \mathcal{K}_\omega^2 \, \hat{\Sigma}_z^{(2)} = k^2 \, \hat{\Sigma}_0^{(2)} +  \mathfrak{A}_1 \, \left(k_x^2 - k_y^2 \right)  \,  \hat{\Sigma}_z^{(2)} \, + \\
\label{Ell110}
& + & \rho \,
\left( k_x \,  \hat{\Sigma}_y^{(2)} - k_y \,  \hat{\Sigma}_x^{(2)} \right) + \frac{1+\beta^2}{2} \, \mathcal{K}_\omega^2 \, \hat{\Sigma}_0^{(2)} +
\frac{1-\beta^2}{2} \, \mathcal{K}_\omega^2 \, \hat{\Sigma}_z^{(2)}
 \, . 
\end{eqnarray}
We find that applying an optical driving field with elliptical polarization results in multiple effects: a constant energy shift, which is, however, always offset by a change in the chemical potential (doping level) of the considered structure, an induced energy bandgap, and modifications to both components of the electron group velocities. 
\par 
The bandgap is represented by two independent contributions  

\begin{equation}
\label{gap01}
\Delta  = \Delta_1 (\beta \, \vert \,\rho, \mathcal{K}_\omega ) +  \Delta_2 (\beta \, \vert \, \mathcal{K}_\omega ) = \beta \, \left(  \rho \, \mathcal{K}_\omega \right)^2 +  \frac{1-\beta^2}{2}\, \mathcal{K}_\omega^2 \, , 
\end{equation}
which demonstrates a non-monotonic dependence on parameter $\beta$ and is substantially different for the various types of elliptically polarized irradiation. 

\medskip 
Next, we obtain and analyze the electron-dressed states of a d-wave altermagnet with $d_{xy}$ symmetry, in which the altermagnetic Hamiltonian term is represented by Eq.~\eqref{altt2}. This type of alternation has received less attention because of the lower anisotropy due to spin-orbit coupling. However, this d-wave alternates demonstrate a highly unusual energy Spectrum induced by external, isotropic irradiation. Therefore, we would also like to consider this situation in detail.

\par 
Once the optical dressing field is applied,  we obtain the following effective Floquet time-independent Hamiltonian 
\begin{eqnarray}
\mc{H}_{2}^{(E)}(\vec{\bf k}) &  = & \mc{H}_{[n=0]}^{(F)}(\vec{\bf k})  + \beta \, \big[ \rho \, \mathcal{K}_\omega \big]^2 \, \hat{\Sigma}_z^{(2)} -
\mathfrak{A}_2  \, \rho \, \mathcal{K}_\omega^2 \, \left[ k_y \, \hat{\Sigma}_x^{(2)} + \beta^2 \, k_x \, \hat{\Sigma}_y^{(2)}  \right]
\, , 
\end{eqnarray}
where the zero-order Floquet Hamiltonian $\mc{H}_{[n=0]}^{(F)}(\vec{\bf k})$

\begin{eqnarray}
\nonumber 
\mc{H}_{[n=0]}^{(F)}(\vec{\bf k}) &=& \mc{H}_{2}(\vec{\bf k}) + \frac{1+\beta^2}{2} \, \mathcal{K}_\omega^2 \, \hat{\Sigma}_0^{(2)} +
\frac{1-\beta^2}{2} \, \mathcal{K}_\omega^2 \, \hat{\Sigma}_z^{(2)} = k^2 \, \hat{\Sigma}_0^{(2)} +   \mathfrak{A}_2  \,  k_x k_y  \,  \hat{\Sigma}_z^{(2)} \, + \\
\label{linrt}
& + & \rho \,
\left( k_x \,  \hat{\Sigma}_y^{(2)} - k_y \,  \hat{\Sigma}_x^{(2)} \right) + \frac{1+\beta^2}{2} \, \mathcal{K}_\omega^2 \, \hat{\Sigma}_0^{(2)} +
\frac{1-\beta^2}{2} \, \mathcal{K}_\omega^2 \, \hat{\Sigma}_z^{(2)} \, , 
\end{eqnarray}
is the same as for the previously considered $d_{x^2-y^2}$ type of altermagnets. We see that $\hat{\Sigma}_x^{(2)}$ and $\hat{\Sigma}_y^{(2)}$ terms are affected in non-equivalent ways for $\beta \neq 1$. The same holds for the corresponding components of the Fermi velocity, as we would expect from an anisotropic driving field. 

\subsection{Linearly polarized dressing field}

We now consider the electronic states in a $d_{x^2 - y^2}$ altermagnet with a Hamiltonian \eqref{mainHam} in the presence of a linearly polarized dressing field 

\begin{equation}
\mc{A}^{(L)}(t) =  \frac{E_0}{\omega} \, \left(
\begin{array}{c}
\cos \theta_0 \\
\sin \theta_0
\end{array}
\right) \, \cos (\omega t) \, ,  
\end{equation}
where $\theta_0$ is the fixed direction of the external field polarization. 
Analyzing Hamiltonian \eqref{HamtLin01}, we immediately discern that $ \hat{\mc{H}}_{-1}^{(F)}(\vec{\bf k}) \equiv  \hat{\mc{H}}_{1}^{(F)}(\vec{\bf k})$ and a linear term in Van Vleck expansion could only yield zero. Therefore, this time we must take into the $\backsim 1/\omega^2$ corrections in Eq.~\eqref{AVVexp01}.

\medskip 
Let's first consider the most straightforward case of graphene  with a Hamiltonian $k_x \Sigma_x^{(2)}+k_y \Sigma_y^{(2)}$. The corresponding Floquet expansion  
terms are given as

\begin{eqnarray}
&&  \hat{\mc{H}}_{0}^{(F)}(k) = \mc{H}_{0}(k)  = \hbar v_F \, k_x \Sigma_x^{(2)} +  \hbar v_F \, k_y \Sigma_y^{(2)} \, , \\
\nonumber 
&&  \hat{\mc{H}}_{-1}^{(F)}(\vec{\bf k})  =  \mc{H}_{1}^{(F)}(\vec{\bf k}) = - \frac{e E_0 \, v_F}{2 \omega} \, \left[ 
\cos \theta_0  \, \Sigma_x^{(2)} + \sin \theta_0  \, \Sigma_y^{(2)}   
\right] \, . 
\end{eqnarray}
The first order expansion term in Eq.~\eqref{AVVexp01} is obviously zero since $\mc{H}_{1}^{(F)}(\vec{\bf k}) = \mc{H}_{-1}^{(F)}(\vec{\bf k})$, just like for all the cases of a linearly polarized irradiation. However, an effective Floquet Hamiltonian is obtained from its second order term:

\begin{eqnarray}
 \hat{\mc{H}}_{1}^{(L)}(\vec{\bf k}) & = & \mc{H}_{0}^{(F)}(k) + \frac{1}{2 \hbar} \,  \left( e E_0 \right)^2 \, \left( \frac{v_F^3}{\omega^4} \right) \, \left(\cos \theta_0 k_y - k_x \sin \theta_0 \right)  \, 
\left[ \cos \theta_0 \,\Sigma_x^{(2)} + \sin \theta_0  \, \Sigma_x^{(2)} \right] \, = \\
\nonumber 
& = & \hbar v_F \, k_x \Sigma_x^{(2)} + \hbar v_F \, k_y \Sigma_y^{(2)} + \frac{1}{2 \hbar} \,  \left( e E_0 \right)^2 \, \left( \frac{v_F^3}{\omega^4} \right) \,  \, \left(\cos \theta_0 k_y - k_x \sin \theta_0 \right)  \, 
\left[ \cos \theta_0 \,\Sigma_x^{(2)} + \sin \theta_0  \, \Sigma_x^{(2)} \right] \, . 
\end{eqnarray}
For the polarization direction along the $x-$axis, $\theta_0=0$, we immediately recover the known expression for the energy dispersions

\begin{equation}
\varepsilon^{(L)}(\vec{\bf k}) = \sqrt{\cos^2 (\theta_{\bf k}) + \left[ J_0\left( \frac{2 e E_0\,v_F}{\hbar \omega^2} \right) \, \sin^2 (\theta_{\bf k})  \right]^2 } \, k \backsimeq \sqrt{ 1 - \left( \frac{e E_0 \, v_F}{\hbar \omega^2} \right)^2 \, \sin^2 (\theta_{\bf k})} \, ,
\end{equation}
where $J_0(x) \backsimeq 1  - x^2/4$ is the zero-order Bessel function.

\medskip
\par

Following a similar approach, we derive following Floquet time-independent Hamiltonian for a $d_{x^2-y^2}$ altermagnet

\begin{eqnarray}
\label{Lingen01}
 && \hat{\mc{H}}_{1}^{(L)}(\vec{\bf k}) = \mc{H}_{[n=0]}^{(F)}(\vec{\bf k}) + \sum\limits_{i=(x,y,z)} \mc{F}_{i}^{(L)}(\rho, \omega \, \vert \, \theta_0) \, \hat{\Sigma}_i^{(2)} \, = \\
\nonumber 
& = & \mc{H}_{[n=0]}^{(F)}(\vec{\bf k})  + \mc{F}_{x}^{(L)}(\rho, \omega \, \vert \, \theta_0) \, \hat{\Sigma}_x^{(2)} + \mc{F}_{y}^{(L)}(\rho, \omega \, \vert \, \theta_0) \, \hat{\Sigma}_y^{(2)} + \mc{F}_{z}^{(L)}(\rho, \omega \, \vert \, \theta_0) \, \hat{\Sigma}_z^{(2)}
\, ,
\end{eqnarray}
where

\begin{eqnarray}
&& \mc{F}_{z}^{(L)}(\rho, \omega \, \vert \, \theta_0) = - \frac{\mathfrak{A}_1}{4}  \, \left[ \mu_\omega \,\mathcal{K}_\omega \right]^2 \, \Big\{ 
2 \cos \theta_0 \, \left[k^2 - 2 k_x (k_x + k_y)\right] + 2 \sin \theta_0 \, \left[k^2 + 2 k_y (k_x + k_y) \right] 
\Big\}
, \\
\nonumber 
&& \mc{F}_{x}^{(L)}(\rho, \omega \, \vert \, \theta_0) = \frac{\rho^3}{4}  \, \left[ \mu_\omega \,\mathcal{K}_\omega \right]^2  \,  \cos \theta_0 \left( \cos \theta_0 k_y - \sin \theta_0 k_x \right)  -  \frac{\mathfrak{A}_1}{2}  \, \left[ \mu_\omega \,\mathcal{K}_\omega \right]^2 \, \rho \, 
\left( \cos \theta_0 k_y - \sin \theta_0 k_x \right) \, \times \\
& \times &  \Big[ 2 \cos \theta_0  \, k_x k_y - \sin \theta_0  \left( k_x^2 + 3 k_y^2 \right) \Big] \, , \\
\nonumber 
&&\mc{F}_{y}^{(L)}(\rho, \omega \, \vert \, \theta_0) = - \frac{\rho^3}{4} \, \left[ \mu_\omega \,\mathcal{K}_\omega \right]^2  \cos \theta_0 \left( \cos \theta_0 k_y - \sin \theta_0 k_x \right)   -  \frac{\mathfrak{A}_1}{2}  \, \left[ \mu_\omega \,\mathcal{K}_\omega \right]^2 \,\rho  \, 
\left( \cos \theta_0 k_y - \sin \theta_0 k_x \right) \, \times \\
& \times &  \Big[ 2 \cos \theta_0 \, k_x^2 -\cos \theta_0 \, k^2 - 2 k_x k_y \, \sin \theta_0 \Big] \, .
\end{eqnarray}
Here,  $\mu_\omega = E_{(0)}/(\hbar \omega) \ll 1$.

\par 
The zero-order Floquet Hamiltonian $\mc{H}_{[n=0]}^{(F)}(\vec{\bf k})$

\begin{eqnarray}
\mc{H}_{[n=0]}^{(F)}(\vec{\bf k}) &=& \mc{H}_{1}^{(0)}(\vec{\bf k}) + \frac{1}{2} \, \mathcal{K}_\omega^2 \, \hat{\Sigma}_0^{(2)} + \mathfrak{A}_1 \, \frac{1}{2} \, \mathcal{K}_\omega^2 \,  \hat{\Sigma}_z^{(2)} = 
k^2 \, \hat{\Sigma}_0^{(2)} \, + \\
\nonumber 
& + & \mathfrak{A}_1 \,  (k_x^2 - k_y^2)  \,  \hat{\Sigma}_z^{(2)}  + \rho \,
\left( k_x \,  \Sigma_y^{(2)} - k_y \,  \hat{\Sigma}_x^{(2)} \right) +  \frac{1}{2} \, \mathcal{K}_\omega^2 \, \hat{\Sigma}_0^{(2)} + \mathfrak{A}_1 \, \frac{1}{2} \, \mathcal{K}_\omega^2 \, \hat{\Sigma}_z^{(2)}
 \, , 
\end{eqnarray}
we see that a finite bandgap is created independent of the direction of the polarization of the dressing field. 

\medskip 
For the polarization direction along the $x-$axis, equation \eqref{Lingen01} is simplified as 

\begin{eqnarray}
&& \hat{\mc{H}}_{1}^{\mathfrak{A}}(\vec{\bf k}) = \mc{H}_{[n=0]}^{(F)}(\vec{\bf k})   - \frac{\mathfrak{A}_1}{4}  \, \left[ \mu_\omega \,\mathcal{K}_\omega \right]^2 \, \left[
k^2 - 2  k_x (k_x + k_y) 
\right]^2  \, \Sigma_z^{(2)}
\, + \\
\nonumber 
&+& \left[ \mu_\omega \,\mathcal{K}_\omega \right]^2  \, \left\{  
\frac{\rho^3}{4}\, k_y  +  \frac{\mathfrak{A}_1}{2} \,\rho \, k_x^2 k_y 
\right\} \, \Sigma_x^{(2)} \, - \\
\nonumber 
& - & \left[ \mu_\omega \,\mathcal{K}_\omega \right]^2 \, \left\{  
\frac{\rho^3}{4} \, k_y  +  \frac{\mathfrak{A}_1}{2}  \, \rho \, k_x \, \left( k_x^2 - k_y^2 \right)
\right\} \, \Sigma_y^{(2)}
\, , 
\end{eqnarray}
which demonstrates that the second-order perturbation expansion terms brings small but still important corrections to the band structure of irradiated altermagnets.

\medskip 
Finally, let us briefly address the case of a $d_{xy}$ altermagnet under linearly polarized dressing field. Here, we perform the perturbation expansion up to the $\backsim \omega^{-2}$ order, as we explain in all detail in Appendix \ref{apb}. 

\par
The zero-order Floquet Hamiltonian $\mc{H}_{[n=0]}^{(F)}(\vec{\bf k})$ is given by 

\begin{eqnarray}
 \hat{\mc{H}}_{0}^{(L)}(\vec{\bf k}) & = &
k^2\, \hat{\Sigma}_0^{(2)} + \mathfrak{A}_1 \,  (k_x^2 - k_y^2)  \,  \hat{\Sigma}_z^{(2)}  + \rho \,
\left( k_x \,   \hat{\Sigma}_y^{(2)} - k_y \,   \hat{\Sigma}_x^{(2)} \right) + \frac{\hbar^2}{4 m} \, \mathcal{K}_\omega^2 \hat{\Sigma}_0^{(2)} \, + \\
\nonumber 
&+& \frac{\hbar^2 \, \mathfrak{A}_1}{8 m} \, \sin{2 \theta_0} \, \mathcal{K}_\omega^2 \, \hat{\Sigma}_z^{(2)}
 \, . 
\end{eqnarray}
Therefore, the bandgap becomes $2 \Delta_2^{(L)}$, where 

\begin{equation}
\label{gap2L}
\Delta_2^{(L)} = \frac{\mathfrak{A}_2}{4} \, \sin(2 \theta_0) \, \left( \frac{e E_0}{\omega} \right)^2  = \frac{\mathfrak{A}_2}{4} \, \sin (2 \theta_0) \, K_\omega^2 \, .
\end{equation}
We observe that the bandgap \eqref{gap2L} disappears for certain directions $\theta_0 = \pi n /2$, $n =0, 1,2,3, ...$ (importantly, $\theta_0 = 0$ and $\pi/2$ are included). The other changes to the finite-$k$ dispersions are given as the modifications of the components of its anisotropic Fermi velocity and are shown at $1/\omega^4$ level, which appear to be small but not negligible.

\section{Results and discussion}
\label{sec4}

Now we turn to our discussion of the energy dispersions of the electron dressed states for both $d_{x^2-y^2}$ and $d_{xy}$  symmetry types of the wave alter magnets. The modification of the energy dispersions due to the optical driving drastically depends on the polarization of the applied field. We put the focus on anisotropic optical dressing fields in which not only the type of polarization but its direction applied to essentially anisotropic and non-linear Hamiltonian  \eqref{mainHam} of $d-$wave altermagnets leads to substantially different electronic states.

\medskip 
The energy dispersions of a $d$-wave altermagnets with  $d_{x^2-y^2}$ symmetry, as presented in Figs.~\ref{FIG:2} and  \ref{FIG:3} show the two solutions for the energy subbands, which depend on the spin index $s = \pm 1$. We also demonstrate the horizontal constant-energy cuts of the calculated energy spectrum by solving each dispersion equation for the wave vector magnitude $k= \vert \vec{\bf k} \vert $ at the fixed angle $\Theta_{\vec{\bf k}}$. We also distinguish the two branches by the direction of out-of-plane polarization obtained from equation

\begin{equation}
\label{ava01}
\text{sign}\Big[ \Big\langle \hat{S}_z \Big\rangle \Big] = \text{sign}\Big[ \Big\langle \Psi_s(\vec{\bf k}) \Big\vert \hat{\Sigma}^{(2)}_z \Big\vert \Psi_s(\vec{\bf k}) \Big\rangle \,\Big] =  \frac{\hbar}{2} \, \cos \Theta_{\vec{{\bf k}}} = \frac{\hbar}{2} \, \frac{V_3(\vec{{\bf k}})}{\vert V(\vec{{\bf k}}) \vert} \, ,
\end{equation}
so that we obtain $\text{sign}\Big[ \Big\langle \hat{S}_z \Big\rangle \Big] = \text{sign} \Big[V_3(\vec{{\bf k}}) \Big]$.

\par 

We have found that in the absence of both optical driving field and spin-orbit interaction, we obtain anisotropic energy subbands with nearly elliptical angular dependence. One of them corresponds to the positive out-of-plane spin polarization (upward direction of $\Big\langle \hat{S}_z \Big\rangle $); the other one is related to the negative $\Big\langle \hat{S}_z \Big\rangle$. Thus, in the absence of spin-orbit locking, each separate energy subband corresponds to a fixed direction of the out-of-plane spin polarization. The bandgap in this case is equal to zero, and both energy subbands start with $\epsilon_{s=\pm 1}(\vec{\bf k}) = 0 $ energy level. Once the finite spin-orbit coupling is introduced, we obtain a very unusual star-shaped angular dependence, which reflects the specific type of Rashba spin-orbit interaction. However, the energy bandgap in this case remains zero, as it does for most other materials with spin-orbit coupling.

\medskip 
Next, we expose our considered altermagnets to an off-resonance optical dressing field which already exhibits spin-orbit coupling due to an electrostatic gating. We consider anisotropic types of the light polarization, specifically, elliptical $(\beta < 1)$ and linear $(\beta = 0)$. We begin with $\beta=0.9$, which is rather similar to the circularly polarized fields $(\beta = 1)$. The results for the energy spectrum for an irradiated altermagent with finite and zero spin-orbit coupling are presented in Figures \ref{FIG:4} and \ref{FIG:5}, respectively. Most importantly, a finite bandgap between the two bands is generated in both cases. We verify the existence of both direct and indirect gaps along the $x-$direction, while the indirect gap along the $y-$axis is missing: the two subbands intersect at a finite wavevector $\vec{\bf k}$. The angular dependence of these subbands is generally preserved in the presence of a circularly polarized driving field. However, now we observe some very unusual types of dispersions with a changing concave-convex form with an increasing wave vector $k$. These dispersions are definitely a result of a highly unusual interplay between the driving field and spin-orbit coupling. It is also interesting to notice that the distinction between the directions of out-of-plain polarization is no longer at $\pm \pi/4$ and $\pm 3\pi/4$ directions. The applied optical driving field substantially shifts the corresponding boundaries. These spin textures also become very different from the separation of the two energy subbands corresponding to $s = \pm 1$. 

\par 
As an alternative, we also apply the elliptically polarized optical filled ways with $\beta = 0.2$ (see Fig.~\ref{FIG:6}), which is much closer to linear polarization than the previously considered cases. We observe that a substantial band gap is still opened, which is in stark contrast with all the known Dirac cone materials, including graphene and the $\alpha-\mc{T}_3$ model.\,\cite{Tamang2024OrbitalMagnetization,Tamang2023TopologicalAlphaT3,Nascimento2025ChiralAlphaT3} The angular dependence (anisotropy) of the energy spectrum in this case is also strongly affected, however, this effect is gap-related: the upper bands are much more elevated in the $x-$direction. We need to keep in mind that these calculations are performed within the first-order perturbation expansion, which is in principle not applicable to linearly polarized irradiation, corresponding to $\beta=0$. This allows us to compare what types of additional Hamiltonian terms result in each separate effect (bandgap and anisotropy). 

\medskip 
Finally, we want to take a quick look at an altermagnet with a $d_{xy}$ symmetry and demonstrate the results in Fig.~\ref{FIG:7}. For $\beta=0.9$, we observe a noticeable bandgap and monotonic energy dispersions. However, its angular dependence is rather unusual with the subbands most extended along $\pm \pi/4$ and $\pm 3\pi/2$ directions due to a $k_x k_y$-term, which is present in both in the initial Hamiltonian and in radiation-induced additions. The separation between the positive and negative directions of out-of-plane spin polarizations corresponds to $0$ and $\pi/2$ angles, which have not been observed for any previous situations in $d_{x^2-y^2}$ altermagnet.

\par 
The generated energy gap for the case of elliptically polarized light and $d_{x^2-y^2}$ symmetry is presented in Fig.~\ref{FIG:8}. This gap depends on parameter $\beta$, spin-orbit coupling $\rho$, and obviously the electron-light coupling perimeter $K_\omega$ since the band gap is basically created by the driving field. We see that equation \eqref{gap01} contains two terms, one of which is increasing and the other decreasing with increasing $\beta$. Therefore, the dependence on the anisotropy of the dressing field is not monotonic, and both circularly and linearly polarized light exhibit a substantial bandgap. The bandgap is also a monotonically increasing function of spin-orbit coupling $\rho$. However, this dependence becomes substantial only for $\beta \neq 0$, i,e., next to the limit of circularly polarized irradiation of the dressing field.   

\medskip
Finally, we investigate the energy dispersion, which is angular-dependent for the case of linearly polarized light. As we are now aware, we cannot limit our consideration to the first-order perturbation expansion; second-order corrections are needed. It is important to mention that the allowable intensity of light in the off-resonance regime could create electron light coupling parameter $K_\omega \backsim 5.0 \, k_{(0)}$. However, an accurate application of the perturbation theory requires the electron-light coupling to be a small parameter. Therefore, we limit our consideration to $K_\omega \leq 1.0 k_{(0)}$. The obtained energy depressions are presented in Figs.~\ref{FIG:9} and \ref{FIG:10}, and we see that a linearly polarized dressing field results in a substantial energy bandgap. This has not been the case for Dirac cone materials. Also, the angular dependence of the energy spectrum is modified only in one direction (perpendicular to the light polarization direction $\theta_0 = 0.0$). We also noticed that the lower branch becomes nearly dispersionless along the $x-$axis, and the change of the energy is rather small for the large range of the wave vectors. Thus, we obtain the picture of the irradiation-modified energy dispersions, which are absolutely different from those in graphene. The only feature they have in common is a change in the group velocities in the direction perpendicular to the light polarization.

\subsection{Edelstein spin susceptibilities} 

Next, we turn our attention to studying how the applied optical dressing field affects the Edelstein effect in $d-$wave altermagents. The Edelstein effect is the generation of finite spin polarizations by an applied electric field in non-centrosymmetric materials with spin-orbit coupling (spin-momentum locking). Thus, in-plane spin polarizations could be generated by external electric fields  $\vec{\bf E}= (E_x, E_y)$. 

\par 
The expectation values of the investigated spin polarization are defined by the components $\chi_{ij}$ of the susceptibility tensor using linear response theory

\begin{equation}
\Big\langle \hat{S}_y \Big\rangle = \chi_{yx} E_x + \chi_{xy} E_y =  - \chi_{xy} E_x - \chi_{xx} E_y
\end{equation}

Using the Kubo formula, and following Refs.[\onlinecite{yarmohammadi2026spin}] we apply the following expression for $\chi_{xy}$ component of 
Edelstein susceptibility tensor

\begin{eqnarray}
\frac{- i m}{e} \, \chi_{xy} & = & \int d^2 \vec{\bf k}\, \Big\{ \sum\limits_{\sigma= \pm} \frac{\pr f[\varepsilon^{(E)}_\sigma (\vec{\bf k})]}{\pr \varepsilon^{(E)}_\sigma(\vec{\bf k}}) \, \frac{- i}{\tau_{0,a}} \,  \Big\langle \Psi_\sigma(\vec{\bf k}) \Big\vert \hat{\Sigma}^{(2)}_x \Big\vert \Psi_\sigma (\vec{\bf k}) \Big\rangle \, \Big\langle \Psi_\sigma(\vec{\bf k}) \Big\vert \hat{P}^{(2)}_y \Big\vert \Psi_\sigma (\vec{\bf k}) \Big\rangle \Big\} \, - \\
\nonumber 
& - & \int d^2 \vec{\bf k} \sum\limits_{\sigma_1 = \pm, \sigma_2 = \pm}^{\sigma_1 \neq \sigma_2} \, \left\{ \, \frac{
f[\varepsilon^{(E)}_{\sigma_1}(\vec{\bf k})]-f[\varepsilon^{(E)}_{\sigma_2}(\vec{\bf k})]}{\varepsilon^{(E)}_{\sigma_1}(\vec{\bf k}) - \varepsilon^{(E)}_{\sigma_2}(\sigma, \vec{\bf k})} \, 
\frac{\Big\langle \Psi_\sigma(\vec{\bf k}) \Big\vert \hat{\Sigma}^{(2)}_z \Big\vert \Psi_\sigma (\vec{\bf k}) \Big\rangle \, 
\Big\langle \Psi_\sigma(\vec{\bf k}) \Big\vert \hat{P}^{(2)}_y \Big\vert \Psi_\sigma (\vec{\bf k}) \Big\rangle
}{
\varepsilon^{(E)}_{\sigma_1}(\vec{\bf k}) - \varepsilon^{(E)}_{\sigma_2}(\vec{\bf k}) + i \tau_{0,e}^{-1}
} \right\} \, . 
\end{eqnarray}
Here, $\tau_{0,a} = \tau_{0,e} = 0.5$eV are the the intra- and inter-band lifetime, $f[\varepsilon^{(E)}(\sigma, \vec{\bf k})]$ is the 
Fermi-Dirac distribution function corresponding to the energy subband $\varepsilon^{(E)}_\sigma (\vec{\bf k})$ with the chemical potential $\mu = \varepsilon^{(E)}(\sigma, \vec{\bf k} = 0)$ and a small temperature $k_B T = 0.05\,E_{(0)}$.   

\par 
All the required matrix elements for the electron spin $ \Big\langle \Psi_{\sigma_{1,2}}(\vec{\bf k}) \Big\vert \hat{\Sigma}^{(2)}_x \Big\vert \Psi_{\sigma_{2,1}} (\vec{\bf k}) \Big\rangle$ and momentum components $\Big\langle \Psi_{\sigma_{1,2}}(\vec{\bf k}) \Big\vert \hat{P}^{(2)}_y \Big\vert \Psi_{\sigma_{2,1}} (\vec{\bf k}) \Big\rangle $ ($\backsim m \, \pr/\pr k_{x,y}$)  could be easily evaluated for known wave functions \eqref{apsi01} and \eqref{apsi01} (see equation \eqref{ava01} as an example). A complete list of all the matrix elements needed to calculate susceptibility $\chi_{xy}$ is also provided in Ref.~[\onlinecite{yarmohammadi2026spin}].

\begin{figure} 
\centering
\includegraphics[width=0.49\textwidth]{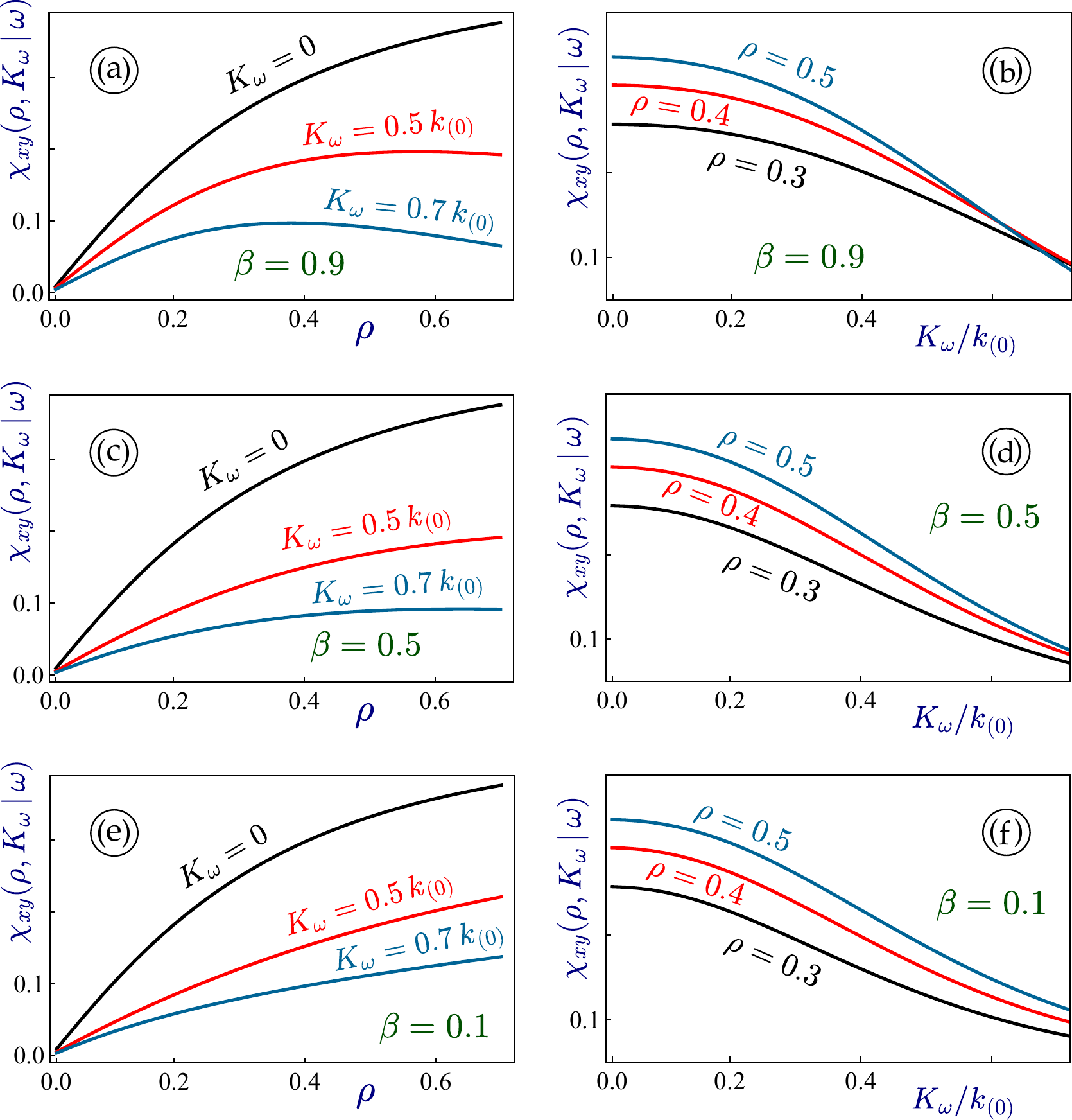}
\caption{(Color online) The transverse components $\chi_{xy}$ of spin-type Edelstein susceptibilities in the units of $e/(4 \pi^2)$ for a $d$-wave altermagnet with $d_{x^2 - y^2}$ pairing symmetry in the presence of the spin-orbit coupling and an off-resonance dressing field. The chemical potential corresponds to the energy $\epsilon_s(\vec{\bf k} = 0)$ for each of these graphs. The left panels $(a)$, $(c)$ and $(e)$  demonstrate how susceptibility $\chi_{xy}$  depends On the strength of Rashba spin-orbit coupling for different cases of electron-photon coupling parameter $K_\omega =0.0$ (zero irradiation, black curves),  $K_\omega =0.5\, k_{(0)}$ (Red curves) and  $K_\omega =0.7\, k_{(0)}$ (blue curves). The right panels $(b)$, $(d)$ and $(f)$ how susceptibility $\chi_{xy}$  depends on the electron-photon coupling parameter $K_\omega$ for different cases of Rashba spin-orbit coupling $\rho=0.3$ (black curves),  $\rho=0.4$ (red curves) and  $\rho=0.5$ (blue curves).}
\label{FIG:11}
\end{figure}
\medskip

\begin{figure} 
\centering
\includegraphics[width=0.49\textwidth]{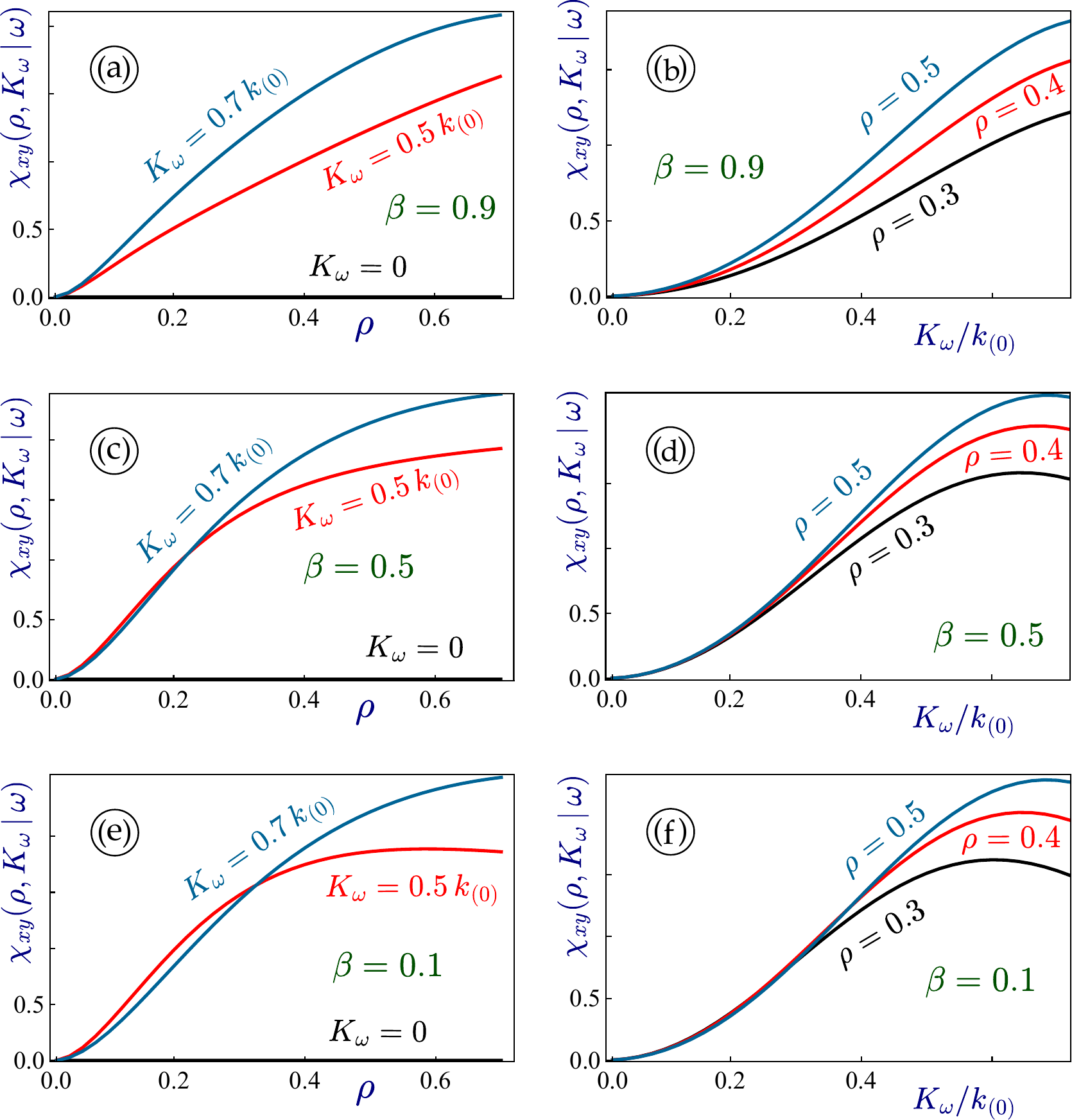}
\caption{(Color online) The transverse components $\chi_{xy}$ of spin-type Edelstein susceptibilities in the units of $e/(4 \pi^2)$ for a $d$-wave altermagnet with $d_{xy}$ pairing symmetry in the presence of the spin-orbit coupling and an off-resonance dressing field. The chemical potential corresponds to the energy $\epsilon_s(\vec{\bf k} = 0)$ for each of these graphs. The left panels $(a)$, $(c)$ and $(e)$  demonstrate how susceptibility $\chi_{xy}$  depends On the strength of Rashba spin-orbit coupling for different cases of electron-photon coupling parameter $K_\omega =0.0$ (zero irradiation, black curves),  $K_\omega =0.5\, k_{(0)}$ (Red curves) and  $K_\omega =0.7\, k_{(0)}$ (blue curves).   The right panels $(b)$, $(d)$ and $(f)$ how susceptibility $\chi_{xy}$  depends on the electron-photon coupling parameter $K_\omega$ for different cases of Rashba spin-orbit coupling $\rho=0.3$ (black curves),  $\rho=0.4$ (red curves) and  $\rho=0.5$ (blue curves).}
\label{FIG:12}
\end{figure}
\medskip

Due to the specific wave symmetry of our model ($d_{x^2-y^2}$), the in-plane electrostatic field cannot create a finite out-of-plane spin polarization. Therefore, we focus on the larger in-plane and off-diagonal components $\chi_{xy} = - \chi_{yx}$ of the susceptibility tensor. In the absence of irradiation, we observe completely antisymmetric susceptibilities and the corresponding spin textures, which arise from inversion and time-reversal (zero bandgap) symmetries.  

\par 
Our current investigation into spin polarization is motivated in part by Ref.[\onlinecite{yarmohammadi2026spin}] and its crucial discussion of how circularly polarized dressing field combined with spin-orbit coupling and external electrostatic fields leads to a unique picture of Edelstein spin polarization patterns, which in turn arise from light-modified band gaps and the components of Fermi velocity. Importantly, the off-diagonal components $\chi_{xy}$ of Edelstein susceptibilities would vanish at a certain frequency and intensity of the applied off-resonance dressing field.

\medskip
Our results for the $\chi_{xy}$ components of the spin polarization tensor for $d-$wave altermagnets with different symmetries are presented in Figs.\,\ref{FIG:11} and \ref{FIG:12}. Specifically, we focus on anisotropic types of external radiation with $\beta < 1$, which are substantially different from circularly polarized radiation. We are mostly interested in observing the Edelstein susceptibility for the case of anisotropic radiation interacting with electrons that also exhibit anisotropic, non-linear energy dispersion.

\par 
We find that the magnitudes of the $\chi_{xy}$ components uniformly increase with spin-orbit coupling strength $\rho$ (and are always equal to zero for $\rho=0$).
However, we also notice that each of these curves has a maximum and starts decreasing at a certain value of $\rho \backsim 0.7$. It is important to keep in mind that $\rho>0.7$ is hardly achievable in a real-life experiment. For an elliptical irradiation with $\beta <1$ -- when the radiation becomes more anisotropic and shifted from the circular polarization type -- it requires a larger value of $\rho$ to reach the maximum for each of the curves, so that the dependence stays monotonic, with a more significant difference between the zero and finite electron-photon coupling parameter $K_\omega$ (added driving field vs. no driving field). The curves corresponding to $K_\omega > 0.5 k_{(0)}$ become nearly identical, while they are very different from each other for the circularly polarized light ($\beta  =1$). The dependence of the $\chi_{xy}$ components on $K_\omega$ is always monotonically decreasing for all types of irradiation. However, for a decreasing $\beta$, the three curves corresponding to the different values of spin-orbit coupling strength $\rho$ become nearly identical, or at least are located a lot closer to each other than what we have observed for the circularly polarized irradiation.

\par 
The situation becomes very different for the altermagnets with $d_{xy}$ type of symmetry, as we show in Fig. \ref{FIG:12}. Most importantly, $\chi_{xy}$ is always zero in the absence of irradiation for any spin-orbit coupling $\rho$. This result stands in stark contrast to previously considered $d_{x^2-y^2}$ alternates, in which the $\chi_{xy}$ tensor components reached their largest values in the absence of the dressing field. Interestingly, the curves for various electron-photon coupling constants demonstrate different increase rates depending on spin-orbit coupling $\rho$ (the curves for the lower electron-light coupling are located higher for the smaller $\rho$). Interestingly,  for a small $ \beta \longrightarrow 0$, all three curves corresponding to different light intensities demonstrate nearly identical dependence (and monotonic increase) on spin-orbit coupling parameter$\rho$, which is not the case for a circularly polarized dressing field. 
 
\medskip 

Transport theory aimed to calculate electric currents, spin currents, and the corresponding conductivities, accounting for an external electric field $\vec{\bf E} = (E_x,E_y)$ that results in a change in the equilibrium Fermi-Dirac distribution function. The corresponding modification to the energy of an electron located at position $\vec{\bf r}$ is given as $\varepsilon(\sigma, \vec{\bf k}) \longrightarrow \varepsilon(\sigma, \vec{\bf k}) + e \vec{\bf E} \cdot \vec{\bf r}$. Since the perturbation due to the external electric field of an electron is considered small compared to its characteristic energy (Fermi energy), we can expand the new distribution function in a power series. 

\begin{equation}
\label{seriesFD}
f[\varepsilon(\sigma, \vec{\bf k})] \,\, \longrightarrow \,\, f[\varepsilon(\sigma, \vec{\bf k})] + \left( e \tau_0 \, \vec{\bf E} \cdot \vec{V}_b  \right) \, \frac{\pr}{\pr \varepsilon} \,
f[\varepsilon(\sigma, \vec{\bf k})] + \, ...  
\end{equation}
where $\vec{V}_b $ is the electron band velocity.

\subsection{Berry curvature and spin currents in altermagnets}

Now a percent to calculating the very curvature in different types of alternating magnets in the presence of an elliptically polarized dressing field. The better car, which is one of the key concepts in electronics, transport, and topology of low-dimensional condensed matter systems, since for a finite number of electrons, there is an extra velocity term, known as anomalous velocity. Also, Berry curvature determines a Chern number, which defines a topological class of the materials under consideration. 

\medskip

As we show in Appendix \ref{apa}, we can calculate the only non-zero component of the Berry curvature

\begin{equation}
\Omega_z (s,\vec{\bf k}) = - \frac{s}{2k} \frac{\pr }{\pr k}  \, \left[ \frac{V_3(\bf \vec{k})}{\vert \vec{\bf V} (\vec{k}) \vert} 
\right]
\, .
\end{equation}
For a gapped graphene with $V_1(k)= \hbar v_F k_x$ and $V_2(k) = \hbar v_F k_y$ and $V_3(k) = \Delta_0$ ($\pr V_3(k)/\pr k = 0$), we immediately obtain 

\begin{equation}
\Omega_z (s,\vec{\bf k}) = - \frac{s}{2k} \frac{\pr }{\pr k}  \, \left[  
\frac{\Delta_0}{\sqrt{
(\hbar v_F\, k)^2 + \Delta_0^2 
}} 
\right] = \frac{s (\hbar v_F)^2 \Delta_0}{2 \left[ (\hbar v_F\, k)^2 + \Delta_0^2 \right]^{3/2}}
\, ,
\end{equation}
which reaches its maximum $(\hbar v_F)^2/(2 \Delta_0^2)$ at $\vec{\bf k} = 0$.

For an altermagnet with $d_{x^2-y^2}$ pairing symmetry, we obtain

\begin{equation}
\label{om01}
\Omega_{z,1} (s,\vec{\bf k}) = \frac{s \, \mathfrak{A}_1 \,\rho^2 \, \cos(2 \phi_{\vec{\bf k}})}{2 k\, \left[
\rho^2 + \left\{\mathfrak{A}_1 k \cos(2 \phi_{\vec{\bf k}})\right\}^2 
\right]^{3/2}
}
\end{equation}

and for $d_{xy}$ pairing symmetry

\begin{equation}
\label{om02}
\Omega_{z,2} (s,\vec{\bf k}) = \frac{2 s \, \mathfrak{A}_2 \,\rho^2 \, \sin(2 \phi_{\vec{\bf k}})}{k \, \left[
(2\rho)^2 + \left\{\mathfrak{A}_1 k \sin(2 \phi_{\vec{\bf k}})\right\}^2 
\right]^{3/2}
} \, . 
\end{equation}

The Berry curvature could be also obtained analytically for an altermagent in the presence of elliptically polarized light,  even though the expression is quite lengthy and complicated so it's better to use a numerically.

\par 
At the same time,  we can provide the result for circular polarized light corresponding to $\beta = 1$ 

\begin{equation}
\label{om03}
\Omega^{(C)}_{z,1} (s,\vec{\bf k})  = - \frac{
\sqrt{2}\, s\, \rho^2 \left[ \mc{K}_\omega^2 \rho^2 - \mathfrak{A}_1 k^2 \cos(2\Theta_{\vec{\bf k}}) \right]
\left\{ 1 + \mathfrak{A}_1^2 \mc{K}_\omega^4 - 2 \mathfrak{A}_1 \mc{K}_\omega^2 \cos(2\theta) \right\}
}{
\left\{
\mathfrak{A}_1^2 k^4
+ 2 k^2 \left[1 + \mathfrak{A}_1^2 \mc{K}_\omega^4 \right] \rho^2
+ 2 \mc{K}_\omega^4 \rho^4
+ \mathfrak{A}_1^2 k^4 \cos(4 \Theta_{\vec{\bf k}})
\right\}^{3/2}
} \, .
\end{equation}
For the vanishing irradiation $\mc{K}_\omega \longrightarrow 0$, this results is immediately reduced to Eq.~\eqref{om01}. Interestingly, the Berry phase for each case (both in the absence or presence, Eqs.~\eqref{om01}-\eqref{om03}) of the dressing field is always zero if there is no spin orbit coupling $\rho \rightarrow 0$. However, the Berry curvature still exists if the altermagnetic order is removed $\mathfrak{A}_1 \rightarrow 0$ but the spin-orbit coupling is kept

\begin{equation}
\label{om04}
\Omega^{(C)}_{z,1} (s,\vec{\bf k})  = - \frac{
s\, \rho \, \mc{K}_\omega^2  
\left\{ 1 + \mathfrak{A}_1^2 \mc{K}_\omega^4 - 2 \mathfrak{A}_1 \mc{K}_\omega^2 \cos(2\theta) \right\}
}{2\,
\left[
k^2 
+ \mc{K}_\omega^4 \rho^2
\right]^{3/2}
} \, .
\end{equation}

For the altermagnets with $d_{xy}$ symmetry, the Berry curvature in the presence of circulary polarized light is obtained as 

\begin{equation}
\label{om05}
\Omega^{(C)}_{z,2} (s,\vec{\bf k}) = - 
\frac{
s \rho^2 \left(1 + \mathfrak{A}_2^2 \mc{K}_\omega^4 - 2 \mathfrak{A}_2 \mc{K}_\omega^2 \cos(2\theta)\right)
\left(2 \mc{K}_\omega^2 \rho^2 - A_1 k^2 \sin(2\theta)\right)
}{
4 \left[
\left(k_x - \mathfrak{A}_2 k_x \mc{K}_\omega^2\right)^2 \,\rho^2
+ \left(k_y + \mathfrak{A}_2 k_y \mc{K}_\omega^2 \right)^2 \,\rho^2
+ \left(\mc{K}_\omega^2 \rho^2 + \mathfrak{A}_2 k_x k_y\right)^2
\right]^{3/2}
}
\end{equation}

Similarly, the Berry curvature still exists if the altermagnetic order is removed $\mathfrak{A}_1 \rightarrow 0$ but the spin-orbit coupling is kept 

\begin{equation}
\label{om06}
\Omega^{(C)}_{z,2} (s,\vec{\bf k}) = -
\frac{\mc{K}_\omega^2 \, s \, \rho}{2 \left(\mc{K}_\omega^4 \rho^2 + k^2 \right)^{3/2}} \, . 
\end{equation}

\medskip

Once we know the Berry curvature for each of the considered altermagnets, we can immediately proceed to the spin current calculations. 
The total spin current can be expressed as

\begin{equation}
\label{sc01}
I_i^{(S),j} = \frac{\hbar}{2} \, \sum\limits_{\sigma = \pm }\int \frac{d^2 \vec{\bf k}}{(2 \pi)^2} \,   \Big\langle \Psi_{\sigma_{1,2}}(\vec{\bf k}) \Big\vert \hat{V}_i^{j} \Big\vert \Psi_{\sigma_{2,1}} (\vec{\bf k}) \Big\rangle\, \left\{ 
f[\varepsilon(\sigma, \vec{\bf k})] + \left( e \tau_0 \, \vec{\bf E} \cdot \vec{V}_b  \right) \, \frac{\pr}{\pr \varepsilon} \,
f[\varepsilon(\sigma, \vec{\bf k})] 
\right\} \, , 
\end{equation}
where $i$ represents the components of the spin current (propagation direction) and $j$ corresponds to spin component of the considered electrons. Here, 
the spin current operator $\hat{V}_i^{j}$, or the total spin current, is obtained from $\hat{V}_i^{j} = 1/2 ( \hat{V}_i \hat{\Sigma}^{(2)}_j + \hat{\Sigma}^{(2)}_j \hat{V}_i  ) = 1/2 \left\{  \hat{V}_i, \, \hat{\Sigma}^{(2)}_j \right\}$. Spin current operator contains both standard band and anomalous velocity contributions $V_i^{j} = \hat{V}^{(b),\,j}_i + \hat{V}^{(a),\,j}_i$. Importantly,  the band velocity operator contributes to all the orders of spin current including the zero order, while the anomalous velocity only effects first and higher orders of the series expansion of the Fermi distribution function in an external electric field \eqref{seriesFD}. The additional anomalous velocity, perpendicular to the direction of
the electric field, is acuqired by two-dimensional materials with a finite Berry curvature. 

\medskip 

The components of the band velocity components for an altermagnet are given by

\begin{equation}
\hbar \Big\langle  \hat{V}^{(b)}_{
 \left\{ \begin{array}{c}
x \\
y
\end{array}
\right\}
} (s,\vec{\bf k})  \Big\rangle = 2 k  \left\{ \begin{array}{c}
\cos \phi_{\vec{\bf k}} \\
\sin \phi_{\vec{\bf k}}
\end{array}
\right\} +  s \mathfrak{A}_1 \, k\, \cos \Theta_{\vec{\bf k}}\, \left\{ \begin{array}{c}
\cos \phi_{\vec{\bf k}} \\
- \sin \phi_{\vec{\bf k}}
\end{array}
\right\} +  s \rho \, \sin \Theta_{\vec{\bf k}}\, \left\{ \begin{array}{c}
\cos \phi_{\vec{\bf k}} \\
\sin \phi_{\vec{\bf k}}
\end{array}
\right\} \, . 
\end{equation}

For the band velocity of an altermagnet with $d_{xy}$ symmetries, only altermagnetic term is changed 

\begin{equation}
s \mathfrak{A}_1 \, k\, \cos \Theta_{\vec{\bf k}}\, \left\{ \begin{array}{c}
\cos \phi_{\vec{\bf k}} \\
- \sin \phi_{\vec{\bf k}}
\end{array}
\right\} \,\,\, \Longrightarrow \,\,\,\,  2 s \mathfrak{A}_2 \, k\, \cos \Theta_{\vec{\bf k}}\, \left\{ \begin{array}{c}
 \sin \phi_{\vec{\bf k}} \\
 \cos \phi_{\vec{\bf k}}
\end{array}
\right\} \, . 
\end{equation}
The components of the anomalous velocity are given by 

\begin{equation}
\frac{\hbar}{e} \Big\langle  \hat{V}^{(a)}_{
 \left\{ \begin{array}{c}
x \\
y
\end{array}
\right\} 
} 
(\vec{\bf E} \, \vert \, s,\vec{\bf k}) 
\Big\rangle =  \left\{ \begin{array}{c}
- E_y \\
E_x
\end{array}
\right\} \, \Omega_z(s,\vec{\bf k})  \, 
\end{equation}
where $(E_x, E_y)$ are the components of electric field and $\Omega_z(s,\vec{\bf k})$ is the Berry curvature obtained above. Now we have all the ingredients for the current \eqref{sc01}, which could be calculated directly. However, this substantial amount of material would definitely fall off the scope of the present paper.

%
%
%
%
%
%
\section{Summary and Concluding remarks}
\label{sec5}

In this paper, we investigated the electron energy spectrum and bandgaps for $d$-wave altermagnets in the presence of an off-resonant optical dressing field and gate-induced spin-orbit coupling. The materials with both $d_{x^2-y^2}$ and $d_{xy}$  symmetries have been studied. The optical driving field leads to noticeable modifications of the electron energy dispersions, resulting in substantially different electronic and collective behavior of the considered materials. Specifically, a modification of the Fermi velocity components leads to a completely different situation in quantum transport, specifically, the Boltzmann conductivity. The results of applying external irradiation depend substantially on its polarization: circularly polarized light opens a bandgap, significantly modifying the material's topological properties, and linearly polarized light creates or modifies anisotropy of the material's energy spectrum.

\medskip
\par
We have uncovered a highly unusual and physically rich picture of the electron dressed states in both $d_{x^2-y^2}$ and $d_{xy}$ altermagnets, which is completely different from that in the previously investigated Dirac cone materials. We have found that a finite bandgap between the two bands is generated by elliptically polarized light, disregarding of the spin-orbit coupling parameter. We observe both direct and indirect gaps along the $ x$-direction, whereas the indirect gap along the $ y$-axis is missing. It is also interesting to notice that the distinction between the directions of out-of-plain polarization is no longer at $\pm \pi/4$ and $\pm 3\pi/4$ directions. When we apply a strongly anisotropic elliptically polarized optical field with $\beta = 0.2$, we observe that a substantial band gap is still opened, which is in stark contrast with all the known Dirac cone materials. We also look at an altermagnet with a $d_{xy}$ symmetry: its separation between the positive and negative directions of out-of-plane spin polarizations corresponds to $0$ and $\pi/2$ angles; such a behavior has not been found in $d_{x^2-y^2}$ altermagnet. We observe that a linearly polarized dressing field yields a substantial energy bandgap in contrast to all known Dirac materials. Also, the angular dependence of the energy spectrum is modified only in one direction (perpendicular to the light polarization direction $\theta_0 = 0.0$).

\medskip
\par 
We also investigated the generation of in-plane spin polarization due to external magnetic fields in various types of altermagnets, known as Edelstein susceptibilities, in the presence of an off-resonance dressing field. We emphasize anisotropic polarization of the external irradiation to explore this effect in the case of anisotropic radiation interacting with electrons, which also exhibit anisotropic, non-linear energy dispersion. We have investigated how susceptibilities depend on two crucial parameters -- the electron-light interaction and the electron spin-orbit coupling strengt. We have found out that even though the general trends and dependence remain unaltered,  introducing an anisotropic driving field leads to several subtle but important changes of $\chi_{xy}$, which allows for the fine-tuning of spin polarizations in a $d$-wave altermagents. This situation is very different for the altermagnets with $d_{xy}$- type symmetry. Specifically, $\chi_{xy}$ is always zero in the absence of irradiation for any spin-orbit coupling $\rho$. This result stands in stark contrast to previously considered $d_{x^2-y^2}$ alternates, in which the $\chi_{xy}$ tensor components reach their largest values in the absence of the dressing field.

\medskip
\par
We obtained closed-form analytical expression for the Berry curvature for altermagnets with $d_{x^2-y^2}$ and $d_{xy}$ types of symmetry, both with and without a dressing field. An analytical expression for the case of elliptically polarized light could also be obtained; however, it is too lengthy and cumbersome to present here. We demonstrate that the Berry curvature is non-zero only for a finite spin-orbit coupling $\rho$; however, it could be finite in the presence of a dressing field even if the altermagnetic order is removed $\mathfrak{A}_{1,2} \longrightarrow 0$. Therefore, the optical driving field plays a similar role in creating a finite Berry curvature, as does spin-orbit coupling $\rho$. The Berry curvature is crucial for determining an additional anomalous current transverse to the direction of the applied electric field. It is also directly related to the Chern number and is used for the topological classification of two-dimensional materials. We have also calculated exact analytical expressions for each components of the band velocity and anomalous velocity for all types of altermagnets, and outlined the calculation of the spin currents in the presence of an optical driving field.

\medskip
\par
Altermagnets are considered most promising candidates for spintronics \,\cite{cortie2020two,vzutic2004spintronics} because they merge the most prominent characteristics of ferromagnets and antiferromagnets, allowing efficient spin current control without stray fields that hinder device scaling.\,\cite{bai2024altermagnetism,vsmejkal2020crystal,vsmejkal2022beyond,fu2025all,jungwirth2016antiferromagnetic,krempasky2024altermagnetic} They also display strong spin-dependent transport effects, such as anisotropic spin Hall responses and magnetoresistance, enabling straightforward electrical detection.\,\cite{takahashi2003spin,mamin2003detection} We strongly believe that our findings on the unusual electronics states, energy spectrum, dispersions and bandgaps, response functions and spin currents for irradiated altermagnets which could be dynamically tuned, are very important for a number of practical applications, including spintronics and device physics.

\begin{acknowledgements}
A.I. was supported by the funding received from TradB-56-75, PSC-CUNY Award \# 68386-00 56. We acknowledge the support from CUNY Research Scholars Program (CRSP); Tiyhearah Danner-Jackson received a fellowship from CRSP funded by the CUNY Office of Research and the New York City Mayor's Office.
\end{acknowledgements}
\medskip

\appendix

\section{Unified expression for the energy dispersions for a spin-1/2 Hamiltonian}
\label{apa}

The considered electronic states for a $d$-wave altermagnet are described by the following Hamiltonian 

\begin{equation}
\label{APHam1}
\mc{H}_1^{(0)}(\vec{k}) = \sum\limits_{i=0}^{3} \vec{\bf {V}}_i (\vec{k}) \, \vec{\Sigma}^{(2)}_i = 
 V_0(\vec{k})  \Sigma^{(2)}_0 + \sum\limits_{i=1}^{3} \vec{\bf {V}_i} (\vec{k})  \,  \vec{\Sigma}^{(2)}_i \,  
\end{equation}
based on all Pauli matrices $\vec{\Sigma}^{(2)}_i$, $i=0,x,y,z$. 
\par 
We can present any three-dimensional vector $\vec{\bf {V}_i} (\vec{k}) = \Big[ V_1(\vec{k}), V_2(\vec{k}), V_3(\vec{k}) \Big]$ as 

\begin{equation}
\vec{\bf {V}}_i (\vec{\bf k})  = \frac{\hat{{\bf n}}({\bf k})}{\vert {\bf V}_i (\vec{k}) \vert} 
 = \frac{{\hat {\bf n}}({\bf k})}{\sqrt{\sum\limits_{i=1}^{3}  {V}_i^2 (\vec{\bf k})}} \, . 
\end{equation} 
Here, $ \hat{\bf n}(\vec{\bf k})  = \Big[ \sin \Theta_{\vec{\bf k}} \, \cos \phi_{\vec{\bf k}}, \, 
 \sin \Theta_{\vec{\bf k}} \, \sin \phi_{\vec{\bf k}}, \,
 \cos \Theta_{\vec{\bf k}} 
 \Big] $ is a unit vector and

\begin{equation}
\label{av3cos}
\Theta_{\bf \vec{k}} = \cos^{-1} \left[ \frac{V_3(\bf \vec{k})}{\vert \vec{\bf V} (\vec{k}) \vert}  \right]
\end{equation} 
and

\begin{equation}
\phi_{\bf \vec{k}} = \tan^{-1} \left[ \frac{V_2(\vec{\bf k}) }{V_1(\vec{\bf k})} \right]
\end{equation} 
are the spherical angles which define our unit vector $\hat{\bf n}(\vec{\bf k})$.  

\medskip 
The eigenvalues of the Hamiltonian \eqref{APHam1} which could be also presented as 

\begin{eqnarray}
\mc{H}_1^{(0)}(\vec{k})  & = &
\left[
\begin{array}{cc}
V_0(\vec{k}) & 0 \\
0 &  V_0(\vec{k}) 
\end{array}
\right] +  \vert \vec{\bf V} (\vec{k}) \vert \, \left(  \hat{\bf n}(\vec{\bf k}) \cdot \vec{\hat{\Sigma}}^{(2)} \right)
\, = \\ 
\nonumber 
 & = & \left[
\begin{array}{cc}
V_0(\vec{k}) & 0 \\
0 &  V_0(\vec{k}) 
\end{array}
\right] +  \vert \vec{\bf V} (\vec{k}) \vert \, 
\left\{
\begin{array}{cc}
\cos \Theta_{\vec{\bf k}} & \sin \Theta_{\vec{\bf k}} \, \tet{e}^{-i \phi_{\vec{\bf k}}} \\
\sin \Theta_{\vec{\bf k}} \, \tet{e}^{i \phi_{\vec{\bf k}}} &  -\cos \Theta_{\vec{\bf k}} \, . 
\end{array}
\right\}
\end{eqnarray}
are easily obtained in the following form

\begin{equation}
\varepsilon_{s=\pm 1}(\vec{\bf k}) = V_0(\vec{k}) + s\,  \vert \vec{\bf V} (\vec{k}) \vert = V_0(\vec{k}) + 
s\, \sqrt{\sum\limits_{i=1}^{3}  {V}_i^2 (\vec{\bf k})} \, . 
\end{equation}
The corresponding spinor eigenstates $\Psi^{(0)}_{[s=\pm 1]}(\vec{\bf k}) = (a,b)^\dagger$ obviously do not depend on $V_0(\vec{k})$ or even $\vert \vec{\bf V} (\vec{k}) \vert$ and are calculated from the following equation 

\begin{equation}
\label{AQ}
\left[ \cos \Theta_{\vec{\bf k}}  \mp 1 \right] a + \tet{e}^{i \phi_{\vec{\bf k}}} \, \sin \Theta_{\vec{\bf k}} \, b = 0 \,  . 
\end{equation}
Solving Eq.~\eqref{AQ} and normalization condition, the spinor eigenstates of a two-level system are obtained in the form of a qubit state  

\begin{equation}
\label{apsi01}
\Psi^{(0)}_{[s = 1]}(\vec{\bf k}) =  \left\{
\begin{array}{c}
\cos \left( \frac{\Theta_{\vec{\bf k}}}{2} \right) \\[0.3cm]
\sin \left( \frac{\Theta_{\vec{\bf k}}}{2} \right) \, \tet{e}^{i \phi_{\vec{\bf k}}} 
\end{array}
\right\} 
\, . 
\end{equation}
and 

\begin{equation}
\label{apsi02}
\Psi^{(0)}_{[s=- 1]}(\vec{\bf k}) =  \left\{
\begin{array}{c}
\sin \left( \frac{\Theta_{\vec{\bf k}}}{2} \right) \\[0.3cm]
- \, \cos \left( \frac{\Theta_{\vec{\bf k}}}{2} \right) \, \tet{e}^{i \phi_{\vec{\bf k}}} 
\end{array}
\right\} 
\, , 
\end{equation}
while the total phase of both components of the spinor is irrelevant, as expected.

\subsection{Berry curvature for irradiated altermagnets}

The Berry curvature is defined as

\begin{equation}
\label{abcur}
\vec{\bf \Omega} (s,\vec{\bf k}) = \vec{\nabla}_{(\vec{\bf k})} \times \vec{\bf A}(\vec{\bf k}) \, , 
\end{equation}
where 

\begin{equation}
\vec{\bf A}(s,\vec{\bf k})  = \Big \langle \Psi_s (\vec{\bf k}) \Big \vert \vec{\nabla}_{\vec{\bf k}} \Psi_s (\vec{\bf k}) \Big \rangle
\,  
\end{equation}
is the Berry connection. $\Big \vert  \Psi_s (\vec{\bf k}) \Big \rangle$ represents on of two wave functions given by Eqs.~\eqref{apsi01} and ~\eqref{apsi02}. We expand vector $\vec{\bf A}(s,\vec{\bf k})$ in polar coordinates $(\hat{\bf k},\hat{\phi})$ as $\vec{\bf A}(s,\vec{\bf k}) = A_k(s,\vec{\bf k}) \,\hat{\bf k} + A_\phi(s,\vec{\bf k}) \,\hat{\phi}$. Therefore, we can directly calculate each of its components

\begin{equation}
\vec{\bf A}(s=1,\vec{\bf k})  = \Big \langle \Psi_s (\vec{\bf k}) \Big \vert \frac{\pr}{\pr k} \Psi_s (\vec{\bf k}) \Big \rangle = 0 
\,  
\end{equation}
since 

\begin{equation}
\vert \frac{\pr}{\pr k} \Psi_s (\vec{\bf k}) \Big \rangle = \frac{1}{2}
 \left\{
\begin{array}{c}
- \sin \left( \frac{\Theta_{\vec{\bf k}}}{2} \right) \\[0.3cm]
\cos \left( \frac{\Theta_{\vec{\bf k}}}{2} \right) \, \tet{e}^{i \phi_{\vec{\bf k}}} 
\end{array}
\right\} 
\, \frac{\pr \Theta_{\vec{\bf k}}}{\pr k}
\, . 
\end{equation}
The other components $\vec{\bf A}(s,\vec{\bf k})$ is calculated as

\begin{equation}
A_{\hat{\phi}} (s=1,\vec{\bf k})  = \Big \langle \Psi_s (\vec{\bf k}) \Big \vert \frac{1}{k} \frac{\pr}{\pr \phi} \Psi_s (\vec{\bf k}) \Big \rangle = 
\frac{1}{k} \,  \sin^2 \left( \frac{\Theta_{\vec{\bf k}}}{2} \right) = \frac{1}{2k} \left[  
1- \cos \left( \Theta_{\vec{\bf k}} \right) 
\right]
\, .
\end{equation}

Thus, we derive

\begin{equation}
\vec{\bf A}(s=1,\vec{\bf k}) = 0 \hat{\bf k} + \frac{1}{2k} \left[  
1- \cos \left( \Theta_{\vec{\bf k}} \right) 
\right] \, \hat{\phi}
\, .
\end{equation}

The calculation for $s = -1$ results in 

\begin{equation}
\vec{\bf A}(s=-1,\vec{\bf k}) = 0 \hat{\bf k} - \frac{1}{2k} \left[  
1- \cos \left( \Theta_{\vec{\bf k}} \right) 
\right] \, \hat{\phi}
\, 
\end{equation}
and we obtain the expression for Berry connection as 

\begin{equation}
\label{bcon}
\vec{\bf A}(s,\vec{\bf k}) = \frac{s}{2k} \left[  
1- \cos \left( \Theta_{\vec{\bf k}} \right) 
\right] \, \hat{\phi} = \frac{s}{2k} \left[  
1- \frac{V_3(\bf \vec{k})}{\vert \vec{\bf V} (\vec{k}) \vert} 
\right] \, \hat{\phi}
\, ,
\end{equation}
where we used Eq.~\eqref{av3cos}.

\medskip 
Using Eq.~\eqref{abcur}, we can now calculate the only non-zero component of the Berry curvature

\begin{equation}
\vec{\bf \Omega}_z (s,\vec{\bf k}) = \left[ \vec{\nabla}_{(\vec{\bf k})} \times \vec{\bf A}(\vec{\bf k})\right]_z = 
\frac{1}{k} \frac{\pr }{\pr k} \, \left( k A_{\hat{\phi}} \right) = \frac{1}{2 k} \frac{\pr }{\pr k} \, \left[
- \cos \left( \Theta_{\vec{\bf k}} \right) 
\right] \, \hat{\phi} = - \frac{s}{2k} \frac{\pr }{\pr k}  \, \left[ \frac{V_3(\bf \vec{k})}{\vert \vec{\bf V} (\vec{k}) \vert} 
\right]
\, .
\end{equation}
For the case of a gapped graphene with $V_1(k)= \hbar v_F k_x$ and $V_2(k) = \hbar v_F k_y$ and $V_3(k) = \Delta_0$ ($\pr V_3(k)/\pr k = 0$), we immediately obtain 

\begin{equation}
\vec{\bf \Omega}_z (s,\vec{\bf k}) = - \frac{s}{2k} \frac{\pr }{\pr k}  \, \left[  
\frac{\Delta_0}{\sqrt{
(\hbar v_F\, k)^2 + \Delta_0^2 
}} 
\right] = \frac{s (\hbar v_F)^2 \Delta_0}{2 \left[ (\hbar v_F\, k)^2 + \Delta_0^2 \right]^{3/2}}
\, ,
\end{equation}
which reaches its maximum $(\hbar v_F)^2/(2 \Delta_0^2)$ at $\vec{\bf k} = 0$.

For an altermagnet with $d_{x^2-y^2}$ pairing symmetry, we obtain

\begin{equation}
\vec{\bf \Omega}_{z,1} (s,\vec{\bf k}) = \frac{s \, \mathfrak{A}_1 \,\rho^2 \, \cos(2 \phi_{\vec{\bf k}})}{2 k\, \left[
\rho^2 + \left\{\mathfrak{A}_1 k \cos(2 \phi_{\vec{\bf k}})\right\}^2 
\right]^{3/2}
} \, , 
\end{equation}

and 

\begin{equation}
\vec{\bf \Omega}_{z,2} (s,\vec{\bf k}) = \frac{2 s \, \mathfrak{A}_2 \,\rho^2 \, \sin(2 \phi_{\vec{\bf k}})}{k \, \left[
(2\rho)^2 + \left\{\mathfrak{A}_1 k \sin(2 \phi_{\vec{\bf k}})\right\}^2 
\right]^{3/2}
} \, 
\end{equation}
for $d_{xy}$ pairing symmetry . It is important to keep in mind that non-zero Berry curvature leads to a finite Chern number of a considered electronic state.

\section{Finding the electron states in the presence of off-resonance dressing fields}
\label{apb}

Now we address the problem of calculating the electronic states of a $d$-wave altermagnet in the presence of an off-resonance dressing field. We consider how each of the term of the Hamiltonian \eqref{mainHam} 

\begin{equation}
\label{AMainHam01}
\hat{\mc{H}}_{1,2}(\vec{\bf k}) = \hat{\mc{H}}^{M}(\vec{\bf k}) + \hat{\mc{H}}_{1,2}^{\mathfrak{A}}(\vec{\bf k}) + \hat{\mc{H}}^{R}(\vec{\bf k})
\end{equation}

where 

\begin{equation}
 \hat{\mc{H}}^{M}(\vec{\bf k})  = \frac{(\hbar k)^2}{2 m} \, \hat{\Sigma}_0^{(2)} =  \frac{\hbar^2}{2 m} \, \left( k_x^2 + k_y^2\right) \,\hat{\Sigma}_0^{(2)} 
\end{equation}
is a standard diagonal and isotropic mass term. The two types $\mc{H}_{1,2}$ of altermagnets corresponding to $d_{x^2 - y^2}$ and $d_{xy}$ pairing symmetries are distinguished by the $\hat{\Sigma}_z^{(2)}$ term:

\begin{equation}
 \hat{\mc{H}}_{1}^{\mathfrak{A}}(\vec{\bf k})  =   \frac{\hbar^2 \mathfrak{A}_1}{2 m}  \, \left( k_x^2 - k_y^2 \right)  \,  \hat{\Sigma}_z^{(2)} \, . 
\end{equation}

and
\begin{equation}
 \hat{\mc{H}}_{2}^{\mathfrak{A}}(\vec{\bf k})  =   \frac{\hbar^2 \mathfrak{A}_2}{2 m}  \, \left( k_x k_y \right)  \,  \hat{\Sigma}_z^{(2)} \, . 
\end{equation}
corresponding to $d_{xy}$ type of symmetry of an altermagnet.

Finally, 

\begin{equation}
 \hat{\mc{H}}_{1,2}^{R}(\vec{\bf k}) = r_{SO} \,
\left( k_x \,  \Sigma_y^{(2)} - k_y \,  \Sigma_x^{(2)} \right) \, , 
\end{equation}
is the Rashba spin-orbit coupling Hamiltonian induced by the external gates voltage. 

\medskip 
We are going to calculate the effect of the off-resonance dressing field is given by a canonical substitute of the electron momentum $k_{x,y} \longrightarrow k_{x,y} - e/\hbar A_{x,y}$. As a result, we obtain time-dependent vector potential. 

\par
Let us first address the most general elliptical polarization with the following vector potential

\begin{equation}
\mc{A}^{(E)}(t) = \frac{E_0}{\omega} \,  \left(
\begin{array}{c}
\cos (\omega t) \\
\beta \sin (\omega t)
\end{array}
\right) \,   \, .
\end{equation}

\medskip 
Importantly, there are several approach to calculating the dressed states. The non-linear $k$-terms are normally treated using van Vleck expansion

\begin{equation}
\label{AVVexp01}
 \hat{\mc{H}}^{(D)}(\vec{\bf k}) =  \hat{\mc{H}}_0^{(F)}(\vec{\bf k}) + (\hbar \omega)^{-1} \, \left[ \hat{\mc{H}}_{-1}^{(F)}(\vec{\bf k}),  \hat{\mc{H}}_{+1}^{(F)}(\vec{\bf k})  \right] + 
1/2 \, (\hbar \omega)^{-2} \, \left[ \hat{\mc{H}}_{-1}^{(F)}(\vec{\bf k}), \,\left[  \hat{\mc{H}}_{0}^{(F)}(\vec{\bf k}),  \hat{\mc{H}}_{+1}^{(F)}(\vec{\bf k}) \right] \, \right] ... \, .
\end{equation}
In most cases, taking only the linear expansion term $ (\hbar \omega)^{-1} \, \left[ \hat{\mc{H}}_{-1}^{(F)}(\vec{\bf k}),  \hat{\mc{H}}_{+1}^{(F)}(\vec{\bf k})  \right] $  into account would be sufficient. 

\par 
Here, the Floquet components $ \hat{\mc{H}}_{[n=0,1,-1]}^{(F)}(\vec{\bf k})$ are time-independent and are given as the time-integrals 

\begin{equation}
\label{Aint01}
 \hat{\mc{H}}_{n}^{(F)}(\vec{\bf k}) = \frac{ \omega}{2 \pi} \, \int\limits_0^{2 \pi/\omega} \,  \hat{\mc{H}}_{1,2}^{(0)}\left[\vec{\bf k}-\frac{e}{\hbar}\vec{\bf A}(t) \right] \, \tet{exp}[i n \, \omega t] \, d t \, ,
\end{equation}
over the period $T = 2 \pi/\omega$ of the time-dependent and periodic terms $ \hat{\mc{H}}_{1,2}^{(0)}\left[\vec{\bf k}-\frac{e}{\hbar}\vec{\bf A}(t) \right] $ obtained from initial non-interacting Hamiltonian \eqref{AMainHam01} in which wave vector $\vec{\bf k}$ is replaced by a canonical substitution $k_{x,y} \longrightarrow k_{x,y} - e/\hbar A_{x,y}$.

\par 
We now consider the case of elliptically polarized irradiation and most common $d_{x^2-y^2}$ type of altermagnets and derive the exact equations for each part of the Hamiltonian affected by the irradiation. The canonical substitution leads to the following time-dependent Hamiltonian

\begin{equation}
 \hat{\mc{H}}^{M}(\vec{\bf k},t)  = \frac{\hbar^2}{2 m} \left\{ k^2 - \frac{2 e E_0}{\hbar \omega} \, [k_x  \cos (\omega t) + k_y \beta \sin (\omega t)] + 
\left( \frac{e E_0}{\hbar \omega} \right)^2  \left[\frac{1+\beta^2}{2} + \frac{1-\beta^2}{2} \cos(2 \omega t) \, \right] \,
\right\} \, \hat{\Sigma}_0^{(2)}
\end{equation}

\begin{eqnarray}
 \hat{\mc{H}}_{1}^{\mathfrak{A}}(\vec{\bf k},t) & =  &  \frac{\hbar^2 \mathfrak{A}_1}{2 m}\left\{ \left( k_x^2 - k_y^2 \right) 
- \frac{e E_0}{\hbar \omega} \, \, [k_x  \cos (\omega t) - \beta k_y \, \sin (\omega t)] + \right. \\
\nonumber 
& + & \left. \left( \frac{e E_0}{\hbar \omega} \right)^2 \left[
\frac{1-\beta^2}{2} + \frac{1+\beta^2}{2} \cos(2 \omega t) 
\right] \,
 \right\} \,  \hat{\Sigma}_z^{(2)} \, . 
\end{eqnarray}

The Rashba spin-orbit coupling Hamiltonian is transformed in the most straightforward way

\begin{equation}
 \hat{\mc{H}}_{1,2}^{R}(\vec{\bf k}) = r_{SO} \,
\left( k_x \,  \Sigma_y^{(2)} - k_y \,  \Sigma_x^{(2)} \right)  - r_{SO} \,\frac{e E_0}{\hbar \omega} \,
\left[ \cos (\omega t) \,  \Sigma_y^{(2)} - \beta \sin (\omega t)  \,  \Sigma_x^{(2)} \right] \, . 
\end{equation}
Next, we evaluate the integrals in Eq. \eqref{Aint01} and obtain

\begin{eqnarray}
 \hat{\mc{H}}_{1}^{(E)}(\vec{\bf k}) &  = & \mc{H}_{[n=0]}^{(F)}(\vec{\bf k})  + \beta \, \left( \frac{e E_0 r}{\hbar \omega} \right)^2  \Sigma_z^{(2)} -
\beta \,  k_y  \, \frac{\mathfrak{A}_1 \, r_{SO}}{2 m} \, \left(\frac{e E_0}{\omega} \right)^2 \, \Sigma_x^{(2)}
\, - \\
\nonumber 
&-&  
\beta \,  k_x  \, \frac{\mathfrak{A}_1 \, r_{SO} }{2 m} \left(\frac{e E_0}{\omega} \right)^2 \, \Sigma_y^{(2)} \, , 
\end{eqnarray}
where the zero-order Floquet Hamiltonian $\mc{H}_{[n=0]}^{(F)}(\vec{\bf k})$

\begin{eqnarray}
\mc{H}_{[n=0]}^{(F)}(\vec{\bf k}) &=& \mc{H}_{1,2}^{(0)}(\vec{\bf k}) + \frac{1+\beta^2}{4 m} \, \left( \frac{e E_0}{\omega} \right)^2 = \frac{(\hbar k)^2}{2 m} \, \hat{\Sigma}_0^{(2)} +  \frac{\hbar^2 \mathfrak{A}_1}{2 m}  \, \left(k_x^2 - k_y^2 \right)  \,  \hat{\Sigma}_z^{(2)} \, + \\
\nonumber 
& + & r_{SO} \,
\left( k_x \,  \Sigma_y^{(2)} - k_y \,  \Sigma_x^{(2)} \right) + \frac{1+\beta^2}{4 m} \, \left( \frac{e E_0}{\omega} \right)^2 \, \Sigma_0^{(2)} +
\frac{1-\beta^2}{4 m} \, \left( \frac{e E_0}{\omega} \right)^2 \, \Sigma_z^{(2)}
 \, , 
\end{eqnarray}
is equal to the initial Hamiltonian \eqref{mainHam}, an additional constant energy shift $\hbar^2/(4 m) \,\left[ e E_0/(\hbar \omega) \right]^2\,(1+\beta^2)$ and an extra bandgap $(1-\beta^2)/(2 m) \, \left[e E_0/(\hbar \omega) \right]^2 $.  

\par 
Here, we used the following relations 

\begin{eqnarray}
\int\limits_0^{2 \pi/\omega} \cos (n \omega t) \, \tet{e}^{\pm i \omega t} & = & \frac{\pi}{\omega} \, \delta_{n,1} \, , \\
\nonumber 
\int\limits_0^{2 \pi/\omega} \sin(n \omega t) \, \tet{e}^{\pm i \omega t} & =  & \frac{\pm i \pi}{\omega} \, \delta_{n,1} \, ,
\end{eqnarray}
as well as 

\begin{eqnarray}
&& \left[\, \sum\limits_{\lambda = 0}^{3} C_{p,\lambda}  \, \hat{\Sigma}_\lambda^{(2)},  \,\,
\sum\limits_{\mu = 0}^{3} C_{m,\mu}  \, \hat{\Sigma}_\mu^{(2)} \,
 \right]  \, =  \\
\nonumber 
& = & \left\{
\begin{array}{cc}
2 i (C_{m,2} C_{p,1} - C_{m,1} C_{p,2}) & -2 C_{m,3} ( C_{p,1} - 2 i C_{p,2}) + ( 2 C_{m,1} - 2 i C_{m,2}) C_{p,3} \\
2 C_{m,3} (C_{p,1} + i C_{p,2}) - 2 (C_{m,1} + i C_{m,2}) C_{p,3} & 2 i (C_{m,1} C_{p,2} - C_{m,2} C_{p,1})
\end{array}
\right\} \, . 
\end{eqnarray}

\subsection{Altermagnets with $d_{xy}$ symmetry under elliptically polarized dressing field}

Now we calculate the electron dressed states for an altermagnet with $d_{xy}$ symmetry. We notice that for $\mc{H}_{2}$ Hamiltonian the altermagnetic term is the only term which needs to be calculated again

\begin{eqnarray}
 \mc{H}_{2}^{\mathfrak{A}}(\vec{\bf k},t) & =  &  \frac{\hbar^2 \mathfrak{A}_1}{2 m}\left\{ k_x k_y  
- \frac{e E_0}{\hbar \omega} \, \, [k_x  \beta \sin (\omega t) + \beta k_y \, \cos (\omega t)] + \right. \\
\nonumber 
& + & \left.\frac{\beta}{2} \, \left( \frac{e E_0}{\hbar \omega} \right)^2 \, \sin(2 \omega t) 
\right\} \,  \hat{\Sigma}_z^{(2)} \, . 
\end{eqnarray}

As a result, we have the following effective Floquet time-independent Hamiltonian 

\begin{eqnarray}
 \hat{\mc{H}}_{2}^{(E)}(\vec{\bf k}) &  = & \mc{H}_{[n=0]}^{(F)}(\vec{\bf k})  + \beta \, \left( \frac{e E_0 r}{\hbar \omega} \right)^2  \Sigma_z^{(2)} -
 k_y  \, \frac{\mathfrak{A}_2 \, r_{SO}}{2 m} \, \left(\frac{e E_0}{\omega} \right)^2 \, \Sigma_x^{(2)}
\, - \\
\nonumber 
&-&  
\beta^2 \,  k_x  \, \frac{\mathfrak{A}_2 \, r_{SO} }{2 m} \left(\frac{e E_0}{\omega} \right)^2 \, \Sigma_y^{(2)} \, , 
\end{eqnarray}
where the zero-order Floquet Hamiltonian $\mc{H}_{[n=0]}^{(F)}(\vec{\bf k})$

\begin{eqnarray}
 \hat{\mc{H}}_{[n=0]}^{(F)}(\vec{\bf k}) &=& \mc{H}_{1,2}^{(0)}(\vec{\bf k}) + \frac{1+\beta^2}{4 m} \, \left( \frac{e E_0}{\omega} \right)^2 = \frac{(\hbar k)^2}{2 m} \, \hat{\Sigma}_0^{(2)} +  \frac{\hbar^2 \mathfrak{A}_2}{2 m}  \,  k_x k_y  \,  \hat{\Sigma}_z^{(2)} \, + \\
\nonumber 
& + & r_{SO} \,
\left( k_x \,  \Sigma_y^{(2)} - k_y \,  \Sigma_x^{(2)} \right) + \frac{1+\beta^2}{4 m} \, \left( \frac{e E_0}{\omega} \right)^2 \, \Sigma_0^{(2)} +
\frac{1-\beta^2}{4 m} \, \left( \frac{e E_0}{\omega} \right)^2 \, \Sigma_z^{(2)}
 \, , 
\end{eqnarray}
is the same as for the previously considered $d_{x^2-y^2}$ type of symmetry. We see that $\Sigma_x^{(2)}$ and $\Sigma_y^{(2)}$ terms are affected in non-equivalent ways for $\beta \neq 1$.

\subsection{Linearly polarized irradiation with an arbitrary polarization direction $\theta_0$}

We now consider the electronic states in a $d_{x^2 - y^2}$ altermagnet with a Hamiltonian \eqref{mainHam} in the presence of a linearly polarized dressing field 

\begin{equation}
\mc{A}^{(L)}(t) =  \frac{E_0}{\omega} \, \left(
\begin{array}{c}
\cos \theta_0 \\
\sin \theta_0
\end{array}
\right) \, \cos (\omega t) \, ,  
\end{equation}
where $\theta_0$ is the fixed direction of the external field polarization. 

\par 
The components of the wave vector is modified as $k_x \longrightarrow k_x  -  e E_0/( \hbar\omega) \, \cos \theta_0 \, \cos(\omega t)$ and  $k_y \longrightarrow k_y  -  e E_0/( \hbar\omega) \, \sin \theta_0 \, \cos(\omega t)$.

\medskip 
The time-dependent Hamiltonian now becomes

\begin{equation}
\label{HamtLin01}
 \hat{\mc{H}}_1(\vec{\bf k},t)  = \mc{H}^{M}(\vec{\bf k},t) +  \mc{H}_{1,2}^{\mathfrak{A}}(\vec{\bf k},t) + \mc{H}^{R}(\vec{\bf k})
\end{equation}

\begin{equation}
 \hat{\mc{H}}^{M}(\vec{\bf k},t)  = \frac{\hbar^2}{2 m} \left\{ k^2 - \frac{2 e E_0}{\hbar \omega} \, [k_x  \cos\theta_0 + k_y \sin \theta_0] \,  \cos (\omega t) + 
\frac{1}{2} \, \left( \frac{e E_0}{\hbar \omega} \right)^2 \, \left[1 +  \cos(2 \omega t) \, \right] \,
\right\} \, \hat{\Sigma}_0^{(2)}
\end{equation}

\begin{eqnarray}
 \hat{\mc{H}}_{1,2}^{\mathfrak{A}}(\vec{\bf k},t) & =  &  \frac{\hbar^2 \mathfrak{A}_1}{2 m}\left\{ \left( k_x^2 - k_y^2 \right) 
- \frac{2 e E_0}{\hbar \omega} \, [k_x  \cos\theta_0 - k_y \sin \theta_0] \,  \cos (\omega t) + \right. \\
\nonumber 
& + & \left. \frac{1}{2} \, \left( \frac{e E_0}{\hbar \omega} \right)^2 \, \left[1 +  \cos(2 \omega t) \, \right]  \,
 \right\} \,  \hat{\Sigma}_z^{(2)} \, . 
\end{eqnarray}

The Rashba spin-orbit coupling Hamiltonian is transformed in the most straightforward way

\begin{equation}
 \hat{\mc{H}}_{1,2}^{R}(\vec{\bf k}) = r_{SO} \,
\left( k_x \,  \Sigma_y^{(2)} - k_y \,  \Sigma_x^{(2)} \right)  - r_{SO} \,\frac{e E_0}{\hbar \omega} \,
\left[ \cos \theta_0 \,  \Sigma_y^{(2)} - \sin \theta_0  \,  \Sigma_x^{(2)} \right] \,   \cos (\omega t) . 
\end{equation}
Analyzing Hamiltonian \eqref{BHamtLin01}, we discern that $ \hat{\mc{H}}_{-1}^{(F)}(\vec{\bf k}) \equiv  \hat{\mc{H}}_{1}^{(F)}(\vec{\bf k})$ and a linear term in Van Vleck expansion could only yield zero. Therefore, this time we must take into account the $\backsim 1/\omega^2$ corrections in Eq.~\eqref{AVVexp01}.

\medskip
\par

Now we can derive following Floquet time-independent Hamiltonian for a $d_{x^2-y^2}$ altermagnet

\begin{eqnarray}
\label{BLingen01}
 \hat{\mc{H}}_{2}^{\mathfrak{A}}(\vec{\bf k},t) =\mc{H}_{[n=0]}^{(F)}(\vec{\bf k})  - \mc{F}^{(z)}_{\mathfrak{A}}(r_{SO}, \omega \, \vert \, \theta_0) \Sigma_z^{(2)} + \mc{F}^{(x)}_{\mathfrak{A}}(r_{SO}, \omega \, \vert \, \theta_0) \Sigma_x^{(2)} + \mc{F}^{(y)}_{\mathfrak{A}}(r_{SO}, \omega \, \vert \, \theta_0) \Sigma_y^{(2)}
\, ,
\end{eqnarray}
where

\begin{eqnarray}
&& \mc{F}^{(z)}_{\mathfrak{A}}(r_{SO}, \omega \, \vert \, \theta_0) = \frac{\mathfrak{A}_1}{8 m \omega^4} \, \left(e E_0 \, r_{SO} \right)^2 \, \left\{ 
2 \cos \theta_0 \, \left[k^2 - 2 k_x (k_x + k_y)\right] + 2 \sin \theta_0 \, \left[k^2 + 2 k_y (k_x + k_y) \right] 
\right\}
, \\
\nonumber 
&& \mc{F}^{(x)}_{\mathfrak{A}}(r_{SO}, \omega \, \vert \, \theta_0) = \frac{r^3 (e E_0)^2}{2 (\hbar \omega)^4} \cos \theta_0 \left( \cos \theta_0 k_y - \sin \theta_0 k_x \right)  -  \left( \frac{\mathfrak{A}_1 \, e E_0}{ 2 m \hbar \omega^2} \right)^2 \, r_{SO} \, 
\left( \cos \theta_0 k_y - \sin \theta_0 k_x \right) \, \times \\
& \times &  \Big[ 2 \cos \theta_0  \, k_x k_y - \sin \theta_0  \left( k_x^2 + 3 k_y^2 \right) \Big] \, , \\
\nonumber 
&& \mc{F}^{(y)}_{\mathfrak{A}}(r_{SO}, \omega \, \vert \, \theta_0) = -\frac{r^3 (e E_0)^2}{2 (\hbar \omega)^4} \cos \theta_0 \left( \cos \theta_0 k_y - \sin \theta_0 k_x \right)  -  \left( \frac{\mathfrak{A}_1 \, e E_0}{ 2 m \hbar \omega^2} \right)^2 \, r_{SO} \, 
\left( \cos \theta_0 k_y - \sin \theta_0 k_x \right) \, \times \\
& \times &  \Big[ 2 \cos \theta_0 \, k_x^2 -\cos \theta_0 \, k^2 - 2 k_x k_y \, \sin \theta_0 \Big]
\end{eqnarray}
The zero-order Floquet Hamiltonian $\mc{H}_{[n=0]}^{(F)}(\vec{\bf k})$

\begin{eqnarray}
\mc{H}_{[n=0]}^{(F)}(\vec{\bf k}) &=& \mc{H}_{1}^{(0)}(\vec{\bf k}) + \frac{\hbar^2}{4 m} \, \left( \frac{e E_0}{\omega} \right)^2 \hat{\Sigma}_0^{(2)} + \frac{\hbar^2 \, \mathfrak{A}_1}{4 m} \, \left( \frac{e E_0}{\omega} \right)^2 \hat{\Sigma}_z^{(2)}= 
\frac{(\hbar k)^2}{2 m} \, \hat{\Sigma}_0^{(2)} \, + \\
\nonumber 
& + & \frac{\hbar^2 \, \mathfrak{A}_1}{2 m}  \,  (k_x^2 - k_y^2)  \,  \hat{\Sigma}_z^{(2)}  + r_{SO} \,
\left( k_x \,  \Sigma_y^{(2)} - k_y \,  \Sigma_x^{(2)} \right) + \frac{\hbar^2}{4 m} \, \left( \frac{e E_0}{\omega} \right)^2 \hat{\Sigma}_0^{(2)} + \frac{\hbar^2 \, \mathfrak{A}_1}{4 m} \, \left( \frac{e E_0}{\omega} \right)^2 \hat{\Sigma}_z^{(2)}
 \, , 
\end{eqnarray}
we see that a finite bandgap is created independent of the direction of the polarization of the dressing field. 

\medskip 
For the polarization direction along the $x-$axis, equation \eqref{BLingen01} is simplified as 

\begin{eqnarray}
&& \hat{\mc{H}}_{1}^{\mathfrak{A}}(\vec{\bf k}) = \mc{H}_{[n=0]}^{(F)}(\vec{\bf k})   - \frac{\mathfrak{A}_1 (e E_0\, r_SO)^2}{4 m \hbar^2 \omega^4} \, \left[
k^2 - 2k_x (k_x + k_y) 
\right]^2  \, \Sigma_z^{(2)}
\, + \\
\nonumber 
&+& \left\{  
\frac{(e E_0)^2}{2 (\hbar \omega)^4} r_{SO}^3 \, k_y  + r_{SO} \, \left( \frac{\hbar \mathfrak{A}_1 \,e E_0}{2 m \omega^2}  \right)^2 \, k_x^2 k_y 
\right\} \, \Sigma_x^{(2)} \, - \\
\nonumber 
& - & \left\{  
\frac{(e E_0)^2}{2 (\hbar \omega)^4} r_{SO}^3 \, k_y  + r_{SO} \, \left( \frac{\hbar \mathfrak{A}_1 \,e E_0}{2 m \omega^2}  \right)^2 \, k_x \, \left( k_x^2 - k_y^2 \right)
\right\} \, \Sigma_y^{(2)}
\, , 
\end{eqnarray}
which demonstrates that the second-order perturbation expansion terms brings small but still important corrections to the band structure of irradiated altermagnets. 

\medskip 
Finally, let us briefly address the case of a $d_{xy}$ altermagnet under linearly polarized dressing field. We perform the perturbation expansion up to the $\backsim \omega^{-2}$ order. The time-dependent Hamiltonian now becomes

\begin{equation}
\label{BHamtLin01}
 \hat{\mc{H}}_1^{(0)}(\vec{\bf k},t)  = \mc{H}^{(0),\,M}(\vec{\bf k},t) +  \mc{H}_{1}^{\mathfrak{A}}(\vec{\bf k},t) + \mc{H}_{1,2}^{(0),\,R}(\vec{\bf k})
\end{equation}
where $\mc{H}^{(0),\,M}(\vec{\bf k},t)$ and $\mc{H}_{1,2}^{(0),\,R}(\vec{\bf k}) $ remain unchanged and 

\begin{eqnarray}
 \hat{\mc{H}}_{1}^{\mathfrak{A}}(\vec{\bf k},t) & =  &  \frac{\hbar^2 \mathfrak{A}_1}{2 m}\left\{  k_x k_y
- \frac{2 e E_0}{\hbar \omega} \, [k_x  \sin\theta_0 + k_y \cos \theta_0] \,  \cos (\omega t) + \right. \\
\nonumber 
& + & \left. \frac{1}{4} \, \left( \frac{e E_0}{\hbar \omega} \right)^2 \,\sin (2 \theta_0) \,  \left[1 +  \cos(2 \omega t) \, \right]  \,
 \right\} \,  \hat{\Sigma}_z^{(2)} \, . 
\end{eqnarray}
The zero-order Floquet Hamiltonian $\mc{H}_{[n=0]}^{(F)}(\vec{\bf k})$ is given by 

\begin{eqnarray}
 \hat{\mc{H}}_{[n=0]}^{(F)}(\vec{\bf k}) & = &
\frac{(\hbar k)^2}{2 m} \, \hat{\Sigma}_0^{(2)} + \frac{\hbar^2 \, \mathfrak{A}_1}{2 m}  \,  (k_x^2 - k_y^2)  \,  \hat{\Sigma}_z^{(2)}  + r_{SO} \,
\left( k_x \,  \Sigma_y^{(2)} - k_y \,  \Sigma_x^{(2)} \right) + \frac{\hbar^2}{4 m} \, \left( \frac{e E_0}{\omega} \right)^2 \hat{\Sigma}_0^{(2)} \, + \\
\nonumber 
&+& \frac{\hbar^2 \, \mathfrak{A}_1}{8 m} \, \sin{2 \theta_0} \, \left( \frac{e E_0}{\omega} \right)^2 \hat{\Sigma}_z^{(2)}
 \, . 
\end{eqnarray}
Therefore, the bandgap $2 \Delta_2^{(L)}$ is obtained as 

\begin{equation}
\Delta_2^{(L)} = \frac{\hbar^2 \, \mathfrak{A}_2}{8 m} \, \sin(2 \theta_0) \, \left( \frac{e E_0}{\omega} \right)^2  = \frac{\mathfrak{A}_2}{4} \, \sin (2 \theta_0) \, K_\omega^2 \, ,
\end{equation}
and disappears at $\theta_0 = \pi n /2$, $n = 1,2,3, ...$. The changes to the finite-$k$ dispersions are given as the modifications of the components of its anisotropic Fermi velocity and are shown at $ \backsim 1/\omega^4$ level, which are are small but not negligible.

\bibliography{DLP}

\begin{thebibliography}{105}
\expandafter\ifx\csname natexlab\endcsname\relax\def\natexlab#1{#1}\fi
\expandafter\ifx\csname bibnamefont\endcsname\relax
  \def\bibnamefont#1{#1}\fi
\expandafter\ifx\csname bibfnamefont\endcsname\relax
  \def\bibfnamefont#1{#1}\fi
\expandafter\ifx\csname citenamefont\endcsname\relax
  \def\citenamefont#1{#1}\fi
\expandafter\ifx\csname url\endcsname\relax
  \def\url#1{\texttt{#1}}\fi
\expandafter\ifx\csname urlprefix\endcsname\relax\def\urlprefix{URL }\fi
\providecommand{\bibinfo}[2]{#2}
\providecommand{\eprint}[2][]{\url{#2}}

\bibitem[{\citenamefont{{\v{S}}mejkal et~al.}(2020)\citenamefont{{\v{S}}mejkal,
  Gonz{\'a}lez-Hern{\'a}ndez, Jungwirth, and Sinova}}]{vsmejkal2020crystal}
\bibinfo{author}{\bibfnamefont{L.}~\bibnamefont{{\v{S}}mejkal}},
  \bibinfo{author}{\bibfnamefont{R.}~\bibnamefont{Gonz{\'a}lez-Hern{\'a}ndez}},
  \bibinfo{author}{\bibfnamefont{T.}~\bibnamefont{Jungwirth}},
  \bibnamefont{and} \bibinfo{author}{\bibfnamefont{J.}~\bibnamefont{Sinova}},
  \bibinfo{journal}{Science advances} \textbf{\bibinfo{volume}{6}},
  \bibinfo{pages}{eaaz8809} (\bibinfo{year}{2020}).

\bibitem[{\citenamefont{Negi et~al.}(2025)\citenamefont{Negi, Aradhyula, Dutta,
  and Roychowdhury}}]{negi2025mnte}
\bibinfo{author}{\bibfnamefont{P.}~\bibnamefont{Negi}},
  \bibinfo{author}{\bibfnamefont{S.~S. S.~A.} \bibnamefont{Aradhyula}},
  \bibinfo{author}{\bibfnamefont{S.}~\bibnamefont{Dutta}}, \bibnamefont{and}
  \bibinfo{author}{\bibfnamefont{S.}~\bibnamefont{Roychowdhury}},
  \bibinfo{journal}{Chemistry of Materials} \textbf{\bibinfo{volume}{37}},
  \bibinfo{pages}{6097} (\bibinfo{year}{2025}).

\bibitem[{\citenamefont{{\v{S}}mejkal et~al.}(2022)\citenamefont{{\v{S}}mejkal,
  Sinova, and Jungwirth}}]{vsmejkal2022beyond}
\bibinfo{author}{\bibfnamefont{L.}~\bibnamefont{{\v{S}}mejkal}},
  \bibinfo{author}{\bibfnamefont{J.}~\bibnamefont{Sinova}}, \bibnamefont{and}
  \bibinfo{author}{\bibfnamefont{T.}~\bibnamefont{Jungwirth}},
  \bibinfo{journal}{Physical Review X} \textbf{\bibinfo{volume}{12}},
  \bibinfo{pages}{031042} (\bibinfo{year}{2022}).

\bibitem[{\citenamefont{Tamang et~al.}(2025)\citenamefont{Tamang, Gurung, Rai,
  Brahimi, and Lounis}}]{tamang2025altermagnetism}
\bibinfo{author}{\bibfnamefont{R.}~\bibnamefont{Tamang}},
  \bibinfo{author}{\bibfnamefont{S.}~\bibnamefont{Gurung}},
  \bibinfo{author}{\bibfnamefont{D.~P.} \bibnamefont{Rai}},
  \bibinfo{author}{\bibfnamefont{S.}~\bibnamefont{Brahimi}}, \bibnamefont{and}
  \bibinfo{author}{\bibfnamefont{S.}~\bibnamefont{Lounis}},
  \bibinfo{journal}{Magnetism} \textbf{\bibinfo{volume}{5}},
  \bibinfo{pages}{17} (\bibinfo{year}{2025}).

\bibitem[{\citenamefont{Liu et~al.}(2026{\natexlab{a}})\citenamefont{Liu, Ma,
  Zhang, Jing, Liu, and Shen}}]{liu2026symmetry}
\bibinfo{author}{\bibfnamefont{J.}~\bibnamefont{Liu}},
  \bibinfo{author}{\bibfnamefont{X.}~\bibnamefont{Ma}},
  \bibinfo{author}{\bibfnamefont{X.}~\bibnamefont{Zhang}},
  \bibinfo{author}{\bibfnamefont{W.}~\bibnamefont{Jing}},
  \bibinfo{author}{\bibfnamefont{Z.}~\bibnamefont{Liu}}, \bibnamefont{and}
  \bibinfo{author}{\bibfnamefont{D.}~\bibnamefont{Shen}},
  \bibinfo{journal}{Nano Convergence} \textbf{\bibinfo{volume}{13}},
  \bibinfo{pages}{6} (\bibinfo{year}{2026}{\natexlab{a}}).

\bibitem[{\citenamefont{Fender et~al.}(2025)\citenamefont{Fender, Gonzalez, and
  Bediako}}]{fender2025altermagnetism}
\bibinfo{author}{\bibfnamefont{S.~S.} \bibnamefont{Fender}},
  \bibinfo{author}{\bibfnamefont{O.}~\bibnamefont{Gonzalez}}, \bibnamefont{and}
  \bibinfo{author}{\bibfnamefont{D.~K.} \bibnamefont{Bediako}},
  \bibinfo{journal}{Journal of the American Chemical Society}
  \textbf{\bibinfo{volume}{147}}, \bibinfo{pages}{2257} (\bibinfo{year}{2025}).

\bibitem[{\citenamefont{Herasymchuk et~al.}(2025)\citenamefont{Herasymchuk,
  Hallberg, Hodt, Linder, Gorbar, and Sukhachov}}]{herasymchuk2025electric}
\bibinfo{author}{\bibfnamefont{A.}~\bibnamefont{Herasymchuk}},
  \bibinfo{author}{\bibfnamefont{K.~B.} \bibnamefont{Hallberg}},
  \bibinfo{author}{\bibfnamefont{E.~W.} \bibnamefont{Hodt}},
  \bibinfo{author}{\bibfnamefont{J.}~\bibnamefont{Linder}},
  \bibinfo{author}{\bibfnamefont{E.}~\bibnamefont{Gorbar}}, \bibnamefont{and}
  \bibinfo{author}{\bibfnamefont{P.}~\bibnamefont{Sukhachov}},
  \bibinfo{journal}{Physical Review B} \textbf{\bibinfo{volume}{112}},
  \bibinfo{pages}{L220404} (\bibinfo{year}{2025}).

\bibitem[{\citenamefont{Song et~al.}(2025)\citenamefont{Song, Bai, Zhou, Han,
  Reichlova, Dil, Liu, Chen, and Pan}}]{song2025altermagnets}
\bibinfo{author}{\bibfnamefont{C.}~\bibnamefont{Song}},
  \bibinfo{author}{\bibfnamefont{H.}~\bibnamefont{Bai}},
  \bibinfo{author}{\bibfnamefont{Z.}~\bibnamefont{Zhou}},
  \bibinfo{author}{\bibfnamefont{L.}~\bibnamefont{Han}},
  \bibinfo{author}{\bibfnamefont{H.}~\bibnamefont{Reichlova}},
  \bibinfo{author}{\bibfnamefont{J.~H.} \bibnamefont{Dil}},
  \bibinfo{author}{\bibfnamefont{J.}~\bibnamefont{Liu}},
  \bibinfo{author}{\bibfnamefont{X.}~\bibnamefont{Chen}}, \bibnamefont{and}
  \bibinfo{author}{\bibfnamefont{F.}~\bibnamefont{Pan}},
  \bibinfo{journal}{Nature Reviews Materials} \textbf{\bibinfo{volume}{10}},
  \bibinfo{pages}{473} (\bibinfo{year}{2025}).

\bibitem[{\citenamefont{Gomonay et~al.}(2024)\citenamefont{Gomonay, Kravchuk,
  Jaeschke-Ubiergo, Yershov, Jungwirth, {\v{S}}mejkal, Brink, and
  Sinova}}]{gomonay2024structure}
\bibinfo{author}{\bibfnamefont{O.}~\bibnamefont{Gomonay}},
  \bibinfo{author}{\bibfnamefont{V.~P.} \bibnamefont{Kravchuk}},
  \bibinfo{author}{\bibfnamefont{R.}~\bibnamefont{Jaeschke-Ubiergo}},
  \bibinfo{author}{\bibfnamefont{K.~V.} \bibnamefont{Yershov}},
  \bibinfo{author}{\bibfnamefont{T.}~\bibnamefont{Jungwirth}},
  \bibinfo{author}{\bibfnamefont{L.}~\bibnamefont{{\v{S}}mejkal}},
  \bibinfo{author}{\bibfnamefont{J.~v.~d.} \bibnamefont{Brink}},
  \bibnamefont{and} \bibinfo{author}{\bibfnamefont{J.}~\bibnamefont{Sinova}},
  \bibinfo{journal}{npj Spintronics} \textbf{\bibinfo{volume}{2}},
  \bibinfo{pages}{35} (\bibinfo{year}{2024}).

\bibitem[{\citenamefont{Hayami et~al.}(2019)\citenamefont{Hayami, Yanagi, and
  Kusunose}}]{hayami2019momentum}
\bibinfo{author}{\bibfnamefont{S.}~\bibnamefont{Hayami}},
  \bibinfo{author}{\bibfnamefont{Y.}~\bibnamefont{Yanagi}}, \bibnamefont{and}
  \bibinfo{author}{\bibfnamefont{H.}~\bibnamefont{Kusunose}},
  \bibinfo{journal}{journal of the physical society of japan}
  \textbf{\bibinfo{volume}{88}}, \bibinfo{pages}{123702}
  (\bibinfo{year}{2019}).

\bibitem[{\citenamefont{Hayami et~al.}(2020)\citenamefont{Hayami, Yanagi, and
  Kusunose}}]{hayami2020bottom}
\bibinfo{author}{\bibfnamefont{S.}~\bibnamefont{Hayami}},
  \bibinfo{author}{\bibfnamefont{Y.}~\bibnamefont{Yanagi}}, \bibnamefont{and}
  \bibinfo{author}{\bibfnamefont{H.}~\bibnamefont{Kusunose}},
  \bibinfo{journal}{Physical Review B} \textbf{\bibinfo{volume}{102}},
  \bibinfo{pages}{144441} (\bibinfo{year}{2020}).

\bibitem[{\citenamefont{He et~al.}(2025)\citenamefont{He, Wen, Okabayashi,
  Miura, Ma, Ohkubo, Seki, Sukegawa, and Mitani}}]{he2025evidence}
\bibinfo{author}{\bibfnamefont{C.}~\bibnamefont{He}},
  \bibinfo{author}{\bibfnamefont{Z.}~\bibnamefont{Wen}},
  \bibinfo{author}{\bibfnamefont{J.}~\bibnamefont{Okabayashi}},
  \bibinfo{author}{\bibfnamefont{Y.}~\bibnamefont{Miura}},
  \bibinfo{author}{\bibfnamefont{T.}~\bibnamefont{Ma}},
  \bibinfo{author}{\bibfnamefont{T.}~\bibnamefont{Ohkubo}},
  \bibinfo{author}{\bibfnamefont{T.}~\bibnamefont{Seki}},
  \bibinfo{author}{\bibfnamefont{H.}~\bibnamefont{Sukegawa}}, \bibnamefont{and}
  \bibinfo{author}{\bibfnamefont{S.}~\bibnamefont{Mitani}},
  \bibinfo{journal}{Nature Communications} \textbf{\bibinfo{volume}{16}},
  \bibinfo{pages}{8235} (\bibinfo{year}{2025}).

\bibitem[{\citenamefont{Tschirner et~al.}(2023)\citenamefont{Tschirner,
  Ke{\ss}ler, Gonzalez~Betancourt, Kotte, Kriegner, B{\"u}chner, Dufouleur,
  Kamp, Jovic, Smejkal et~al.}}]{tschirner2023saturation}
\bibinfo{author}{\bibfnamefont{T.}~\bibnamefont{Tschirner}},
  \bibinfo{author}{\bibfnamefont{P.}~\bibnamefont{Ke{\ss}ler}},
  \bibinfo{author}{\bibfnamefont{R.~D.} \bibnamefont{Gonzalez~Betancourt}},
  \bibinfo{author}{\bibfnamefont{T.}~\bibnamefont{Kotte}},
  \bibinfo{author}{\bibfnamefont{D.}~\bibnamefont{Kriegner}},
  \bibinfo{author}{\bibfnamefont{B.}~\bibnamefont{B{\"u}chner}},
  \bibinfo{author}{\bibfnamefont{J.}~\bibnamefont{Dufouleur}},
  \bibinfo{author}{\bibfnamefont{M.}~\bibnamefont{Kamp}},
  \bibinfo{author}{\bibfnamefont{V.}~\bibnamefont{Jovic}},
  \bibinfo{author}{\bibfnamefont{L.}~\bibnamefont{Smejkal}},
  \bibnamefont{et~al.}, \bibinfo{journal}{Apl Materials}
  \textbf{\bibinfo{volume}{11}} (\bibinfo{year}{2023}).

\bibitem[{\citenamefont{Zhou et~al.}(2025)\citenamefont{Zhou, Cheng, Hu, Chu,
  Bai, Han, Liu, Pan, and Song}}]{zhou2025manipulation}
\bibinfo{author}{\bibfnamefont{Z.}~\bibnamefont{Zhou}},
  \bibinfo{author}{\bibfnamefont{X.}~\bibnamefont{Cheng}},
  \bibinfo{author}{\bibfnamefont{M.}~\bibnamefont{Hu}},
  \bibinfo{author}{\bibfnamefont{R.}~\bibnamefont{Chu}},
  \bibinfo{author}{\bibfnamefont{H.}~\bibnamefont{Bai}},
  \bibinfo{author}{\bibfnamefont{L.}~\bibnamefont{Han}},
  \bibinfo{author}{\bibfnamefont{J.}~\bibnamefont{Liu}},
  \bibinfo{author}{\bibfnamefont{F.}~\bibnamefont{Pan}}, \bibnamefont{and}
  \bibinfo{author}{\bibfnamefont{C.}~\bibnamefont{Song}},
  \bibinfo{journal}{Nature} \textbf{\bibinfo{volume}{638}},
  \bibinfo{pages}{645} (\bibinfo{year}{2025}).

\bibitem[{\citenamefont{Choi et~al.}(2026)\citenamefont{Choi, Jeong, Jalan, and
  Lee}}]{choi2026exploring}
\bibinfo{author}{\bibfnamefont{I.~H.} \bibnamefont{Choi}},
  \bibinfo{author}{\bibfnamefont{S.~G.} \bibnamefont{Jeong}},
  \bibinfo{author}{\bibfnamefont{B.}~\bibnamefont{Jalan}}, \bibnamefont{and}
  \bibinfo{author}{\bibfnamefont{J.~S.} \bibnamefont{Lee}},
  \bibinfo{journal}{Nano Convergence} \textbf{\bibinfo{volume}{13}},
  \bibinfo{pages}{1} (\bibinfo{year}{2026}).

\bibitem[{\citenamefont{Fedchenko et~al.}(2024)\citenamefont{Fedchenko,
  Min{\'a}r, Akashdeep, D’souza, Vasilyev, Tkach, Odenbreit, Nguyen,
  Kutnyakhov, Wind et~al.}}]{fedchenko2024observation}
\bibinfo{author}{\bibfnamefont{O.}~\bibnamefont{Fedchenko}},
  \bibinfo{author}{\bibfnamefont{J.}~\bibnamefont{Min{\'a}r}},
  \bibinfo{author}{\bibfnamefont{A.}~\bibnamefont{Akashdeep}},
  \bibinfo{author}{\bibfnamefont{S.~W.} \bibnamefont{D’souza}},
  \bibinfo{author}{\bibfnamefont{D.}~\bibnamefont{Vasilyev}},
  \bibinfo{author}{\bibfnamefont{O.}~\bibnamefont{Tkach}},
  \bibinfo{author}{\bibfnamefont{L.}~\bibnamefont{Odenbreit}},
  \bibinfo{author}{\bibfnamefont{Q.}~\bibnamefont{Nguyen}},
  \bibinfo{author}{\bibfnamefont{D.}~\bibnamefont{Kutnyakhov}},
  \bibinfo{author}{\bibfnamefont{N.}~\bibnamefont{Wind}}, \bibnamefont{et~al.},
  \bibinfo{journal}{Science advances} \textbf{\bibinfo{volume}{10}},
  \bibinfo{pages}{eadj4883} (\bibinfo{year}{2024}).

\bibitem[{\citenamefont{Lee et~al.}(2024)\citenamefont{Lee, Lee, Jung, Jung,
  Kim, Lee, Seok, Kim, Park, {\v{S}}mejkal et~al.}}]{lee2024broken}
\bibinfo{author}{\bibfnamefont{S.}~\bibnamefont{Lee}},
  \bibinfo{author}{\bibfnamefont{S.}~\bibnamefont{Lee}},
  \bibinfo{author}{\bibfnamefont{S.}~\bibnamefont{Jung}},
  \bibinfo{author}{\bibfnamefont{J.}~\bibnamefont{Jung}},
  \bibinfo{author}{\bibfnamefont{D.}~\bibnamefont{Kim}},
  \bibinfo{author}{\bibfnamefont{Y.}~\bibnamefont{Lee}},
  \bibinfo{author}{\bibfnamefont{B.}~\bibnamefont{Seok}},
  \bibinfo{author}{\bibfnamefont{J.}~\bibnamefont{Kim}},
  \bibinfo{author}{\bibfnamefont{B.~G.} \bibnamefont{Park}},
  \bibinfo{author}{\bibfnamefont{L.}~\bibnamefont{{\v{S}}mejkal}},
  \bibnamefont{et~al.}, \bibinfo{journal}{Physical review letters}
  \textbf{\bibinfo{volume}{132}}, \bibinfo{pages}{036702}
  (\bibinfo{year}{2024}).

\bibitem[{\citenamefont{Reimers et~al.}(2024)\citenamefont{Reimers, Odenbreit,
  {\v{S}}mejkal, Strocov, Constantinou, Hellenes, Jaeschke~Ubiergo, Campos,
  Bharadwaj, Chakraborty et~al.}}]{reimers2024direct}
\bibinfo{author}{\bibfnamefont{S.}~\bibnamefont{Reimers}},
  \bibinfo{author}{\bibfnamefont{L.}~\bibnamefont{Odenbreit}},
  \bibinfo{author}{\bibfnamefont{L.}~\bibnamefont{{\v{S}}mejkal}},
  \bibinfo{author}{\bibfnamefont{V.~N.} \bibnamefont{Strocov}},
  \bibinfo{author}{\bibfnamefont{P.}~\bibnamefont{Constantinou}},
  \bibinfo{author}{\bibfnamefont{A.~B.} \bibnamefont{Hellenes}},
  \bibinfo{author}{\bibfnamefont{R.}~\bibnamefont{Jaeschke~Ubiergo}},
  \bibinfo{author}{\bibfnamefont{W.~H.} \bibnamefont{Campos}},
  \bibinfo{author}{\bibfnamefont{V.~K.} \bibnamefont{Bharadwaj}},
  \bibinfo{author}{\bibfnamefont{A.}~\bibnamefont{Chakraborty}},
  \bibnamefont{et~al.}, \bibinfo{journal}{Nature Communications}
  \textbf{\bibinfo{volume}{15}}, \bibinfo{pages}{2116} (\bibinfo{year}{2024}).

\bibitem[{\citenamefont{Amin et~al.}(2024)\citenamefont{Amin, Dal~Din, Golias,
  Niu, Zakharov, Fromage, Fields, Heywood, Cousins, Maccherozzi
  et~al.}}]{amin2024nanoscale}
\bibinfo{author}{\bibfnamefont{O.}~\bibnamefont{Amin}},
  \bibinfo{author}{\bibfnamefont{A.}~\bibnamefont{Dal~Din}},
  \bibinfo{author}{\bibfnamefont{E.}~\bibnamefont{Golias}},
  \bibinfo{author}{\bibfnamefont{Y.}~\bibnamefont{Niu}},
  \bibinfo{author}{\bibfnamefont{A.}~\bibnamefont{Zakharov}},
  \bibinfo{author}{\bibfnamefont{S.}~\bibnamefont{Fromage}},
  \bibinfo{author}{\bibfnamefont{C.}~\bibnamefont{Fields}},
  \bibinfo{author}{\bibfnamefont{S.}~\bibnamefont{Heywood}},
  \bibinfo{author}{\bibfnamefont{R.}~\bibnamefont{Cousins}},
  \bibinfo{author}{\bibfnamefont{F.}~\bibnamefont{Maccherozzi}},
  \bibnamefont{et~al.}, \bibinfo{journal}{Nature}
  \textbf{\bibinfo{volume}{636}}, \bibinfo{pages}{348} (\bibinfo{year}{2024}).

\bibitem[{\citenamefont{Gonzalez~Betancourt
  et~al.}(2023)\citenamefont{Gonzalez~Betancourt, Zub{\'a}{\v{c}},
  Gonzalez-Hernandez, Geishendorf, {\v{S}}ob{\'a}{\v{n}}, Springholz,
  Olejn{\'\i}k, {\v{S}}mejkal, Sinova, Jungwirth
  et~al.}}]{gonzalez2023spontaneous}
\bibinfo{author}{\bibfnamefont{R.}~\bibnamefont{Gonzalez~Betancourt}},
  \bibinfo{author}{\bibfnamefont{J.}~\bibnamefont{Zub{\'a}{\v{c}}}},
  \bibinfo{author}{\bibfnamefont{R.}~\bibnamefont{Gonzalez-Hernandez}},
  \bibinfo{author}{\bibfnamefont{K.}~\bibnamefont{Geishendorf}},
  \bibinfo{author}{\bibfnamefont{Z.}~\bibnamefont{{\v{S}}ob{\'a}{\v{n}}}},
  \bibinfo{author}{\bibfnamefont{G.}~\bibnamefont{Springholz}},
  \bibinfo{author}{\bibfnamefont{K.}~\bibnamefont{Olejn{\'\i}k}},
  \bibinfo{author}{\bibfnamefont{L.}~\bibnamefont{{\v{S}}mejkal}},
  \bibinfo{author}{\bibfnamefont{J.}~\bibnamefont{Sinova}},
  \bibinfo{author}{\bibfnamefont{T.}~\bibnamefont{Jungwirth}},
  \bibnamefont{et~al.}, \bibinfo{journal}{Physical Review Letters}
  \textbf{\bibinfo{volume}{130}}, \bibinfo{pages}{036702}
  (\bibinfo{year}{2023}).

\bibitem[{\citenamefont{Chen et~al.}(2025)\citenamefont{Chen, James, and
  Dugdale}}]{chen2025exposing}
\bibinfo{author}{\bibfnamefont{W.}~\bibnamefont{Chen}},
  \bibinfo{author}{\bibfnamefont{A.~D.} \bibnamefont{James}}, \bibnamefont{and}
  \bibinfo{author}{\bibfnamefont{S.~B.} \bibnamefont{Dugdale}},
  \bibinfo{journal}{arXiv preprint arXiv:2511.01094}  (\bibinfo{year}{2025}).

\bibitem[{\citenamefont{Yu et~al.}(2025)}]{yu2025neel}
\bibinfo{author}{\bibfnamefont{T.}~\bibnamefont{Yu}} \bibnamefont{et~al.}
  (\bibinfo{year}{2025}).

\bibitem[{\citenamefont{Galitski and Spielman}(2013)}]{galitski2013spin}
\bibinfo{author}{\bibfnamefont{V.}~\bibnamefont{Galitski}} \bibnamefont{and}
  \bibinfo{author}{\bibfnamefont{I.~B.} \bibnamefont{Spielman}},
  \bibinfo{journal}{Nature} \textbf{\bibinfo{volume}{494}}, \bibinfo{pages}{49}
  (\bibinfo{year}{2013}).

\bibitem[{\citenamefont{Usman et~al.}(2026)\citenamefont{Usman, Ghosh, and
  Islam}}]{usman2026resonant}
\bibinfo{author}{\bibfnamefont{M.}~\bibnamefont{Usman}},
  \bibinfo{author}{\bibfnamefont{T.~K.} \bibnamefont{Ghosh}}, \bibnamefont{and}
  \bibinfo{author}{\bibfnamefont{S.~F.} \bibnamefont{Islam}},
  \bibinfo{journal}{Journal of Physics: Condensed Matter}
  (\bibinfo{year}{2026}).

\bibitem[{\citenamefont{Pandita and Islam}(2026)}]{pandita2026magnetotransport}
\bibinfo{author}{\bibfnamefont{A.}~\bibnamefont{Pandita}} \bibnamefont{and}
  \bibinfo{author}{\bibfnamefont{S.~F.} \bibnamefont{Islam}},
  \bibinfo{journal}{Physical Review B} \textbf{\bibinfo{volume}{113}},
  \bibinfo{pages}{115419} (\bibinfo{year}{2026}).

\bibitem[{\citenamefont{Mojarro and Ulloa}(2025)}]{Mojarro2025MajoranaKagome}
\bibinfo{author}{\bibfnamefont{M.~A.} \bibnamefont{Mojarro}} \bibnamefont{and}
  \bibinfo{author}{\bibfnamefont{S.~E.} \bibnamefont{Ulloa}},
  \bibinfo{journal}{2D Materials} \textbf{\bibinfo{volume}{12}},
  \bibinfo{pages}{035015} (\bibinfo{year}{2025}).

\bibitem[{\citenamefont{Naka et~al.}(2025)\citenamefont{Naka, Motome, and
  Seo}}]{naka2025altermagnetic}
\bibinfo{author}{\bibfnamefont{M.}~\bibnamefont{Naka}},
  \bibinfo{author}{\bibfnamefont{Y.}~\bibnamefont{Motome}}, \bibnamefont{and}
  \bibinfo{author}{\bibfnamefont{H.}~\bibnamefont{Seo}}, \bibinfo{journal}{npj
  Spintronics} \textbf{\bibinfo{volume}{3}}, \bibinfo{pages}{1}
  (\bibinfo{year}{2025}).

\bibitem[{\citenamefont{Cheong and Huang}(2025)}]{cheong2025altermagnetism}
\bibinfo{author}{\bibfnamefont{S.-W.} \bibnamefont{Cheong}} \bibnamefont{and}
  \bibinfo{author}{\bibfnamefont{F.-T.} \bibnamefont{Huang}},
  \bibinfo{journal}{npj Quantum Materials} \textbf{\bibinfo{volume}{10}},
  \bibinfo{pages}{38} (\bibinfo{year}{2025}).

\bibitem[{\citenamefont{Wang et~al.}(2025)\citenamefont{Wang, Zhang, Zhang,
  Sun, Dagotto, Xu, and Hu}}]{wang2025spin}
\bibinfo{author}{\bibfnamefont{Z.-M.} \bibnamefont{Wang}},
  \bibinfo{author}{\bibfnamefont{Y.}~\bibnamefont{Zhang}},
  \bibinfo{author}{\bibfnamefont{S.-B.} \bibnamefont{Zhang}},
  \bibinfo{author}{\bibfnamefont{J.-H.} \bibnamefont{Sun}},
  \bibinfo{author}{\bibfnamefont{E.}~\bibnamefont{Dagotto}},
  \bibinfo{author}{\bibfnamefont{D.-H.} \bibnamefont{Xu}}, \bibnamefont{and}
  \bibinfo{author}{\bibfnamefont{L.-H.} \bibnamefont{Hu}},
  \bibinfo{journal}{Physical Review Letters} \textbf{\bibinfo{volume}{135}},
  \bibinfo{pages}{176705} (\bibinfo{year}{2025}).

\bibitem[{\citenamefont{Roig et~al.}(2025)\citenamefont{Roig, Yu, Ekman,
  Kreisel, Andersen, and Agterberg}}]{roig2025quasisymmetry}
\bibinfo{author}{\bibfnamefont{M.}~\bibnamefont{Roig}},
  \bibinfo{author}{\bibfnamefont{Y.}~\bibnamefont{Yu}},
  \bibinfo{author}{\bibfnamefont{R.~C.} \bibnamefont{Ekman}},
  \bibinfo{author}{\bibfnamefont{A.}~\bibnamefont{Kreisel}},
  \bibinfo{author}{\bibfnamefont{B.~M.} \bibnamefont{Andersen}},
  \bibnamefont{and} \bibinfo{author}{\bibfnamefont{D.~F.}
  \bibnamefont{Agterberg}}, \bibinfo{journal}{Physical Review Letters}
  \textbf{\bibinfo{volume}{135}}, \bibinfo{pages}{016703}
  (\bibinfo{year}{2025}).

\bibitem[{\citenamefont{Islam and Basu}(2023)}]{islam2023properties}
\bibinfo{author}{\bibfnamefont{M.}~\bibnamefont{Islam}} \bibnamefont{and}
  \bibinfo{author}{\bibfnamefont{S.}~\bibnamefont{Basu}},
  \bibinfo{journal}{Journal of Physics: Condensed Matter}
  (\bibinfo{year}{2023}).

\bibitem[{\citenamefont{Sarkar and Agarwal}(2025)}]{sarkar2025spin}
\bibinfo{author}{\bibfnamefont{S.}~\bibnamefont{Sarkar}} \bibnamefont{and}
  \bibinfo{author}{\bibfnamefont{A.}~\bibnamefont{Agarwal}},
  \bibinfo{journal}{Physical Review B} \textbf{\bibinfo{volume}{112}},
  \bibinfo{pages}{195420} (\bibinfo{year}{2025}).

\bibitem[{\citenamefont{Kapri}(2025)}]{kapri2025spin}
\bibinfo{author}{\bibfnamefont{P.}~\bibnamefont{Kapri}},
  \bibinfo{journal}{Physical Review B} \textbf{\bibinfo{volume}{112}},
  \bibinfo{pages}{155422} (\bibinfo{year}{2025}).

\bibitem[{\citenamefont{Tamang et~al.}(2023)\citenamefont{Tamang, Verma, and
  Biswas}}]{tamang2023orbital}
\bibinfo{author}{\bibfnamefont{L.}~\bibnamefont{Tamang}},
  \bibinfo{author}{\bibfnamefont{S.}~\bibnamefont{Verma}}, \bibnamefont{and}
  \bibinfo{author}{\bibfnamefont{T.}~\bibnamefont{Biswas}},
  \bibinfo{journal}{arXiv preprint arXiv:2309.07074}  (\bibinfo{year}{2023}).

\bibitem[{\citenamefont{Mazin}(2023)}]{mazin2023altermagnetism}
\bibinfo{author}{\bibfnamefont{I.}~\bibnamefont{Mazin}},
  \bibinfo{journal}{Physical Review B} \textbf{\bibinfo{volume}{107}},
  \bibinfo{pages}{L100418} (\bibinfo{year}{2023}).

\bibitem[{\citenamefont{Sato and Ando}(2017)}]{sato2017topological}
\bibinfo{author}{\bibfnamefont{M.}~\bibnamefont{Sato}} \bibnamefont{and}
  \bibinfo{author}{\bibfnamefont{Y.}~\bibnamefont{Ando}},
  \bibinfo{journal}{Reports on Progress in Physics}
  \textbf{\bibinfo{volume}{80}}, \bibinfo{pages}{076501}
  (\bibinfo{year}{2017}).

\bibitem[{\citenamefont{Leraand et~al.}(2025)\citenamefont{Leraand, M{\ae}land,
  and Sudb{\o}}}]{leraand2025phonon}
\bibinfo{author}{\bibfnamefont{K.}~\bibnamefont{Leraand}},
  \bibinfo{author}{\bibfnamefont{K.}~\bibnamefont{M{\ae}land}},
  \bibnamefont{and} \bibinfo{author}{\bibfnamefont{A.}~\bibnamefont{Sudb{\o}}},
  \bibinfo{journal}{Physical Review B} \textbf{\bibinfo{volume}{112}},
  \bibinfo{pages}{104510} (\bibinfo{year}{2025}).

\bibitem[{\citenamefont{Rasmussen et~al.}(2025)\citenamefont{Rasmussen,
  Gondolf, Barkman, Roig, Agterberg, Kreisel, and
  Andersen}}]{rasmussen2025inherent}
\bibinfo{author}{\bibfnamefont{C.~L.} \bibnamefont{Rasmussen}},
  \bibinfo{author}{\bibfnamefont{J.}~\bibnamefont{Gondolf}},
  \bibinfo{author}{\bibfnamefont{M.}~\bibnamefont{Barkman}},
  \bibinfo{author}{\bibfnamefont{M.}~\bibnamefont{Roig}},
  \bibinfo{author}{\bibfnamefont{D.~F.} \bibnamefont{Agterberg}},
  \bibinfo{author}{\bibfnamefont{A.}~\bibnamefont{Kreisel}}, \bibnamefont{and}
  \bibinfo{author}{\bibfnamefont{B.~M.} \bibnamefont{Andersen}},
  \bibinfo{journal}{arXiv preprint arXiv:2509.03247}  (\bibinfo{year}{2025}).

\bibitem[{\citenamefont{Fukaya et~al.}(2025)\citenamefont{Fukaya, Lu, Yada,
  Tanaka, and Cayao}}]{fukaya2025superconducting}
\bibinfo{author}{\bibfnamefont{Y.}~\bibnamefont{Fukaya}},
  \bibinfo{author}{\bibfnamefont{B.}~\bibnamefont{Lu}},
  \bibinfo{author}{\bibfnamefont{K.}~\bibnamefont{Yada}},
  \bibinfo{author}{\bibfnamefont{Y.}~\bibnamefont{Tanaka}}, \bibnamefont{and}
  \bibinfo{author}{\bibfnamefont{J.}~\bibnamefont{Cayao}},
  \bibinfo{journal}{Journal of Physics: Condensed Matter}
  \textbf{\bibinfo{volume}{37}}, \bibinfo{pages}{313003}
  (\bibinfo{year}{2025}).

\bibitem[{\citenamefont{Monkman et~al.}(2026)\citenamefont{Monkman, Weng,
  Heinsdorf, Nocera, Barlas, and Franz}}]{monkman2026persistent}
\bibinfo{author}{\bibfnamefont{K.}~\bibnamefont{Monkman}},
  \bibinfo{author}{\bibfnamefont{J.}~\bibnamefont{Weng}},
  \bibinfo{author}{\bibfnamefont{N.}~\bibnamefont{Heinsdorf}},
  \bibinfo{author}{\bibfnamefont{A.}~\bibnamefont{Nocera}},
  \bibinfo{author}{\bibfnamefont{Y.}~\bibnamefont{Barlas}}, \bibnamefont{and}
  \bibinfo{author}{\bibfnamefont{M.}~\bibnamefont{Franz}},
  \bibinfo{journal}{Physical Review X} \textbf{\bibinfo{volume}{16}},
  \bibinfo{pages}{011057} (\bibinfo{year}{2026}).

\bibitem[{\citenamefont{Hadjipaschalis
  et~al.}(2025)\citenamefont{Hadjipaschalis, Ghorashi, and
  Cano}}]{hadjipaschalis2025majoranas}
\bibinfo{author}{\bibfnamefont{A.}~\bibnamefont{Hadjipaschalis}},
  \bibinfo{author}{\bibfnamefont{S.~A.~A.} \bibnamefont{Ghorashi}},
  \bibnamefont{and} \bibinfo{author}{\bibfnamefont{J.}~\bibnamefont{Cano}},
  \bibinfo{journal}{Physical Review B} \textbf{\bibinfo{volume}{112}},
  \bibinfo{pages}{214430} (\bibinfo{year}{2025}).

\bibitem[{\citenamefont{Oka and
  Aoki}(2009{\natexlab{a}})}]{Oka2009PhotovoltaicHall}
\bibinfo{author}{\bibfnamefont{T.}~\bibnamefont{Oka}} \bibnamefont{and}
  \bibinfo{author}{\bibfnamefont{H.}~\bibnamefont{Aoki}},
  \bibinfo{journal}{Physical Review B} \textbf{\bibinfo{volume}{79}},
  \bibinfo{pages}{081406} (\bibinfo{year}{2009}{\natexlab{a}}).

\bibitem[{\citenamefont{Oka and Kitamura}(2019)}]{oka2019floquet}
\bibinfo{author}{\bibfnamefont{T.}~\bibnamefont{Oka}} \bibnamefont{and}
  \bibinfo{author}{\bibfnamefont{S.}~\bibnamefont{Kitamura}},
  \bibinfo{journal}{Annual Review of Condensed Matter Physics}
  \textbf{\bibinfo{volume}{10}}, \bibinfo{pages}{387} (\bibinfo{year}{2019}).

\bibitem[{\citenamefont{Topp et~al.}(2019)\citenamefont{Topp, Jotzu, McIver,
  Xian, Rubio, and Sentef}}]{topp2019topological}
\bibinfo{author}{\bibfnamefont{G.~E.} \bibnamefont{Topp}},
  \bibinfo{author}{\bibfnamefont{G.}~\bibnamefont{Jotzu}},
  \bibinfo{author}{\bibfnamefont{J.~W.} \bibnamefont{McIver}},
  \bibinfo{author}{\bibfnamefont{L.}~\bibnamefont{Xian}},
  \bibinfo{author}{\bibfnamefont{A.}~\bibnamefont{Rubio}}, \bibnamefont{and}
  \bibinfo{author}{\bibfnamefont{M.~A.} \bibnamefont{Sentef}},
  \bibinfo{journal}{Physical Review Research} \textbf{\bibinfo{volume}{1}},
  \bibinfo{pages}{023031} (\bibinfo{year}{2019}).

\bibitem[{\citenamefont{Mojarro
  et~al.}(2020{\natexlab{a}})\citenamefont{Mojarro, Ibarra-Sierra,
  Sandoval-Santana, Carrillo-Bastos, and Naumis}}]{Mojarro2020FloquetKekule}
\bibinfo{author}{\bibfnamefont{M.~A.} \bibnamefont{Mojarro}},
  \bibinfo{author}{\bibfnamefont{V.~G.} \bibnamefont{Ibarra-Sierra}},
  \bibinfo{author}{\bibfnamefont{J.~C.} \bibnamefont{Sandoval-Santana}},
  \bibinfo{author}{\bibfnamefont{R.}~\bibnamefont{Carrillo-Bastos}},
  \bibnamefont{and} \bibinfo{author}{\bibfnamefont{G.~G.}
  \bibnamefont{Naumis}}, \bibinfo{journal}{Physical Review B}
  \textbf{\bibinfo{volume}{102}}, \bibinfo{pages}{165301}
  (\bibinfo{year}{2020}{\natexlab{a}}).

\bibitem[{\citenamefont{Goldman and Dalibard}(2014)}]{goldman2014periodically}
\bibinfo{author}{\bibfnamefont{N.}~\bibnamefont{Goldman}} \bibnamefont{and}
  \bibinfo{author}{\bibfnamefont{J.}~\bibnamefont{Dalibard}},
  \bibinfo{journal}{Physical review X} \textbf{\bibinfo{volume}{4}},
  \bibinfo{pages}{031027} (\bibinfo{year}{2014}).

\bibitem[{\citenamefont{Mojarro et~al.}(2021)\citenamefont{Mojarro,
  Carrillo-Bastos, and Maytorena}}]{mojarro2021optical}
\bibinfo{author}{\bibfnamefont{M.}~\bibnamefont{Mojarro}},
  \bibinfo{author}{\bibfnamefont{R.}~\bibnamefont{Carrillo-Bastos}},
  \bibnamefont{and} \bibinfo{author}{\bibfnamefont{J.~A.}
  \bibnamefont{Maytorena}}, \bibinfo{journal}{Physical Review B}
  \textbf{\bibinfo{volume}{103}}, \bibinfo{pages}{165415}
  (\bibinfo{year}{2021}).

\bibitem[{\citenamefont{Iurov et~al.}(2020{\natexlab{a}})\citenamefont{Iurov,
  Zhemchuzhna, Fekete, Gumbs, and Huang}}]{iurov2020klein}
\bibinfo{author}{\bibfnamefont{A.}~\bibnamefont{Iurov}},
  \bibinfo{author}{\bibfnamefont{L.}~\bibnamefont{Zhemchuzhna}},
  \bibinfo{author}{\bibfnamefont{P.}~\bibnamefont{Fekete}},
  \bibinfo{author}{\bibfnamefont{G.}~\bibnamefont{Gumbs}}, \bibnamefont{and}
  \bibinfo{author}{\bibfnamefont{D.}~\bibnamefont{Huang}},
  \bibinfo{journal}{Physical Review Research} \textbf{\bibinfo{volume}{2}},
  \bibinfo{pages}{043245} (\bibinfo{year}{2020}{\natexlab{a}}).

\bibitem[{\citenamefont{Kibis}(2010)}]{kibis2010metal}
\bibinfo{author}{\bibfnamefont{O.}~\bibnamefont{Kibis}},
  \bibinfo{journal}{Physical Review B} \textbf{\bibinfo{volume}{81}},
  \bibinfo{pages}{165433} (\bibinfo{year}{2010}).

\bibitem[{\citenamefont{Mojarro
  et~al.}(2020{\natexlab{b}})\citenamefont{Mojarro, Ibarra-Sierra,
  Sandoval-Santana, Carrillo-Bastos, and Naumis}}]{mojarro2020dynamical}
\bibinfo{author}{\bibfnamefont{M.}~\bibnamefont{Mojarro}},
  \bibinfo{author}{\bibfnamefont{V.}~\bibnamefont{Ibarra-Sierra}},
  \bibinfo{author}{\bibfnamefont{J.}~\bibnamefont{Sandoval-Santana}},
  \bibinfo{author}{\bibfnamefont{R.}~\bibnamefont{Carrillo-Bastos}},
  \bibnamefont{and} \bibinfo{author}{\bibfnamefont{G.~G.}
  \bibnamefont{Naumis}}, \bibinfo{journal}{Physical Review B}
  \textbf{\bibinfo{volume}{102}}, \bibinfo{pages}{165301}
  (\bibinfo{year}{2020}{\natexlab{b}}).

\bibitem[{\citenamefont{Ibarra-Sierra et~al.}(2019)\citenamefont{Ibarra-Sierra,
  Sandoval-Santana, Kunold, and Naumis}}]{ibarra2019dynamical}
\bibinfo{author}{\bibfnamefont{V.}~\bibnamefont{Ibarra-Sierra}},
  \bibinfo{author}{\bibfnamefont{J.}~\bibnamefont{Sandoval-Santana}},
  \bibinfo{author}{\bibfnamefont{A.}~\bibnamefont{Kunold}}, \bibnamefont{and}
  \bibinfo{author}{\bibfnamefont{G.~G.} \bibnamefont{Naumis}},
  \bibinfo{journal}{Physical Review B} \textbf{\bibinfo{volume}{100}},
  \bibinfo{pages}{125302} (\bibinfo{year}{2019}).

\bibitem[{\citenamefont{Iurov et~al.}(2024)\citenamefont{Iurov, Mattis,
  Zhemchuzhna, Gumbs, and Huang}}]{iurov2024floquet}
\bibinfo{author}{\bibfnamefont{A.}~\bibnamefont{Iurov}},
  \bibinfo{author}{\bibfnamefont{M.}~\bibnamefont{Mattis}},
  \bibinfo{author}{\bibfnamefont{L.}~\bibnamefont{Zhemchuzhna}},
  \bibinfo{author}{\bibfnamefont{G.}~\bibnamefont{Gumbs}}, \bibnamefont{and}
  \bibinfo{author}{\bibfnamefont{D.}~\bibnamefont{Huang}},
  \bibinfo{journal}{Applied Sciences} \textbf{\bibinfo{volume}{14}},
  \bibinfo{pages}{6027} (\bibinfo{year}{2024}).

\bibitem[{\citenamefont{Tamang et~al.}(2021)\citenamefont{Tamang, Nag, and
  Biswas}}]{Tamang2021Floquet}
\bibinfo{author}{\bibfnamefont{L.}~\bibnamefont{Tamang}},
  \bibinfo{author}{\bibfnamefont{T.}~\bibnamefont{Nag}}, \bibnamefont{and}
  \bibinfo{author}{\bibfnamefont{T.}~\bibnamefont{Biswas}},
  \bibinfo{journal}{Physical Review B} \textbf{\bibinfo{volume}{104}},
  \bibinfo{pages}{174308} (\bibinfo{year}{2021}).

\bibitem[{\citenamefont{Iurov et~al.}(2017{\natexlab{a}})\citenamefont{Iurov,
  Zhemchuzhna, Gumbs, and Huang}}]{iurov2017exploring}
\bibinfo{author}{\bibfnamefont{A.}~\bibnamefont{Iurov}},
  \bibinfo{author}{\bibfnamefont{L.}~\bibnamefont{Zhemchuzhna}},
  \bibinfo{author}{\bibfnamefont{G.}~\bibnamefont{Gumbs}}, \bibnamefont{and}
  \bibinfo{author}{\bibfnamefont{D.}~\bibnamefont{Huang}},
  \bibinfo{journal}{Journal of Applied Physics} \textbf{\bibinfo{volume}{122}}
  (\bibinfo{year}{2017}{\natexlab{a}}).

\bibitem[{\citenamefont{Zhou et~al.}(2023)\citenamefont{Zhou, Bao, Fan, Wang,
  Zhong, Zhang, Tang, Duan, and Zhou}}]{zhou2023floquet}
\bibinfo{author}{\bibfnamefont{S.}~\bibnamefont{Zhou}},
  \bibinfo{author}{\bibfnamefont{C.}~\bibnamefont{Bao}},
  \bibinfo{author}{\bibfnamefont{B.}~\bibnamefont{Fan}},
  \bibinfo{author}{\bibfnamefont{F.}~\bibnamefont{Wang}},
  \bibinfo{author}{\bibfnamefont{H.}~\bibnamefont{Zhong}},
  \bibinfo{author}{\bibfnamefont{H.}~\bibnamefont{Zhang}},
  \bibinfo{author}{\bibfnamefont{P.}~\bibnamefont{Tang}},
  \bibinfo{author}{\bibfnamefont{W.}~\bibnamefont{Duan}}, \bibnamefont{and}
  \bibinfo{author}{\bibfnamefont{S.}~\bibnamefont{Zhou}},
  \bibinfo{journal}{Physical Review Letters} \textbf{\bibinfo{volume}{131}},
  \bibinfo{pages}{116401} (\bibinfo{year}{2023}).

\bibitem[{\citenamefont{Roslyak et~al.}(2010)\citenamefont{Roslyak, Iurov,
  Gumbs, and Huang}}]{roslyak2010unimpeded}
\bibinfo{author}{\bibfnamefont{O.}~\bibnamefont{Roslyak}},
  \bibinfo{author}{\bibfnamefont{A.}~\bibnamefont{Iurov}},
  \bibinfo{author}{\bibfnamefont{G.}~\bibnamefont{Gumbs}}, \bibnamefont{and}
  \bibinfo{author}{\bibfnamefont{D.}~\bibnamefont{Huang}},
  \bibinfo{journal}{Journal of Physics: Condensed Matter}
  \textbf{\bibinfo{volume}{22}}, \bibinfo{pages}{165301}
  (\bibinfo{year}{2010}).

\bibitem[{\citenamefont{Iurov et~al.}(2013)\citenamefont{Iurov, Gumbs, Roslyak,
  and Huang}}]{iurov2013photon}
\bibinfo{author}{\bibfnamefont{A.}~\bibnamefont{Iurov}},
  \bibinfo{author}{\bibfnamefont{G.}~\bibnamefont{Gumbs}},
  \bibinfo{author}{\bibfnamefont{O.}~\bibnamefont{Roslyak}}, \bibnamefont{and}
  \bibinfo{author}{\bibfnamefont{D.}~\bibnamefont{Huang}},
  \bibinfo{journal}{Journal of Physics: Condensed Matter}
  \textbf{\bibinfo{volume}{25}}, \bibinfo{pages}{135502}
  (\bibinfo{year}{2013}).

\bibitem[{\citenamefont{Iurov et~al.}(2020{\natexlab{b}})\citenamefont{Iurov,
  Zhemchuzhna, Dahal, Gumbs, and Huang}}]{iurov2020quantum}
\bibinfo{author}{\bibfnamefont{A.}~\bibnamefont{Iurov}},
  \bibinfo{author}{\bibfnamefont{L.}~\bibnamefont{Zhemchuzhna}},
  \bibinfo{author}{\bibfnamefont{D.}~\bibnamefont{Dahal}},
  \bibinfo{author}{\bibfnamefont{G.}~\bibnamefont{Gumbs}}, \bibnamefont{and}
  \bibinfo{author}{\bibfnamefont{D.}~\bibnamefont{Huang}},
  \bibinfo{journal}{Physical Review B} \textbf{\bibinfo{volume}{101}},
  \bibinfo{pages}{035129} (\bibinfo{year}{2020}{\natexlab{b}}).

\bibitem[{\citenamefont{Islam and Saha}(2018)}]{islam2018driven}
\bibinfo{author}{\bibfnamefont{S.~F.} \bibnamefont{Islam}} \bibnamefont{and}
  \bibinfo{author}{\bibfnamefont{A.}~\bibnamefont{Saha}},
  \bibinfo{journal}{Physical Review B} \textbf{\bibinfo{volume}{98}},
  \bibinfo{pages}{235424} (\bibinfo{year}{2018}).

\bibitem[{\citenamefont{Iurov et~al.}(2022{\natexlab{a}})\citenamefont{Iurov,
  Zhemchuzhna, Gumbs, Huang, and Fekete}}]{iurov2022optically}
\bibinfo{author}{\bibfnamefont{A.}~\bibnamefont{Iurov}},
  \bibinfo{author}{\bibfnamefont{L.}~\bibnamefont{Zhemchuzhna}},
  \bibinfo{author}{\bibfnamefont{G.}~\bibnamefont{Gumbs}},
  \bibinfo{author}{\bibfnamefont{D.}~\bibnamefont{Huang}}, \bibnamefont{and}
  \bibinfo{author}{\bibfnamefont{P.}~\bibnamefont{Fekete}},
  \bibinfo{journal}{Physical Review B} \textbf{\bibinfo{volume}{105}},
  \bibinfo{pages}{115309} (\bibinfo{year}{2022}{\natexlab{a}}).

\bibitem[{\citenamefont{Kristinsson et~al.}(2016)\citenamefont{Kristinsson,
  Kibis, Morina, and Shelykh}}]{kristinsson2016control}
\bibinfo{author}{\bibfnamefont{K.}~\bibnamefont{Kristinsson}},
  \bibinfo{author}{\bibfnamefont{O.~V.} \bibnamefont{Kibis}},
  \bibinfo{author}{\bibfnamefont{S.}~\bibnamefont{Morina}}, \bibnamefont{and}
  \bibinfo{author}{\bibfnamefont{I.~A.} \bibnamefont{Shelykh}},
  \bibinfo{journal}{Scientific reports} \textbf{\bibinfo{volume}{6}},
  \bibinfo{pages}{1} (\bibinfo{year}{2016}).

\bibitem[{\citenamefont{Iurov et~al.}(2017{\natexlab{b}})\citenamefont{Iurov,
  Huang, and Zhemchuzhna}}]{iurov2017controlling}
\bibinfo{author}{\bibfnamefont{G.}~\bibnamefont{Iurov},
  \bibfnamefont{Andrii~and}},
  \bibinfo{author}{\bibfnamefont{D.}~\bibnamefont{Huang}}, \bibnamefont{and}
  \bibinfo{author}{\bibfnamefont{L.}~\bibnamefont{Zhemchuzhna}},
  \bibinfo{journal}{Journal of Applied Physics} \textbf{\bibinfo{volume}{121}}
  (\bibinfo{year}{2017}{\natexlab{b}}).

\bibitem[{\citenamefont{Horing et~al.}(2016)\citenamefont{Horing, Iurov, Gumbs,
  Politano, and Chiarello}}]{horing2016low}
\bibinfo{author}{\bibfnamefont{N.}~\bibnamefont{Horing}},
  \bibinfo{author}{\bibfnamefont{A.}~\bibnamefont{Iurov}},
  \bibinfo{author}{\bibfnamefont{G.}~\bibnamefont{Gumbs}},
  \bibinfo{author}{\bibfnamefont{A.}~\bibnamefont{Politano}}, \bibnamefont{and}
  \bibinfo{author}{\bibfnamefont{G.}~\bibnamefont{Chiarello}}
  (\bibinfo{year}{2016}).

\bibitem[{\citenamefont{Ross-Harvey et~al.}(2025)\citenamefont{Ross-Harvey,
  Iurov, Zhemchuzhna, Gumbs, Huang, and Fekete}}]{ross2025dynamical}
\bibinfo{author}{\bibfnamefont{G.}~\bibnamefont{Ross-Harvey}},
  \bibinfo{author}{\bibfnamefont{A.}~\bibnamefont{Iurov}},
  \bibinfo{author}{\bibfnamefont{L.}~\bibnamefont{Zhemchuzhna}},
  \bibinfo{author}{\bibfnamefont{G.}~\bibnamefont{Gumbs}},
  \bibinfo{author}{\bibfnamefont{D.}~\bibnamefont{Huang}}, \bibnamefont{and}
  \bibinfo{author}{\bibfnamefont{P.}~\bibnamefont{Fekete}},
  \bibinfo{journal}{Physical Review B} \textbf{\bibinfo{volume}{111}},
  \bibinfo{pages}{045413} (\bibinfo{year}{2025}).

\bibitem[{\citenamefont{Constant et~al.}(2016)\citenamefont{Constant, Hornett,
  Chang, and Hendry}}]{constant2016all}
\bibinfo{author}{\bibfnamefont{T.~J.} \bibnamefont{Constant}},
  \bibinfo{author}{\bibfnamefont{S.~M.} \bibnamefont{Hornett}},
  \bibinfo{author}{\bibfnamefont{D.~E.} \bibnamefont{Chang}}, \bibnamefont{and}
  \bibinfo{author}{\bibfnamefont{E.}~\bibnamefont{Hendry}},
  \bibinfo{journal}{Nature Physics} \textbf{\bibinfo{volume}{12}},
  \bibinfo{pages}{124} (\bibinfo{year}{2016}).

\bibitem[{\citenamefont{Oriekhov and Gusynin}(2020)}]{oriekhov2020rkky}
\bibinfo{author}{\bibfnamefont{D.}~\bibnamefont{Oriekhov}} \bibnamefont{and}
  \bibinfo{author}{\bibfnamefont{V.}~\bibnamefont{Gusynin}},
  \bibinfo{journal}{Physical Review B} \textbf{\bibinfo{volume}{101}},
  \bibinfo{pages}{235162} (\bibinfo{year}{2020}).

\bibitem[{\citenamefont{Yarmohammadi et~al.}(2025)\citenamefont{Yarmohammadi,
  Z{\"u}licke, Berakdar, Linder, and Freericks}}]{yarmohammadi2025anisotropic}
\bibinfo{author}{\bibfnamefont{M.}~\bibnamefont{Yarmohammadi}},
  \bibinfo{author}{\bibfnamefont{U.}~\bibnamefont{Z{\"u}licke}},
  \bibinfo{author}{\bibfnamefont{J.}~\bibnamefont{Berakdar}},
  \bibinfo{author}{\bibfnamefont{J.}~\bibnamefont{Linder}}, \bibnamefont{and}
  \bibinfo{author}{\bibfnamefont{J.~K.} \bibnamefont{Freericks}},
  \bibinfo{journal}{Physical Review B} \textbf{\bibinfo{volume}{111}},
  \bibinfo{pages}{224412} (\bibinfo{year}{2025}).

\bibitem[{\citenamefont{Wang et~al.}(2026)\citenamefont{Wang, Cai, Tang, Lu,
  Chen, Sheng, Feng, Zhong, Zhang, Yu et~al.}}]{wang2026observation}
\bibinfo{author}{\bibfnamefont{F.}~\bibnamefont{Wang}},
  \bibinfo{author}{\bibfnamefont{X.}~\bibnamefont{Cai}},
  \bibinfo{author}{\bibfnamefont{X.}~\bibnamefont{Tang}},
  \bibinfo{author}{\bibfnamefont{J.}~\bibnamefont{Lu}},
  \bibinfo{author}{\bibfnamefont{W.}~\bibnamefont{Chen}},
  \bibinfo{author}{\bibfnamefont{T.}~\bibnamefont{Sheng}},
  \bibinfo{author}{\bibfnamefont{R.}~\bibnamefont{Feng}},
  \bibinfo{author}{\bibfnamefont{H.}~\bibnamefont{Zhong}},
  \bibinfo{author}{\bibfnamefont{H.}~\bibnamefont{Zhang}},
  \bibinfo{author}{\bibfnamefont{P.}~\bibnamefont{Yu}}, \bibnamefont{et~al.},
  \bibinfo{journal}{Nature Materials} pp. \bibinfo{pages}{1--6}
  (\bibinfo{year}{2026}).

\bibitem[{\citenamefont{Merboldt et~al.}(2025)\citenamefont{Merboldt,
  Sch{\"u}ler, Schmitt, Bange, Bennecke, Gadge, Pierz, Schumacher, Momeni,
  Steil et~al.}}]{merboldt2025observation}
\bibinfo{author}{\bibfnamefont{M.}~\bibnamefont{Merboldt}},
  \bibinfo{author}{\bibfnamefont{M.}~\bibnamefont{Sch{\"u}ler}},
  \bibinfo{author}{\bibfnamefont{D.}~\bibnamefont{Schmitt}},
  \bibinfo{author}{\bibfnamefont{J.~P.} \bibnamefont{Bange}},
  \bibinfo{author}{\bibfnamefont{W.}~\bibnamefont{Bennecke}},
  \bibinfo{author}{\bibfnamefont{K.}~\bibnamefont{Gadge}},
  \bibinfo{author}{\bibfnamefont{K.}~\bibnamefont{Pierz}},
  \bibinfo{author}{\bibfnamefont{H.~W.} \bibnamefont{Schumacher}},
  \bibinfo{author}{\bibfnamefont{D.}~\bibnamefont{Momeni}},
  \bibinfo{author}{\bibfnamefont{D.}~\bibnamefont{Steil}},
  \bibnamefont{et~al.}, \bibinfo{journal}{Nature Physics}
  \textbf{\bibinfo{volume}{21}}, \bibinfo{pages}{1093} (\bibinfo{year}{2025}).

\bibitem[{\citenamefont{Chen et~al.}(2026)\citenamefont{Chen, Wang, Li, and
  Du}}]{Chen2026few}
\bibinfo{author}{\bibfnamefont{Y.-X.} \bibnamefont{Chen}},
  \bibinfo{author}{\bibfnamefont{G.}~\bibnamefont{Wang}},
  \bibinfo{author}{\bibfnamefont{M.}~\bibnamefont{Li}}, \bibnamefont{and}
  \bibinfo{author}{\bibfnamefont{T.-Y.} \bibnamefont{Du}},
  \bibinfo{journal}{Physical Review B} \textbf{\bibinfo{volume}{113}},
  \bibinfo{pages}{035110} (\bibinfo{year}{2026}).

\bibitem[{\citenamefont{Wang et~al.}(2013)\citenamefont{Wang, Steinberg,
  Jarillo-Herrero, and Gedik}}]{wang2013observation}
\bibinfo{author}{\bibfnamefont{Y.}~\bibnamefont{Wang}},
  \bibinfo{author}{\bibfnamefont{H.}~\bibnamefont{Steinberg}},
  \bibinfo{author}{\bibfnamefont{P.}~\bibnamefont{Jarillo-Herrero}},
  \bibnamefont{and} \bibinfo{author}{\bibfnamefont{N.}~\bibnamefont{Gedik}},
  \bibinfo{journal}{Science} \textbf{\bibinfo{volume}{342}},
  \bibinfo{pages}{453} (\bibinfo{year}{2013}).

\bibitem[{\citenamefont{Sentef et~al.}(2015)\citenamefont{Sentef, Claassen,
  Kemper, Moritz, Oka, Freericks, and Devereaux}}]{sentef2015theory}
\bibinfo{author}{\bibfnamefont{M.}~\bibnamefont{Sentef}},
  \bibinfo{author}{\bibfnamefont{M.}~\bibnamefont{Claassen}},
  \bibinfo{author}{\bibfnamefont{A.}~\bibnamefont{Kemper}},
  \bibinfo{author}{\bibfnamefont{B.}~\bibnamefont{Moritz}},
  \bibinfo{author}{\bibfnamefont{T.}~\bibnamefont{Oka}},
  \bibinfo{author}{\bibfnamefont{J.}~\bibnamefont{Freericks}},
  \bibnamefont{and}
  \bibinfo{author}{\bibfnamefont{T.}~\bibnamefont{Devereaux}},
  \bibinfo{journal}{Nature communications} \textbf{\bibinfo{volume}{6}},
  \bibinfo{pages}{7047} (\bibinfo{year}{2015}).

\bibitem[{\citenamefont{H{\"u}bener et~al.}(2017)\citenamefont{H{\"u}bener,
  Sentef, De~Giovannini, Kemper, and Rubio}}]{hubener2017creating}
\bibinfo{author}{\bibfnamefont{H.}~\bibnamefont{H{\"u}bener}},
  \bibinfo{author}{\bibfnamefont{M.~A.} \bibnamefont{Sentef}},
  \bibinfo{author}{\bibfnamefont{U.}~\bibnamefont{De~Giovannini}},
  \bibinfo{author}{\bibfnamefont{A.~F.} \bibnamefont{Kemper}},
  \bibnamefont{and} \bibinfo{author}{\bibfnamefont{A.}~\bibnamefont{Rubio}},
  \bibinfo{journal}{Nature communications} \textbf{\bibinfo{volume}{8}},
  \bibinfo{pages}{13940} (\bibinfo{year}{2017}).

\bibitem[{\citenamefont{Beaulieu et~al.}(2024)\citenamefont{Beaulieu, Dong,
  Christiansson, Werner, Pincelli, Ziegler, Taniguchi, Watanabe, Chernikov,
  Wolf et~al.}}]{beaulieu2024berry}
\bibinfo{author}{\bibfnamefont{S.}~\bibnamefont{Beaulieu}},
  \bibinfo{author}{\bibfnamefont{S.}~\bibnamefont{Dong}},
  \bibinfo{author}{\bibfnamefont{V.}~\bibnamefont{Christiansson}},
  \bibinfo{author}{\bibfnamefont{P.}~\bibnamefont{Werner}},
  \bibinfo{author}{\bibfnamefont{T.}~\bibnamefont{Pincelli}},
  \bibinfo{author}{\bibfnamefont{J.~D.} \bibnamefont{Ziegler}},
  \bibinfo{author}{\bibfnamefont{T.}~\bibnamefont{Taniguchi}},
  \bibinfo{author}{\bibfnamefont{K.}~\bibnamefont{Watanabe}},
  \bibinfo{author}{\bibfnamefont{A.}~\bibnamefont{Chernikov}},
  \bibinfo{author}{\bibfnamefont{M.}~\bibnamefont{Wolf}}, \bibnamefont{et~al.},
  \bibinfo{journal}{Science Advances} \textbf{\bibinfo{volume}{10}},
  \bibinfo{pages}{eadk3897} (\bibinfo{year}{2024}).

\bibitem[{\citenamefont{Lindner et~al.}(2011)\citenamefont{Lindner, Refael, and
  Galitski}}]{lindner2011floquet}
\bibinfo{author}{\bibfnamefont{N.~H.} \bibnamefont{Lindner}},
  \bibinfo{author}{\bibfnamefont{G.}~\bibnamefont{Refael}}, \bibnamefont{and}
  \bibinfo{author}{\bibfnamefont{V.}~\bibnamefont{Galitski}},
  \bibinfo{journal}{Nature Physics} \textbf{\bibinfo{volume}{7}},
  \bibinfo{pages}{490} (\bibinfo{year}{2011}).

\bibitem[{\citenamefont{Oka and Aoki}(2009{\natexlab{b}})}]{oka2009floquet}
\bibinfo{author}{\bibfnamefont{T.}~\bibnamefont{Oka}} \bibnamefont{and}
  \bibinfo{author}{\bibfnamefont{H.}~\bibnamefont{Aoki}},
  \bibinfo{journal}{Phys. Rev. B} \textbf{\bibinfo{volume}{79}},
  \bibinfo{pages}{081406} (\bibinfo{year}{2009}{\natexlab{b}}).

\bibitem[{\citenamefont{Zeng and Cui}(2015)}]{zeng2015optical}
\bibinfo{author}{\bibfnamefont{H.}~\bibnamefont{Zeng}} \bibnamefont{and}
  \bibinfo{author}{\bibfnamefont{X.}~\bibnamefont{Cui}},
  \bibinfo{journal}{Chemical Society Reviews} \textbf{\bibinfo{volume}{44}},
  \bibinfo{pages}{2629} (\bibinfo{year}{2015}).

\bibitem[{\citenamefont{Choe et~al.}(2016)\citenamefont{Choe, Sung, and
  Chang}}]{choe2016understanding}
\bibinfo{author}{\bibfnamefont{D.-H.} \bibnamefont{Choe}},
  \bibinfo{author}{\bibfnamefont{H.-J.} \bibnamefont{Sung}}, \bibnamefont{and}
  \bibinfo{author}{\bibfnamefont{K.~J.} \bibnamefont{Chang}},
  \bibinfo{journal}{Physical Review B} \textbf{\bibinfo{volume}{93}},
  \bibinfo{pages}{125109} (\bibinfo{year}{2016}).

\bibitem[{\citenamefont{Cayssol et~al.}(2013)\citenamefont{Cayssol, D{\'o}ra,
  Simon, and Moessner}}]{cayssol2013floquet}
\bibinfo{author}{\bibfnamefont{J.}~\bibnamefont{Cayssol}},
  \bibinfo{author}{\bibfnamefont{B.}~\bibnamefont{D{\'o}ra}},
  \bibinfo{author}{\bibfnamefont{F.}~\bibnamefont{Simon}}, \bibnamefont{and}
  \bibinfo{author}{\bibfnamefont{R.}~\bibnamefont{Moessner}},
  \bibinfo{journal}{physica status solidi (RRL)--Rapid Research Letters}
  \textbf{\bibinfo{volume}{7}}, \bibinfo{pages}{101} (\bibinfo{year}{2013}).

\bibitem[{\citenamefont{Iurov et~al.}(2019)\citenamefont{Iurov, Gumbs, and
  Huang}}]{iurov2019peculiar}
\bibinfo{author}{\bibfnamefont{A.}~\bibnamefont{Iurov}},
  \bibinfo{author}{\bibfnamefont{G.}~\bibnamefont{Gumbs}}, \bibnamefont{and}
  \bibinfo{author}{\bibfnamefont{D.}~\bibnamefont{Huang}},
  \bibinfo{journal}{Physical Review B} \textbf{\bibinfo{volume}{99}},
  \bibinfo{pages}{205135} (\bibinfo{year}{2019}).

\bibitem[{\citenamefont{Dey and Ghosh}(2018)}]{dey2018photoinduced}
\bibinfo{author}{\bibfnamefont{B.}~\bibnamefont{Dey}} \bibnamefont{and}
  \bibinfo{author}{\bibfnamefont{T.~K.} \bibnamefont{Ghosh}},
  \bibinfo{journal}{Physical Review B} \textbf{\bibinfo{volume}{98}},
  \bibinfo{pages}{075422} (\bibinfo{year}{2018}).

\bibitem[{\citenamefont{Tamang and
  Biswas}(2023{\natexlab{a}})}]{tamang2023probing}
\bibinfo{author}{\bibfnamefont{L.}~\bibnamefont{Tamang}} \bibnamefont{and}
  \bibinfo{author}{\bibfnamefont{T.}~\bibnamefont{Biswas}},
  \bibinfo{journal}{Physical Review B} \textbf{\bibinfo{volume}{107}},
  \bibinfo{pages}{085408} (\bibinfo{year}{2023}{\natexlab{a}}).

\bibitem[{\citenamefont{Paul et~al.}(2026)\citenamefont{Paul, Lahiri,
  Bhattacharyya, and Basu}}]{paul2026emergent}
\bibinfo{author}{\bibfnamefont{G.}~\bibnamefont{Paul}},
  \bibinfo{author}{\bibfnamefont{S.}~\bibnamefont{Lahiri}},
  \bibinfo{author}{\bibfnamefont{K.}~\bibnamefont{Bhattacharyya}},
  \bibnamefont{and} \bibinfo{author}{\bibfnamefont{S.}~\bibnamefont{Basu}},
  \bibinfo{journal}{Physical Review B} \textbf{\bibinfo{volume}{113}},
  \bibinfo{pages}{035145} (\bibinfo{year}{2026}).

\bibitem[{\citenamefont{Iurov et~al.}(2022{\natexlab{b}})\citenamefont{Iurov,
  Zhemchuzhna, Gumbs, Huang, Tse, Blaise, and Ejiogu}}]{iurov2022floquet}
\bibinfo{author}{\bibfnamefont{A.}~\bibnamefont{Iurov}},
  \bibinfo{author}{\bibfnamefont{L.}~\bibnamefont{Zhemchuzhna}},
  \bibinfo{author}{\bibfnamefont{G.}~\bibnamefont{Gumbs}},
  \bibinfo{author}{\bibfnamefont{D.}~\bibnamefont{Huang}},
  \bibinfo{author}{\bibfnamefont{W.-K.} \bibnamefont{Tse}},
  \bibinfo{author}{\bibfnamefont{K.}~\bibnamefont{Blaise}}, \bibnamefont{and}
  \bibinfo{author}{\bibfnamefont{C.}~\bibnamefont{Ejiogu}},
  \bibinfo{journal}{Scientific Reports} \textbf{\bibinfo{volume}{12}},
  \bibinfo{pages}{21348} (\bibinfo{year}{2022}{\natexlab{b}}).

\bibitem[{\citenamefont{Iorsh et~al.}(2024)\citenamefont{Iorsh, Sedov, Kolodny,
  Sinitskiy, and Kibis}}]{iorsh2024floquet}
\bibinfo{author}{\bibfnamefont{I.}~\bibnamefont{Iorsh}},
  \bibinfo{author}{\bibfnamefont{D.}~\bibnamefont{Sedov}},
  \bibinfo{author}{\bibfnamefont{S.}~\bibnamefont{Kolodny}},
  \bibinfo{author}{\bibfnamefont{R.}~\bibnamefont{Sinitskiy}},
  \bibnamefont{and} \bibinfo{author}{\bibfnamefont{O.}~\bibnamefont{Kibis}},
  \bibinfo{journal}{Physical Review B} \textbf{\bibinfo{volume}{109}},
  \bibinfo{pages}{035104} (\bibinfo{year}{2024}).

\bibitem[{\citenamefont{Mohan and Rao}(2018)}]{mohan2018interplay}
\bibinfo{author}{\bibfnamefont{P.}~\bibnamefont{Mohan}} \bibnamefont{and}
  \bibinfo{author}{\bibfnamefont{S.}~\bibnamefont{Rao}},
  \bibinfo{journal}{Physical Review B} \textbf{\bibinfo{volume}{98}},
  \bibinfo{pages}{165406} (\bibinfo{year}{2018}).

\bibitem[{\citenamefont{Ghorashi and Li}(2025)}]{ghorashi2025dynamical}
\bibinfo{author}{\bibfnamefont{S.~A.~A.} \bibnamefont{Ghorashi}}
  \bibnamefont{and} \bibinfo{author}{\bibfnamefont{Q.}~\bibnamefont{Li}},
  \bibinfo{journal}{Physical Review Letters} \textbf{\bibinfo{volume}{135}},
  \bibinfo{pages}{236702} (\bibinfo{year}{2025}).

\bibitem[{\citenamefont{Fu et~al.}(2026)\citenamefont{Fu, Mondal, Liu, Tanaka,
  and Cayao}}]{fu2026floquet}
\bibinfo{author}{\bibfnamefont{P.-H.} \bibnamefont{Fu}},
  \bibinfo{author}{\bibfnamefont{S.}~\bibnamefont{Mondal}},
  \bibinfo{author}{\bibfnamefont{J.-F.} \bibnamefont{Liu}},
  \bibinfo{author}{\bibfnamefont{Y.}~\bibnamefont{Tanaka}}, \bibnamefont{and}
  \bibinfo{author}{\bibfnamefont{J.}~\bibnamefont{Cayao}},
  \bibinfo{journal}{Physical Review Letters} \textbf{\bibinfo{volume}{136}},
  \bibinfo{pages}{066703} (\bibinfo{year}{2026}).

\bibitem[{\citenamefont{Liu et~al.}(2026{\natexlab{b}})\citenamefont{Liu,
  Zhuang, Zhu, Wu, and Yan}}]{liu2026light}
\bibinfo{author}{\bibfnamefont{D.}~\bibnamefont{Liu}},
  \bibinfo{author}{\bibfnamefont{Z.-Y.} \bibnamefont{Zhuang}},
  \bibinfo{author}{\bibfnamefont{D.}~\bibnamefont{Zhu}},
  \bibinfo{author}{\bibfnamefont{Z.}~\bibnamefont{Wu}}, \bibnamefont{and}
  \bibinfo{author}{\bibfnamefont{Z.}~\bibnamefont{Yan}},
  \bibinfo{journal}{Physical Review B} \textbf{\bibinfo{volume}{113}},
  \bibinfo{pages}{L060409} (\bibinfo{year}{2026}{\natexlab{b}}).

\bibitem[{\citenamefont{Zhu et~al.}(2025)\citenamefont{Zhu, Zhou, Wang, Wei,
  and Ruan}}]{zhu2025floquet}
\bibinfo{author}{\bibfnamefont{T.}~\bibnamefont{Zhu}},
  \bibinfo{author}{\bibfnamefont{D.}~\bibnamefont{Zhou}},
  \bibinfo{author}{\bibfnamefont{H.}~\bibnamefont{Wang}},
  \bibinfo{author}{\bibfnamefont{S.-H.} \bibnamefont{Wei}}, \bibnamefont{and}
  \bibinfo{author}{\bibfnamefont{J.}~\bibnamefont{Ruan}},
  \bibinfo{journal}{arXiv preprint arXiv:2508.02542}  (\bibinfo{year}{2025}).

\bibitem[{\citenamefont{Li et~al.}(2026)\citenamefont{Li, Wang, Deng, and
  Duan}}]{li2026rkky}
\bibinfo{author}{\bibfnamefont{Y.-X.} \bibnamefont{Li}},
  \bibinfo{author}{\bibfnamefont{R.-Q.} \bibnamefont{Wang}},
  \bibinfo{author}{\bibfnamefont{M.-X.} \bibnamefont{Deng}}, \bibnamefont{and}
  \bibinfo{author}{\bibfnamefont{H.-J.} \bibnamefont{Duan}},
  \bibinfo{journal}{arXiv preprint arXiv:2601.09303}  (\bibinfo{year}{2026}).

\bibitem[{\citenamefont{Ke et~al.}(2024)\citenamefont{Ke, Asmar, and
  Tse}}]{ke2024floquet}
\bibinfo{author}{\bibfnamefont{M.}~\bibnamefont{Ke}},
  \bibinfo{author}{\bibfnamefont{M.~M.} \bibnamefont{Asmar}}, \bibnamefont{and}
  \bibinfo{author}{\bibfnamefont{W.-K.} \bibnamefont{Tse}},
  \bibinfo{journal}{Physical Review B} \textbf{\bibinfo{volume}{110}},
  \bibinfo{pages}{035307} (\bibinfo{year}{2024}).

\bibitem[{\citenamefont{Hayami and Motome}(2023)}]{Hayami2023TRSB}
\bibinfo{author}{\bibfnamefont{S.}~\bibnamefont{Hayami}} \bibnamefont{and}
  \bibinfo{author}{\bibfnamefont{Y.}~\bibnamefont{Motome}},
  \bibinfo{journal}{Nature Communications} \textbf{\bibinfo{volume}{14}},
  \bibinfo{pages}{1} (\bibinfo{year}{2023}).

\bibitem[{\citenamefont{Yarmohammadi et~al.}(2026)\citenamefont{Yarmohammadi,
  Berritta, Bukov, {\v{S}}mejkal, Linder, and Oppeneer}}]{yarmohammadi2026spin}
\bibinfo{author}{\bibfnamefont{M.}~\bibnamefont{Yarmohammadi}},
  \bibinfo{author}{\bibfnamefont{M.}~\bibnamefont{Berritta}},
  \bibinfo{author}{\bibfnamefont{M.}~\bibnamefont{Bukov}},
  \bibinfo{author}{\bibfnamefont{L.}~\bibnamefont{{\v{S}}mejkal}},
  \bibinfo{author}{\bibfnamefont{J.}~\bibnamefont{Linder}}, \bibnamefont{and}
  \bibinfo{author}{\bibfnamefont{P.~M.} \bibnamefont{Oppeneer}},
  \bibinfo{journal}{Physical Review B} \textbf{\bibinfo{volume}{113}},
  \bibinfo{pages}{L060403} (\bibinfo{year}{2026}).

\bibitem[{\citenamefont{Tamang et~al.}(2024)\citenamefont{Tamang, Verma, and
  Biswas}}]{Tamang2024OrbitalMagnetization}
\bibinfo{author}{\bibfnamefont{L.}~\bibnamefont{Tamang}},
  \bibinfo{author}{\bibfnamefont{S.}~\bibnamefont{Verma}}, \bibnamefont{and}
  \bibinfo{author}{\bibfnamefont{T.}~\bibnamefont{Biswas}},
  \bibinfo{journal}{Physical Review B} \textbf{\bibinfo{volume}{110}},
  \bibinfo{pages}{165426} (\bibinfo{year}{2024}).

\bibitem[{\citenamefont{Tamang and
  Biswas}(2023{\natexlab{b}})}]{Tamang2023TopologicalAlphaT3}
\bibinfo{author}{\bibfnamefont{L.}~\bibnamefont{Tamang}} \bibnamefont{and}
  \bibinfo{author}{\bibfnamefont{T.}~\bibnamefont{Biswas}},
  \bibinfo{journal}{Physical Review B} \textbf{\bibinfo{volume}{107}},
  \bibinfo{pages}{085408} (\bibinfo{year}{2023}{\natexlab{b}}).

\bibitem[{\citenamefont{Nascimento et~al.}(2025)\citenamefont{Nascimento,
  Cunha, Paz, Costa~Filho, Pereira, Peeters, and
  da~Costa}}]{Nascimento2025ChiralAlphaT3}
\bibinfo{author}{\bibfnamefont{J.~P.~G.} \bibnamefont{Nascimento}},
  \bibinfo{author}{\bibfnamefont{S.~M.} \bibnamefont{Cunha}},
  \bibinfo{author}{\bibfnamefont{M.~L.~A.} \bibnamefont{Paz}},
  \bibinfo{author}{\bibfnamefont{R.~N.} \bibnamefont{Costa~Filho}},
  \bibinfo{author}{\bibfnamefont{J.~M.} \bibnamefont{Pereira}},
  \bibinfo{author}{\bibfnamefont{F.~M.} \bibnamefont{Peeters}},
  \bibnamefont{and} \bibinfo{author}{\bibfnamefont{D.~R.}
  \bibnamefont{da~Costa}}, \bibinfo{journal}{Physical Review B}
  \textbf{\bibinfo{volume}{112}}, \bibinfo{pages}{125410}
  (\bibinfo{year}{2025}).

\bibitem[{\citenamefont{Cortie et~al.}(2020)\citenamefont{Cortie, Causer, Rule,
  Fritzsche, Kreuzpaintner, and Klose}}]{cortie2020two}
\bibinfo{author}{\bibfnamefont{D.~L.} \bibnamefont{Cortie}},
  \bibinfo{author}{\bibfnamefont{G.~L.} \bibnamefont{Causer}},
  \bibinfo{author}{\bibfnamefont{K.~C.} \bibnamefont{Rule}},
  \bibinfo{author}{\bibfnamefont{H.}~\bibnamefont{Fritzsche}},
  \bibinfo{author}{\bibfnamefont{W.}~\bibnamefont{Kreuzpaintner}},
  \bibnamefont{and} \bibinfo{author}{\bibfnamefont{F.}~\bibnamefont{Klose}},
  \bibinfo{journal}{Advanced Functional Materials}
  \textbf{\bibinfo{volume}{30}}, \bibinfo{pages}{1901414}
  (\bibinfo{year}{2020}).

\bibitem[{\citenamefont{{\v{Z}}uti{\'c}
  et~al.}(2004)\citenamefont{{\v{Z}}uti{\'c}, Fabian, and
  Sarma}}]{vzutic2004spintronics}
\bibinfo{author}{\bibfnamefont{I.}~\bibnamefont{{\v{Z}}uti{\'c}}},
  \bibinfo{author}{\bibfnamefont{J.}~\bibnamefont{Fabian}}, \bibnamefont{and}
  \bibinfo{author}{\bibfnamefont{S.~D.} \bibnamefont{Sarma}},
  \bibinfo{journal}{Reviews of modern physics} \textbf{\bibinfo{volume}{76}},
  \bibinfo{pages}{323} (\bibinfo{year}{2004}).

\bibitem[{\citenamefont{Bai et~al.}(2024)\citenamefont{Bai, Feng, Liu,
  {\v{S}}mejkal, Mokrousov, and Yao}}]{bai2024altermagnetism}
\bibinfo{author}{\bibfnamefont{L.}~\bibnamefont{Bai}},
  \bibinfo{author}{\bibfnamefont{W.}~\bibnamefont{Feng}},
  \bibinfo{author}{\bibfnamefont{S.}~\bibnamefont{Liu}},
  \bibinfo{author}{\bibfnamefont{L.}~\bibnamefont{{\v{S}}mejkal}},
  \bibinfo{author}{\bibfnamefont{Y.}~\bibnamefont{Mokrousov}},
  \bibnamefont{and} \bibinfo{author}{\bibfnamefont{Y.}~\bibnamefont{Yao}},
  \bibinfo{journal}{Advanced Functional Materials}
  \textbf{\bibinfo{volume}{34}}, \bibinfo{pages}{2409327}
  (\bibinfo{year}{2024}).

\bibitem[{\citenamefont{Fu et~al.}(2025)\citenamefont{Fu, Lv, Xu, Cayao, Liu,
  and Yu}}]{fu2025all}
\bibinfo{author}{\bibfnamefont{P.-H.} \bibnamefont{Fu}},
  \bibinfo{author}{\bibfnamefont{Q.}~\bibnamefont{Lv}},
  \bibinfo{author}{\bibfnamefont{Y.}~\bibnamefont{Xu}},
  \bibinfo{author}{\bibfnamefont{J.}~\bibnamefont{Cayao}},
  \bibinfo{author}{\bibfnamefont{J.-F.} \bibnamefont{Liu}}, \bibnamefont{and}
  \bibinfo{author}{\bibfnamefont{X.-L.} \bibnamefont{Yu}},
  \bibinfo{journal}{npj Quantum Materials}  (\bibinfo{year}{2025}).

\bibitem[{\citenamefont{Jungwirth et~al.}(2016)\citenamefont{Jungwirth, Marti,
  Wadley, and Wunderlich}}]{jungwirth2016antiferromagnetic}
\bibinfo{author}{\bibfnamefont{T.}~\bibnamefont{Jungwirth}},
  \bibinfo{author}{\bibfnamefont{X.}~\bibnamefont{Marti}},
  \bibinfo{author}{\bibfnamefont{P.}~\bibnamefont{Wadley}}, \bibnamefont{and}
  \bibinfo{author}{\bibfnamefont{J.}~\bibnamefont{Wunderlich}},
  \bibinfo{journal}{Nature nanotechnology} \textbf{\bibinfo{volume}{11}},
  \bibinfo{pages}{231} (\bibinfo{year}{2016}).

\bibitem[{\citenamefont{Krempask{\`y} et~al.}(2024)\citenamefont{Krempask{\`y},
  {\v{S}}mejkal, D’souza, Hajlaoui, Springholz, Uhl{\'\i}{\v{r}}ov{\'a},
  Alarab, Constantinou, Strocov, Usanov et~al.}}]{krempasky2024altermagnetic}
\bibinfo{author}{\bibfnamefont{J.}~\bibnamefont{Krempask{\`y}}},
  \bibinfo{author}{\bibfnamefont{L.}~\bibnamefont{{\v{S}}mejkal}},
  \bibinfo{author}{\bibfnamefont{S.}~\bibnamefont{D’souza}},
  \bibinfo{author}{\bibfnamefont{M.}~\bibnamefont{Hajlaoui}},
  \bibinfo{author}{\bibfnamefont{G.}~\bibnamefont{Springholz}},
  \bibinfo{author}{\bibfnamefont{K.}~\bibnamefont{Uhl{\'\i}{\v{r}}ov{\'a}}},
  \bibinfo{author}{\bibfnamefont{F.}~\bibnamefont{Alarab}},
  \bibinfo{author}{\bibfnamefont{P.}~\bibnamefont{Constantinou}},
  \bibinfo{author}{\bibfnamefont{V.}~\bibnamefont{Strocov}},
  \bibinfo{author}{\bibfnamefont{D.}~\bibnamefont{Usanov}},
  \bibnamefont{et~al.}, \bibinfo{journal}{Nature}
  \textbf{\bibinfo{volume}{626}}, \bibinfo{pages}{517} (\bibinfo{year}{2024}).

\bibitem[{\citenamefont{Takahashi and Maekawa}(2003)}]{takahashi2003spin}
\bibinfo{author}{\bibfnamefont{S.}~\bibnamefont{Takahashi}} \bibnamefont{and}
  \bibinfo{author}{\bibfnamefont{S.}~\bibnamefont{Maekawa}},
  \bibinfo{journal}{Physical Review B} \textbf{\bibinfo{volume}{67}},
  \bibinfo{pages}{052409} (\bibinfo{year}{2003}).

\bibitem[{\citenamefont{Mamin et~al.}(2003)\citenamefont{Mamin, Budakian, Chui,
  and Rugar}}]{mamin2003detection}
\bibinfo{author}{\bibfnamefont{H.~J.} \bibnamefont{Mamin}},
  \bibinfo{author}{\bibfnamefont{R.}~\bibnamefont{Budakian}},
  \bibinfo{author}{\bibfnamefont{B.~W.} \bibnamefont{Chui}}, \bibnamefont{and}
  \bibinfo{author}{\bibfnamefont{D.}~\bibnamefont{Rugar}},
  \bibinfo{journal}{Physical Review Letters} \textbf{\bibinfo{volume}{91}},
  \bibinfo{pages}{207604} (\bibinfo{year}{2003}).

\end{thebibliography}
\end{document}